\documentclass[aps,pra,twocolumn,superscriptaddress,10pt,nofootinbib,longbibliography,floatfix,footinbib]{revtex4-1}

\usepackage[utf8]{inputenc}
\usepackage[TS1,T1]{fontenc}
\usepackage{microtype}
\usepackage[english]{babel}
\usepackage{amsmath}
\usepackage{amssymb}
\usepackage{float}
\usepackage{graphicx}
\usepackage[breaklinks=true, colorlinks, pdfpagelabels, hidelinks]{hyperref} %
\usepackage{cleveref}
\usepackage{dsfont}
\usepackage{fixmath}
\usepackage{tensor}
\usepackage{braket}
\usepackage{paralist}
\usepackage{upgreek}
\usepackage{color}
\usepackage[dvipsnames]{xcolor}
\usepackage{booktabs}
\usepackage{multirow}
\usepackage{dcolumn}
\usepackage{siunitx}
\usepackage[version=4]{mhchem}
\usepackage{makecell}
\usepackage[caption=false]{subfig}
\usepackage{array}
\usepackage{tabularx}

\newcommand{\pare}[1]{\left( {#1} \right)}
\newcommand{\spare}[1]{\left[ {#1} \right]}
\newcommand{\cpare}[1]{\left\{ {#1} \right\}}
\newcommand{\hc}{\text{H.c.}}
\newcommand{\cc}{\text{c.c.}}
\newcommand{\tr}{\textrm{tr}}
\newcommand{\id}{\mathbb{1}}
\newcommand{\curl}{\grad \times }
\newcommand{\grad}{\nabla }
\newcommand{\fd}[2]{\frac{d {#1}}{d {#2}}}
\newcommand{\com}[2]{\left[ {#1},{#2} \right]}
\newcommand{\aop}{\hat a}
\newcommand{\acom}[2]{\left\{ {#1},{#2} \right\}}
\newcommand*\dd{\mathop{}\!\mathrm{d}}
\renewcommand{\vec}[1]{\mathbold{#1}}
\newcommand{\im}{\text{i}}
\let\Im\relax
\DeclareMathOperator{\Im}{Im}
\let\Re\relax
\DeclareMathOperator{\Re}{Re}
\newcommand{\tens}[1]{\vec{#1}}
\newcommand{\op}[1]{\hat{#1}}
\newcommand{\ccsymbol}{*}
\newcommand{\cconj}[1]{{#1}^\ccsymbol}
\newcommand{\hcsymbol}{\dagger}
\newcommand{\hconj}[1]{{#1}^\hcsymbol}

\newcommand{\kronecker}{\delta}

\newcommand{\signum}{\sigma}

\newcommand{\order}{\mathcal{O}}
\newcommand{\besselJ}[1]{J_{#1}}
\newcommand{\besselY}[1]{Y_{#1}}
\newcommand{\besselI}[1]{I_{#1}}
\newcommand{\besselK}[1]{K_{#1}}

\newcommand{\unitvec}{\vec{e}}

\newcommand{\xpos}{x}
\newcommand{\ypos}{y}
\newcommand{\zpos}{z}
\newcommand{\exvec}{\unitvec_\xpos}
\newcommand{\eyvec}{\unitvec_\ypos}
\newcommand{\ezvec}{\unitvec_\zpos}
\newcommand{\rpos}{r}
\newcommand{\phipos}{\varphi}
\newcommand{\ervec}{\unitvec_\rpos}
\newcommand{\ephivec}{\unitvec_\phipos}
\newcommand{\Lagdens}{\mathcal{L}}

\newcommand{\Hamilfunc}{H}

\newcommand{\boltzmann}{k_B}
\newcommand{\mel}{m_\text{el}}
\newcommand{\elcharge}{e}
\newcommand{\body}{B}

\newcommand{\normveccomp}{n}
\newcommand{\normvec}{{\vec{\normveccomp}}}
\newcommand{\zerovec}{\boldsymbol{0}}
\newcommand{\temp}{T}

\newcommand{\atomic}{\text{at}}
\newcommand{\internal}{\text{int}}

\newcommand{\optical}{\text{opt}}

\newcommand{\vibrational}{\text{phn}}
\newcommand{\thermal}{\text{th}}
\newcommand{\atphonint}{{\atomic\text{-}\vibrational}}
\newcommand{\radpress}{\text{dp}}
\newcommand{\strainopt}{\text{st}}

\newcommand{\continuous}{\text{c}}
\newcommand{\discrete}{\text{d}}
\newcommand{\offresonant}{\text{nres}}
\newcommand{\resonant}{\text{res}}
\newcommand{\positionsymbol}{r}
\newcommand{\pos}{\vec{\positionsymbol}}

\newcommand{\tm}{t}

\newcommand{\post}{\pos,\tm}

\newcommand{\rad}{R}
\newcommand{\len}{L}
\newcommand{\clamplen}{\len}

\newcommand{\freq}{\omega}
\newcommand{\wavelen}{\lambda}
\newcommand{\Efieldcomp}{E}
\newcommand{\Bfieldcomp}{B}
\newcommand{\Efield}{\vec{\Efieldcomp}}
\newcommand{\Bfield}{\vec{\Bfieldcomp}}

\newcommand{\vecpotcomp}{A}
\newcommand{\vecpot}{\vec{\vecpotcomp}}

\newcommand{\poynting}{\vec{S}}
\newcommand{\power}{P}
\newcommand{\intensity}{I}
\newcommand{\cvac}{c}
\newcommand{\cbody}{v}
\newcommand{\permitt}{\epsilon}
\newcommand{\permeab}{\mu}
\newcommand{\vacpermitt}{\permitt_0}
\newcommand{\vacpermeab}{\permeab_0}
\newcommand{\relpermitt}{\permitt}
\newcommand{\relpermeab}{\permeab}
\newcommand{\relpermitttenscomp}{\relpermitt}
\newcommand{\relpermitttens}{\tens{\relpermitttenscomp}}
\newcommand{\relpermeabtens}{\tens{\relpermeab}}
\newcommand{\rootrelpermitt}{\rho}
\newcommand{\rootrelpermitttens}{\tens{\rootrelpermitt}}
\newcommand{\amodecomp}{a}
\newcommand{\amode}{\vec{\amodecomp}}
\newcommand{\amodercomp}{a}
\newcommand{\amoder}{\vec{\amodercomp}}
\newcommand{\emodecomp}{e}
\newcommand{\bmodecomp}{b}
\newcommand{\emode}{\vec{\emodecomp}}
\newcommand{\bmode}{\vec{\bmodecomp}}
\newcommand{\emodercomp}{\mathcal{E}}
\newcommand{\bmodercomp}{\mathcal{B}}
\newcommand{\emoder}{\boldsymbol{\emodercomp}}
\newcommand{\bmoder}{\boldsymbol{\bmodercomp}}
\newcommand{\photev}{d}
\newcommand{\photindex}{\eta}

\newcommand{\photindexb}{{\photindex'}}

\newcommand{\photnormvar}{\alpha}

\newcommand{\photfreq}{\omega}

\newcommand{\ufieldcomp}{u}
\newcommand{\ufield}{\vec{\ufieldcomp}}

\newcommand{\udotfield}{\dot{\ufield}}

\newcommand{\uddotfield}{\ddot{\ufield}}
\newcommand{\dens}{\rho}
\newcommand{\elastenscomp}{C}
\newcommand{\elastens}{\tens{\elastenscomp}}
\newcommand{\YoungE}{E}
\newcommand{\Poissonnu}{\nu}
\newcommand{\lamemu}{\mu}
\newcommand{\lamelambda}{\lambda}
\newcommand{\pifieldcomp}{\pi}
\newcommand{\pifield}{\vec{\pifieldcomp}}

\newcommand{\straintenscomp}{S}
\newcommand{\straintens}{\tens{\straintenscomp}}
\newcommand{\stresstenscomp}{T}
\newcommand{\stresstens}{\tens{\stresstenscomp}}

\newcommand{\Dphon}{\mathcal{D}}

\newcommand{\wmodecomp}{w}
\newcommand{\wmode}{\vec{\wmodecomp}}
\newcommand{\wmodercomp}{\mathcal{W}}
\newcommand{\wmoder}{\boldsymbol{\wmodercomp}}
\newcommand{\wmoderA}{\wmoder^a}
\newcommand{\wmoderB}{\wmoder^b}
\newcommand{\wmoderC}{\wmoder^c}
\newcommand{\wmoderAicomp}{\wmodercomp^{ka}}
\newcommand{\wmoderBicomp}{\wmodercomp^{kb}}
\newcommand{\wmoderCicomp}{\wmodercomp^{kc}}

\newcommand{\strainmodecomp}{s}
\newcommand{\strainmode}{\vec{\strainmodecomp}}
\newcommand{\strainmodercomp}{\mathcal{S}}
\newcommand{\strainmoder}{\boldsymbol{\strainmodercomp}}
\newcommand{\stressmodecomp}{t}
\newcommand{\stressmode}{\vec{\stressmodecomp}}
\newcommand{\stressmodercomp}{\mathcal{T}}
\newcommand{\stressmoder}{\boldsymbol{\stressmodercomp}}

\newcommand{\phononBCmatrixcomp}{M}
\newcommand{\phononBCmatrix}{\tens{M}}

\newcommand{\phonindex}{\gamma}

\newcommand{\phonindexb}{{\gamma'}}

\newcommand{\phonfreq}{\omega}
\newcommand{\phonnormvar}{\beta}

\newcommand{\ufieldmodedens}{\mathcal{U}}
\newcommand{\pifieldmodedens}{\Pi}

\newcommand{\trans}{t}
\newcommand{\longitud}{l}
\newcommand{\clong}{c_\longitud}
\newcommand{\ctrans}{c_\trans}
\newcommand{\cFL}{c_h}

\newcommand{\pockelstenscomp}{P}
\newcommand{\photela}{P_1}
\newcommand{\photelb}{P_2}
\newcommand{\photelc}{P_3}
\newcommand{\pockelstens}{\tens{\pockelstenscomp}}

\newcommand{\prtrbd}[1]{\bar{#1}}
\newcommand{\photonoverlap}{\mathcal{A}}

\newcommand{\pot}{V}
\newcommand{\scalarshift}{\text{s}}
\newcommand{\vectorshift}{\text{v}}
\newcommand{\tensorshift}{\text{t}}
\newcommand{\reddetuned}{\text{r}}
\newcommand{\bluedetuned}{\text{b}}
\newcommand{\adsorption}{\text{ad}}

\newcommand{\CPpotstrength}{C}

\newcommand{\trapr}{\atrpos_0}

\newcommand{\trapphi}{\atphipos_0}

\newcommand{\trapz}{\atzpos_0}

\newcommand{\trapy}{\atypos_0}

\newcommand{\trapib}{{\atposbsymbol^i_0}}
\newcommand{\trappos}{\atpos_0}

\newcommand{\trapfreq}{\omega}
\newcommand{\rtrapfreq}{\trapfreq_{\rpos}}
\newcommand{\phitrapfreq}{\trapfreq_{\phipos}}
\newcommand{\ztrapfreq}{\trapfreq_{\zpos}}
\newcommand{\itrapfreq}{\trapfreq_{ i}}
\newcommand{\jtrapfreq}{\trapfreq_{ j}}
\newcommand{\rphitrapfreq}{\trapfreq_{\rpos\phipos}}
\newcommand{\rztrapfreq}{\trapfreq_{\rpos\zpos}}
\newcommand{\phiztrapfreq}{\trapfreq_{\phipos\zpos}}
\newcommand{\ijtrapfreq}{\trapfreq_{ij}}
\newcommand{\ijcrosscoupling}{g_{ij}}
\newcommand{\opticaltrapdepth}{\pot_0}
\newcommand{\atmass}{M}

\newcommand{\atpossymbol}{r}
\newcommand{\atposbsymbol}{x}
\newcommand{\atpos}{\vec{\atpossymbol}}

\newcommand{\atypos}{\ypos}
\newcommand{\atzpos}{\zpos}
\newcommand{\atrpos}{\rpos}
\newcommand{\atphipos}{\phipos}
\newcommand{\atrposb}{\atposbsymbol^\rpos}
\newcommand{\atphiposb}{\atposbsymbol^\phipos}
\newcommand{\atzposb}{\atposbsymbol^\zpos}
\newcommand{\atiposb}{\atposbsymbol^i}

\newcommand{\atmomsymbol}{p}
\newcommand{\atmom}{\vec{\atmomsymbol}}

\newcommand{\polarizab}{\alpha}
\newcommand{\FSpolarizab}{\tilde{\polarizab}}
\newcommand{\HFSpolarizab}{\polarizab}

\newcommand{\HFSscalarpolarizab}{\HFSpolarizab_\scalarshift}
\newcommand{\FSscalarpolarizab}{\FSpolarizab_\scalarshift}
\newcommand{\HFSvectorpolarizab}{\HFSpolarizab_\vectorshift}
\newcommand{\FSvectorpolarizab}{\FSpolarizab_\vectorshift}
\newcommand{\HFStensorpolarizab}{\HFSpolarizab_\tensorshift}
\newcommand{\FStensorpolarizab}{\FSpolarizab_\tensorshift}

\newcommand{\atintn}{n}
\newcommand{\atS}{S}
\newcommand{\atL}{L}
\newcommand{\atintindex}{\lambda}
\newcommand{\atJ}{J}

\newcommand{\atI}{I}
\newcommand{\atF}{F}

\newcommand{\atMF}{M_\atF}
\newcommand{\atFSstate}{\xi}
\newcommand{\atHFSstate}{\atintindex}

\newcommand{\atn}{n}

\newcommand{\ataop}{\hat{a}}

\newcommand{\extBfieldcomp}{B_\text{ext}}
\newcommand{\extBfield}{\vec{B}_\text{ext}}

\newcommand{\zposB}{{\zpos_B}}

\newcommand{\ezvecB}{\vec{z}_B}
\newcommand{\ezvecBcomp}{z_B}
\newcommand{\photk}{k}
\newcommand{\photkb}{\photk'}

\newcommand{\photkR}{\kappa}
\newcommand{\photl}{m}
\newcommand{\photlb}{\photl'}

\newcommand{\photfam}{f}
\newcommand{\photfamb}{\photfam'}

\newcommand{\photn}{n}
\newcommand{\photnb}{n'}

\newcommand{\TEmode}{\text{TE}}
\newcommand{\TMmode}{\text{TM}}
\newcommand{\HEmode}{\text{HE}}
\newcommand{\EHmode}{\text{EH}}

\newcommand{\rposR}{x}
\newcommand{\phota}{a}
\newcommand{\photaR}{\alpha}
\newcommand{\photaa}{\tilde{a}}
\newcommand{\photaaR}{\tilde{\alpha}}
\newcommand{\photb}{b}
\newcommand{\photbR}{\beta}
\newcommand{\photbb}{\tilde{b}}
\newcommand{\photbbR}{\tilde{\beta}}
\newcommand{\photw}{w}

\newcommand{\polarization}{\sigma}

\newcommand{\polindex}{\polarization}
\newcommand{\polindexb}{{\polarization'}}
\newcommand{\phipol}{{\polindex_\phipos}}
\newcommand{\zpol}{{\polindex_\zpos}}
\newcommand{\phipolb}{{\polindex_\phipos'}}
\newcommand{\zpolb}{{\polindex_\zpos'}}
\newcommand{\phiphase}{\theta_\phipos}
\newcommand{\zphase}{\theta_\zpos}
\newcommand{\phonk}{p}

\newcommand{\phonkR}{\varpi}
\newcommand{\phonl}{j}

\newcommand{\phonfam}{f}

\newcommand{\Tmode}{\text{T}}
\newcommand{\Lmode}{\text{L}}
\newcommand{\Fmode}{\text{F}}
\newcommand{\phona}{a}
\newcommand{\phonaR}{\alpha}
\newcommand{\phonaa}{\tilde{\phona}}
\newcommand{\phonaaR}{\tilde{\phonaR}}
\newcommand{\phonb}{b}
\newcommand{\phonbR}{\beta}
\newcommand{\phonbb}{\tilde{\phonb}}
\newcommand{\phonbbR}{\tilde{\phonbR}}
\newcommand{\phonn}{n}

\newcommand{\phondecayrate}{\kappa}
\newcommand{\Hamilop}{\op{\Hamilfunc}}
\newcommand{\potop}{\op\pot}

\newcommand{\dissipator}{\mathcal{D}}
\newcommand{\nbar}{\bar{n}}

\newcommand{\densop}{\op{\rho}}
\newcommand{\densopb}{\op{\sigma}}
\newcommand{\densopc}{\op{\mu}}
\newcommand{\phondensop}{\densopb}
\newcommand{\atdensop}{\densopc}

\newcommand{\phonaop}{\op{b}}
\newcommand{\ufieldop}{\op{\ufield}}

\newcommand{\pifieldop}{\op{\pifield}}

\newcommand{\straintensop}{\op{\straintens}}

\newcommand{\atmomop}{\op\atmom}
\newcommand{\atrmomop}{{\op{\atmomsymbol}^\atrpos}}
\newcommand{\atphimomop}{{\op{\atmomsymbol}^\atphipos}}
\newcommand{\atzmomop}{{\op{\atmomsymbol}^\atzpos}}

\newcommand{\atposop}{\op\atpos}
\newcommand{\atrposop}{\op\atrpos}

\newcommand{\atphiposop}{\op\atphipos}

\newcommand{\atzposop}{\op\atzpos}

\newcommand{\atiposbop}{\op{\atposbsymbol}^i}
\newcommand{\iaop}{\op{a}_i}

\newcommand{\atrpossd}{\Delta \atrpos}
\newcommand{\atrposbsd}{\Delta \atrposb}
\newcommand{\atphipossd}{\Delta \atphipos}
\newcommand{\atphiposbsd}{\Delta \atphiposb}
\newcommand{\atzpossd}{\Delta \atzpos}
\newcommand{\atzposbsd}{\Delta \atzposb}

\newcommand{\atiposbsd}{\Delta \atiposb}

\newcommand{\atphong}{g}
\newcommand{\couplingfunct}{g}

\newcommand{\scatrate}{\Gamma}

\newcommand{\DOS}{\rho}

\newcommand{\frechetD}{D}
\newcommand{\partialfrechetD}{\delta}

\renewcommand{\div}{\nabla \cdot}

\newcommand{\delt}{\partial_t}

\newcommand{\ddelt}{\dot}
\newcommand{\ddeltt}{\ddot}
\newcommand{\Dphot}{\mathcal{D}}
\newcommand{\delx}{{\partial_\xpos}}

\newcommand{\delphi}{{\partial_\phipos}}
\newcommand{\delr}{{\partial_\rpos}}

\newcommand{\dr}{\text{d}\vec{r}}
\newcommand{\C}{\mathds{C}}
\newcommand{\R}{\mathds{R}}
\newcommand{\Z}{\mathds{Z}}
\newcommand{\N}{\mathds{N}}
\newcommand{\inner}[2]{(#1|#2)}

\newcommand{\latin}[1]{\textit{#1}}

\crefname{paragraph}{paragraph}{paragraphs}
\Crefname{equation}{Eq.}{Eqs.}
\newlength{\tmplen}

\newcommand{\subcref}[2]{\cref{#1}#2}
\newcommand{\subCref}[2]{\Cref{#1}#2}
\newcommand{\subcrefand}[3]{\namecrefs{#1}~\ref{#1}#2 and \ref{#1}#3}
\newcommand{\subCrefand}[3]{\nameCrefs{#1}~\ref{#1}#2 and \ref{#1}#3}
\newcommand{\subcrefb}[2]{\namecref{#1}~(#2)}
\newcommand{\subCrefb}[2]{\nameCref{#1}~(#2)}
\newcommand{\subcrefandb}[3]{\namecrefs{#1}~(#2) and (#3)}
\newcommand{\subCrefandb}[3]{\nameCrefs{#1}~(#2) and (#3)}
\newcommand{\subCreftob}[3]{\nameCrefs{#1}~(#2)-(#3)}
\newlength{\widthFigA}
\newlength{\marginLeftFigA}
\newlength{\marginRightFigA}
\newlength{\widthFigB}
\newlength{\marginLeftFigB}
\newlength{\marginRightFigB}
\newlength{\widthFigC}
\newlength{\marginLeftFigC}
\newlength{\marginRightFigC}
\newlength{\widthFigD}
\newlength{\marginLeftFigD}
\newlength{\marginRightFigD}

\hypersetup{final}

\begin{document}
\selectlanguage{english}

\title{Heating in Nanophotonic Traps for Cold Atoms}

\author{Daniel Hümmer}
\affiliation{Institute for Quantum Optics and Quantum Information of the Austrian Academy of Sciences, 6020 Innsbruck, Austria}
\affiliation{Institute for Theoretical Physics, University of Innsbruck, 6020 Innsbruck, Austria}
\author{Philipp Schneeweiss}
\affiliation{Atominstitut, TU Wien, 1020 Vienna, Austria}
\author{Arno Rauschenbeutel}
\affiliation{Atominstitut, TU Wien, 1020 Vienna, Austria}
\affiliation{Department of Physics, Humboldt-Universität zu Berlin, 10099 Berlin, Germany}
\author{Oriol Romero-Isart}
\affiliation{Institute for Quantum Optics and Quantum Information of the Austrian Academy of Sciences, 6020 Innsbruck, Austria}
\affiliation{Institute for Theoretical Physics, University of Innsbruck, 6020 Innsbruck, Austria}
\date{\today}

\begin{abstract}
Laser-cooled atoms that are trapped and optically interfaced with light in nanophotonic waveguides are a powerful platform for fundamental research in quantum optics as well as for applications in quantum communication and quantum-information processing. Ever since the first realization of such a hybrid quantum-nanophotonic system about a decade ago, heating rates of the atomic motion observed in various experimental settings have typically been exceeding those in comparable free-space optical microtraps by about three orders of magnitude. This excessive heating is a roadblock for the implementation of certain protocols and devices. Still, its origin has so far remained elusive and, at the typical atom-surface separations of less than an optical wavelength encountered in nanophotonic traps, numerous effects may potentially contribute to atom heating. Here, we theoretically describe the effect of mechanical vibrations of waveguides on guided light fields and provide a general theory of particle-phonon interaction in nanophotonic traps. We test our theory by applying it to the case of laser-cooled cesium atoms in nanofiber-based two-color optical traps. We find excellent quantitative agreement between the predicted heating rates and experimentally measured values. Our theory predicts that, in this setting, the dominant heating process stems from the optomechanical coupling of the optically trapped atoms to the continuum of thermally occupied flexural mechanical modes of the waveguide structure. Surprisingly, the effect of the high-$Q$ mechanical resonances which have previously been observed in this system can be neglected, even if they coincide with the trap frequencies. Beyond unraveling the long-standing riddle of excessive heating in nanofiber-based atom traps, we also study the dependence of the heating rates on the relevant system parameters and find a strong $R^{-5/2}$ scaling with the inverse waveguide radius. Our findings allow us to propose several strategies for minimizing the heating which also provide guidelines for the design of next-generation nanophotonic cold-atom systems. Finally, given that the predicted heating rate is proportional to the mass of the trapped particle, our findings are also highly relevant for optomechanics experiments with dielectric nanoparticles that are optically trapped close to nanophotonic waveguides.
\end{abstract}

\maketitle

\section*{Introduction}
Small particles, such as laser-cooled atoms or dielectric nanospheres, are nowadays routinely trapped at submicron distances from solids. Structures currently investigated include photonic crystal waveguides~\cite{thompson_coupling_2013,goban_superradiance_2015,magrini_near-field_2018}, optical nanofibers~\cite{vetsch_optical_2010,goban_demonstration_2012,beguin_generation_2014,kato_strong_2015,lee_inhomogeneous_2015,corzo_large_2016}, single carbon-nanotubes~\cite{gierling_cold-atom_2011,schneeweiss_dispersion_2012}, dielectric membranes~\cite{diehl_optical_2018}, and even macroscopic prisms~\cite{hammes_cold-atom_2002,bender_probing_2014}. The opportunities in research and application for systems combining atoms and solids are numerous, including the search for novel fundamental forces~\cite{geraci_improved_2008,geraci_short-range_2010,arkanihamed_hierarchy_1998,dalvit_casimir_2011,klimchitskaya_casimir_2009}, the implementation of quantum metrology and sensing using collective atomic state entanglement~\cite{beguin_observation_2018}, and integrated quantum memories for photons guided in nanoscale waveguides~\cite{sayrin_storage_2015,gouraud_demonstration_2015,corzo_waveguide-coupled_2019}. A rich toolbox is already available for the cooling, trapping, positioning, and probing of atoms and nanoparticles. However, not all techniques commonly used in free-space traps for manipulating trapped particles are compatible with the presence of solid structures in their immediate proximity: Control laser beams, for instance, may be reflected or scattered in undesired ways. Moreover, additional effects such as van der Waals forces or coupling of the atoms or particles to thermal excitations in the solid have to be considered.

Full control at the quantum level over the internal as well as external degrees of freedom of individual atoms coupled to a nanophotonic structure was achieved only recently~\cite{meng_near-ground-state_2018}. A key challenge in this context is the  heating of the atomic motion observed in these systems~\cite{reitz_coherence_2013,albrecht_fictitious_2016} which can reach rates of several hundred motional quanta per second -- about three orders of magnitude larger than in comparable free-space optical traps. Large cooling rates realized, for example, by ultrastrong spin-motion coupling~\cite{schneeweiss_cold-atom-based_2018,dareau_observation_2018}, are required to overcome the heating and prepare atoms close to their motional ground state. In essence, the observed storage times of atoms in nanophotonic traps have fallen short of expectations, both for trapped cesium~\cite{goban_demonstration_2012,beguin_generation_2014,kato_strong_2015,corzo_large_2016,goban_superradiance_2015} and rubidium~\cite{lee_inhomogeneous_2015} atoms, ever since the first implementation of a nanofiber-based trap for laser-cooled atoms~\cite{vetsch_optical_2010}. The origin of the strong heating and the corresponding low lifetimes has so far remained elusive. There is a range of conceivable causes, such as Raman scattering of the trapping light fields in the waveguide material \cite{engelbrecht_nichtlineare_2015}, Brillouin scattering~\cite{beugnot_brillouin_2014,florez_brillouin_2016}, or Johnson-Nyquist noise~\cite{henkel_loss_1999}. However, estimates of their effect, provided as supplemental material %
\footnote{See supplemental material at the end of this article for estimates of the contribution of other mechanisms to the heating of nanofiber-trapped cold atoms}, %
demonstrate that these mechanisms fail to explain heating rates observed in experiments. Additionally, tapered optical fibers, as used for realizing nanofiber-based cold-atom traps, exhibit thermally driven high-$Q$ torsional mechanical resonances which have been considered as a likely candidate for explaining the large heating in these systems~\cite{wuttke_optically_2013}. In contrast, optical traps that are based on the evanescent field above a prism surface seem to feature small heating rates which are compatible, for instance, with Bose-Einstein condensation of cesium atoms~\cite{rychtarik_two-dimensional_2004}. Indeed, even at room temperature, one does not expect thermally excited phonon modes of the macroscopic prism to contribute to the heating of the trapped atoms~\cite{henkel_heating_1999}.

Here, we identify thermally populated flexural phononic modes of the nanoscopic waveguide as the dominant contributor to the large heating rates observed in nanofiber-based cold-atom traps. We give a concise description of the effect of mechanical modes on light guided in optical waveguides and provide a general theory of the resulting atom-phonon interaction in nanophotonic traps. Based on this formalism, we perform a case study for the cesium two-color nanofiber-based trap described in refs.~\cite{vetsch_optical_2010,reitz_coherence_2013,albrecht_fictitious_2016}. Relying on independently measured system properties, we predict heating rates in excellent quantitative agreement with experimental observations. Surprisingly, the effect of the high-$Q$ torsional mechanical resonances that have previously been observed in this system~\cite{wuttke_optically_2013} can be neglected, even if they coincide with the trap frequencies. We then use our model to numerically and analytically infer the scaling of the heating rates with system parameters such as the mechanical properties of the fiber, its temperature, or the trap frequencies. This systematic analysis allows us to outline strategies for minimizing the heating, thereby suggesting a solution to a long-standing problem of nanofiber-based cold-atom systems. While we formulate our theory in terms of atoms near nanofibers, it is indeed applicable to any kind of polarizable object trapped by conservative forces due to the light field surrounding a photonic structure. Building on the agreement obtained in the case study, our quantitative formalism might therefore be used for the faithful description of other nanophotonic cold-atom systems and, more generally, optomechanical systems with small particles, such as dielectric nanospheres \cite{magrini_near-field_2018,chang_cavity_2010,romero-isart_toward_2010,li_millikelvin_2011,gieseler_subkelvin_2012,kiesel_cavity_2013,fonseca_nonlinear_2016,jain_direct_2016}, trapped in close vicinity to hot solid bodies.

This article is structured as follows: In \cref{sec: framework}, we provide a general quantum theory describing atoms trapped in the optical near field of a vibrating photonic structure. In particular, we derive the general form of the atom-phonon interaction and discuss the resulting heating rates of the atomic motion. \Cref{sec: case study} is dedicated to a case study of heating rates expected in a nanofiber-based two-color trap for laser-cooled atoms. In \cref{sec: photon appendix}, we review the concept of photonic eigenmodes and summarize the modes of a nanofiber, while \cref{sec: atom appendix} recapitulates the resulting forces acting on trapped atoms. In \cref{sec: phonon appendix}, we review quantized linear elastodynamics and summarize the phononic eigenmodes of a nanofiber. In \Cref{sec: interaction appendix}, we supply details on how to calculate the atom-phonon coupling constants based on the framework presented in \cref{sec: photon appendix,sec: atom appendix,sec: phonon appendix}. The parameters of the experimental setup considered in the case study are listed in \cref{sec: case study appendix}.

\section{Atoms Trapped near Vibrating Photonic Structures}
\label{sec: framework}
Micro- and nanophotonic traps rely on the optical near fields surrounding a photonic structure to spatially confine laser-cooled atoms in high vacuum. The optical fields are detuned from resonances of the atom such that they do not drive transitions between its internal (electronic) states. Confinement is achieved through gradients in the electric field that result in optical forces acting on the atom, analogous to free-space optical dipole traps \cite{grimm_optical_2000}. In contrast to free-space setups, a dielectric photonic structure is used to pattern laser light in a way that creates local minima suitable for trapping atoms in the optical potential \cite{chang_colloquium_2018}.
The light can either be guided by the structure such that atoms interact with the evanescent fields surrounding it \cite{mabuchi_atom_1994,dowling_evanescent_1996,vernooy_quantum_1997,le_kien_atom_2004,vetsch_optical_2010,christensen_trapping_2008,goban_demonstration_2012,hung_trapped_2013,goban_atom-light_2014}, or scattered by the structure \cite{ovchinnikov_atomic_1991,le_kien_microtraps_2009,thompson_coupling_2013,goban_superradiance_2015}; see \cref{sec: photon appendix}. In either case a fraction of the light is absorbed, which can lead to a bulk temperature of the dielectric of several hundred kelvins due to the weak thermal coupling to its environment \cite{wuttke_thermalization_2013}. In consequence, mechanical modes of the photonic structure are thermally excited. These mechanical modes (phonons) are in turn coupled to the external (motional) state of trapped atoms through the optical forces and other forces acting between the atoms and the structure.

\medskip{}

An individual atom trapped in the optical near field surrounding a mechanically vibrating photonic structure suspended in high vacuum can be modeled by the Hamiltonian
\begin{equation}\label{eqn: total Hamiltonian}
  \Hamilop = \Hamilop_\atomic  + \Hamilop_\vibrational + \Hamilop_\atphonint~.
\end{equation}
The first term describes the dynamics of the trapped atom in the absence of phonons. Atoms are trapped at a distance of a few hundred nanometers from the surface of the structure because the near fields decay on a scale given by the optical wavelength. At such distances, corrections $\potop_\adsorption$ to the optical potential $\potop_\optical$ due to surface effects like dispersion forces become relevant \cite{le_kien_atom_2004,buhmann_dispersion_2012}. Optical forces and dispersion forces are additive to first order \cite{fuchs_nonadditivity_2018}; hence, the total potential experienced by the atom is $\potop_0 \equiv \potop_\optical + \potop_\adsorption$. While the potential in general couples all atomic degrees of freedom \cite{dareau_observation_2018,meng_near-ground-state_2018}, we focus on scenarios without coupling of electronic and motional states and assume that the atom does not change its internal state. In this case $\potop_0 = \pot_0(\atposop)$; that is, the center of mass of the atom is subject to a potential $\pot_0$ which depends on the internal state of the atom (see \cref{sec: atom appendix}). Approximating the potential as harmonic for an atom close to its trapped motional ground state yields the atom Hamiltonian
\begin{equation}\label{eqn: atom Hamiltonian harmonic approximation}
  \Hamilop_\atomic \equiv \sum_i \hbar \itrapfreq \hconj\ataop_i\ataop_i ~,
\end{equation}
where $i$ labels the three orthogonal symmetry axes of the potential in harmonic approximation, $\itrapfreq $ are the trap frequencies, $\hbar$ is the reduced Planck constant, and $\ataop_i$ and $\hconj\ataop_i$ are ladder operators for the harmonic motion of the trapped atom.

The second term $\Hamilop_\vibrational$ in \cref{eqn: total Hamiltonian} describes the free evolution of the phonon field of the photonic structure. Vibrations at frequencies relevant to atom traps can be modeled by linear elasticity theory because the corresponding phonon wavelengths are sufficiently large not to resolve the microscopic structure of the solid. Linear elasticity theory describes the dynamics of elastic deformations of a continuous body around its equilibrium configuration \cite{achenbach_wave_1973,auld_acoustic_1973-1,gurtin_linear_1984}. The deformations are described by the \emph{displacement field} $\ufield$, a real-valued vector field which indicates magnitude and direction of the displacement of each point of the body from equilibrium at a given time. A quantum formulation of linear elasticity theory can be obtained through canonical quantization based on phononic eigenmodes; see \cref{sec: phonon appendix}. The eigenmodes can be labeled by a suitable multi-index $\phonindex$ which may contain both discrete and continuous indices. In terms of ladder operators $\phonaop_\phonindex$ and $\hconj\phonaop_\phonindex$ of the phonon field, the resulting phonon Hamiltonian is
\begin{equation}\label{eqn: Hamiltonian phonons}
  \Hamilop_\vibrational \equiv \sum_\phonindex \hbar\phonfreq_\phonindex \hconj\phonaop_\phonindex\phonaop_\phonindex~,
\end{equation}
where the sum symbolizes an integral in the case of the continuous index components.

The last term $\Hamilop_\atphonint$ in the Hamiltonian \cref{eqn: total Hamiltonian} describes the coupling between the atomic motion and the phonon field. In order to obtain explicit expressions for the atom-phonon coupling, it is necessary to know how the potential experienced by the atom is changed by vibrations. Here, we give an overview of how this dependence can be modeled, while further details as well as explicit expressions for the resulting coupling constants in the case of a nanofiber-based atom trap are provided in \cref{sec: interaction appendix}. The coupling arises both because vibrations displace the photonic structure relative to the atom and because they change the electromagnetic properties of the structure in two ways \cite{zoubi_optomechanical_2016}: First, vibrations deform the surface of the structure, as determined by the displacement field $\ufield$. Second, they locally change the refractive index and introduce birefringence (photoelastic effect), as determined by the \emph{strain tensor} $\straintens$. The strain tensor describes deformations of the solid and has components $\straintenscomp^{ij} \equiv \pare{\partial_i \ufieldcomp^j + \partial_j \ufieldcomp^i}/2$, where $\partial_i$ indicates a spatial derivative. Both effects modify the photonic eigenmodes and hence the optical trapping fields. The optical fields and surface forces adapt to changes caused by vibrations on a timescale that is fast compared to the motion of the trapped atom. We can therefore treat the total potential as a functional $\pot[\ufield,\straintens](\atpos)$ which, in the absence of vibrations, reduces to the potential $\pot[\zerovec,\zerovec](\atpos) \equiv \pot_0(\atpos)$ included in $\Hamilop_\atomic$.

Thermal vibrations only weakly modify the atom trap. In consequence, it is justified to expand the potential to linear order around \mbox{$\ufield = \zerovec$} and \mbox{$\straintens = \zerovec$}, and approximate $\pot[\ufield,\straintens] \simeq \pot_0 + \frechetD\pot_{(\zerovec,\zerovec)}[\ufield,\straintens]$. The first-order term is the functional derivative of $\pot[\ufield',\straintens']$, evaluated at $(\ufield', \straintens') = (\zerovec,\zerovec)$ and in direction $(\ufield$, $\straintens)$; see %
\footnote{%
\label{note: functional derivative}
The Fréchet derivative $\frechetD F$ of a functional $F[\vec{x}]$ evaluated at $\vec{x} = \vec{a}$ and in direction $\vec{n}$ is defined as \cite{yamamuro_differential_1974,werner_funktionalanalysis_2011} %
$$ \frechetD F_{\vec{a}}[\vec{n}] \equiv \lim_{h \to 0} \cpare{ F[\vec{a} + h\vec{n}] - F[\vec{a}] }/h~.$$
The derivative is linear in $\vec{n}$, and can be used in a Taylor \mbox{expansion \cite{werner_funktionalanalysis_2011}}. In particular, it is suitable for the linear-order approximation $F[\vec{x}] \simeq F[\vec{a}] + \frechetD F_{\vec{a}}[\vec{x}]$. The partial Fréchet derivative $\partialfrechetD_{\vec{x}}G$ of a multivariate functional $G[\vec{x},\vec{y}]$ with respect to $\vec{x}$ evaluated at $(\vec{x},\vec{y}) = (\vec{a},\vec{b})$ and in direction $\vec{n}$ is defined as \cite{yamamuro_differential_1974}
$$\partialfrechetD_\vec{x} G_{(\vec{a},\vec{b})}[\vec{n}] \equiv \lim_{h \to 0} \cpare{ G[\vec{a} + h\vec{n},\vec{b}] - G[\vec{a},\vec{b}] }/h~.$$
Partial derivatives can be used to express the total derivative $\frechetD$ of a multivariate functional \cite{yamamuro_differential_1974}, for instance
$$\frechetD G_{(\vec{a},\vec{b})}[\vec{n},\vec{m}] = \partialfrechetD_\vec{x} G_{(\vec{a},\vec{b})}[\vec{n}] + \partialfrechetD_\vec{y} G_{(\vec{a},\vec{b})}[\vec{m}]$$ in the case of a bivariate functional.}.
This term approximates phonon-induced variations of the potential and acts as the atom-phonon interaction Hamiltonian
\begin{equation}\label{eqn: definition atom-phonon interaction Hamiltonian}
  \Hamilop_\atphonint \equiv \frechetD\pot_{(\zerovec,\zerovec)}[\ufieldop,\straintensop](\atposop)~.
\end{equation}
Truncating the expansion at linear order corresponds to assuming that the atom interacts only with single phonons at a time. Since the potential depends on both displacement and strain, there are two contributions to the interaction Hamiltonian, a \emph{displacement coupling} (\radpress) due to the direct dependence of the potential on $\ufield$, and a \emph{strain coupling} (\strainopt) due to the dependence on $\straintens$:
\begin{equation}\label{eqn: atom-phonon interaction Hamiltonian contributions}
  \Hamilop_\atphonint = \partialfrechetD_{\ufield}\pot_{(\zerovec,\zerovec)}[\ufieldop] + \partialfrechetD_{\straintens}\pot_{(\zerovec,\zerovec)}[\straintensop]~,
\end{equation}
Here, $\partialfrechetD$ is the partial functional derivative \cite{Note1}. The interaction Hamiltonian is linear in $\ufieldop$ and $\straintensop$ because the functional derivative is linear. By expanding displacement and strain in terms of phononic eigenmodes, the Hamiltonian can thus be expressed in terms of a position-dependent, complex-valued \emph{coupling function} $\couplingfunct_{\phonindex}(\atpos)$ for each phonon mode $\phonindex$,
\begin{equation}\label{eqn: general form atom-phonon interaction Hamiltonian}
    \Hamilop_\atphonint = \sum_\phonindex\spare{\couplingfunct_{\phonindex}(\atposop) \phonaop_\phonindex + \hc}~,
\end{equation}
where $\couplingfunct_{\phonindex}(\atpos) = \couplingfunct^\radpress_{\phonindex}(\atpos) + \couplingfunct^\strainopt_{\phonindex}(\atpos)$.
The coupling function $\couplingfunct^\radpress_{\phonindex}(\atpos)$ derives from displacement coupling and $\couplingfunct^\strainopt_{\phonindex}(\atpos)$ from strain coupling.

Furthermore, we approximate the phonon-induced forces acting on a trapped atom as linear in the atom position by  expanding \cref{eqn: definition atom-phonon interaction Hamiltonian} to first order around the trap minimum $\trappos$. The interaction Hamiltonian then takes the form %
\footnote{%
The term $\frechetD\pot_{(\zerovec,\zerovec)}[\ufieldop,\straintensop](\trappos)$ at order zero in the expansion describes a light-induced change in the mechanical equilibrium configuration of the photonic structure. We may safely neglect this constant shift, because it is small compared to the dimensions of a nanoscale structure and therefore only weakly modifies its photonic and phononic spectrum.%
}
\begin{equation}\label{eqn: linear force interaction Hamiltonians}
  \begin{split}
    \Hamilop_{\atphonint} &\simeq \sum_{i\phonindex}  \hbar (\ataop_i + \hconj\ataop_i) (\atphong_{\phonindex i}\phonaop_\phonindex + \cconj\atphong_{\phonindex i} \hconj\phonaop_\phonindex)~,
  \end{split}
\end{equation}
where the coupling constants are
\begin{equation}\label{eqn: definition coupling constants}
    \atphong_{\phonindex i} \equiv \frac{\atiposbsd}{\hbar} \partial_i \couplingfunct_{\phonindex}(\trappos)~.
\end{equation}
The length $\atiposbsd \equiv \sqrt{\hbar/(2 \atmass \itrapfreq)}$ is the zero-point motion of the atom of mass $\atmass$ in the trap. The coupling constants quantify the interaction of each phonon mode $\phonindex$ with the motion of the atom in direction $i$. Analogous to the coupling function, there are contributions from both displacement and strain coupling, $\atphong_{\phonindex i} = \atphong^\radpress_{\phonindex i} + \atphong^\strainopt_{\phonindex i}$.

The variation of the optical potential caused by displacement can in general be modeled by perturbatively calculating the new photonic eigenmodes in the presence of shifted boundaries of the nanostructure \cite{johnson_perturbation_2002}. The displacement has two effects: First, it shifts the photonic structure, together with the electromagnetic fields surrounding it, relative to the trapped atom. Second, it deforms the surface of the structure, leading to new photonic eigenmodes and thereby also deforming the electromagnetic fields. The first effect scales with the ratio between the displacement of the surface and the size of the atom trap (the extent of the wave function of the atom). The second effect, on the other hand, scales with the ratio between the displacement and the dimensions of the structure. Since the trap is typically at least one order of magnitude smaller than the photonic structure (see \cref{sec: case study}), we neglect the second effect and assume that both optical and surface potential are displaced as a whole together with the fiber surface \cite{le_kien_phonon-mediated_2007}. This model is particularly useful for structures such as nanofibers which have a simple geometrical shape and highly symmetric mechanical modes. The resulting displacement coupling functions $\atphong_{\phonindex}^\radpress(\atpos)$ for a nanofiber-based atom trap in particular are given in \cref{sec: interaction appendix}.

Strain leads to changes in the optical potential through the photoelastic effect, which can be modeled by a strain-dependent permittivity tensor $\prtrbd\relpermitttens[\straintens]$ \cite{nelson_theory_1971,narasimhamurty_photoelastic_2012,wuttke_optically_2013}. The modified permittivity is then in general neither homogeneous nor isotropic, and results in modified electric fields $\prtrbd\Efield$ surrounding the fiber and thus in a modified optical potential $\pot_\optical[\prtrbd\Efield]$. In consequence, the total potential $\pot[\ufield,\straintens]$ depends on strain. We neglect the influence of strain on the surface forces because they arise from the interaction of the atom with charges in a thin slice at the surface of the fiber and are largely independent of changes in the interior of the fiber \cite{buhmann_dispersion_2012}. The strain coupling function $\atphong_{\phonindex}^\strainopt(\atpos)$ can then be obtained by perturbatively calculating the new photonic eigenmodes in the presence of a modified permittivity; see \cref{sec: interaction appendix}.

\medskip{}

Having obtained the Hamiltonian of the coupled atom-phonon system, we can now describe the resulting evolution of the atomic motion. The cold atom can absorb kinetic energy from the thermally excited phonon field of the photonic structure (\emph{heating} of the atomic motion). Provided that the atom-phonon coupling is weak compared to the trap frequencies and the coherence time of phonon excitations, the phonon field can be adiabatically eliminated. The effective evolution of the density matrix $\atdensop(\tm)$ describing the motional state of the atom is then governed by a master equation \cite{cohen-tannoudji_atom-photon_1998,breuer_theory_2002}; see \cref{sec: interaction appendix}. Heating of the atom is reflected in the increase of the expected number of motional quanta $n_i(\tm) \equiv \tr [\atdensop(\tm) \hconj\ataop_i \ataop_i]$ along a spatial direction $i$. The population grows linearly with heating rate $\scatrate_i^\thermal$ for sufficiently short times,
\begin{equation}\label{eqn: short time population evolution}
 n_i(\tm) \simeq \scatrate_i^\thermal \tm~,
\end{equation}
assuming that the atom is in the motional ground state at $\tm=0$.

The phononic eigenmodes supported by the photonic structure can feature both discrete and continuous frequency spectra. Discrete spectra are observed for phonon modes with a spacing in frequency that is larger than their damping rates. In contrast, if a set of modes has frequency spacings much smaller than their damping rates (e.g., because the mechanical excitation is efficiently transmitted from the structure to its suspension), the discrete mechanical resonances are no longer discernible, and the spectrum is effectively continuous. Hence, we distinguish the contribution $\scatrate^{\discrete}_{i}$ of discrete mechanical resonances from the contribution $\scatrate^{\continuous}_{i}$ of a continuum of phonon modes:
\begin{equation}\label{eqn: heating rate contributions}
 \scatrate^{\thermal}_{i} = \scatrate^{\continuous}_{i} + \scatrate^{\discrete}_{i}~.
\end{equation}

For continuous phonon modes, Fermi's golden rule can be employed to calculate the heating rate $\scatrate^{\continuous}_{i}$ \cite{cohen-tannoudji_atom-photon_1998}:
\begin{equation}\label{eqn: Fermis golden rule}
  \scatrate^{\continuous}_{i} = 2\pi \nbar_i \sum_{\phonindex_i} \DOS_{\phonindex_i} |\atphong_{\phonindex_i i}|^2~.
\end{equation}
The sum runs over the discrete set of continuous phonon modes $\phonindex_i$ that are resonant with the trap, $\phonfreq_{\phonindex_i} = \itrapfreq$. The thermal occupation of the resonant phonon modes is $\bar{n}_i \equiv 1/\spare{\exp\pare{\hbar \itrapfreq/\boltzmann\temp}-1}$, where $\temp$ is the temperature of the photonic structure and $\boltzmann$ is the Boltzmann constant \cite{gerry_introductory_2005}. The phonon density of states is given by the inverse slope of the phonon dispersion relation (band structure), $\DOS_\phonindex \equiv | d\phonfreq_\phonindex / d\phonk |^{-1}$, where $\phonk$ is the propagation constant along the fiber; see \cref{sec: phonon appendix}.

The discrete resonances have finite lifetimes corresponding to decay rates $\phondecayrate_\phonindex$ due to internal losses and nonzero coupling to the suspension. Adiabatic elimination of these discrete mechanical modes in general leads to the heating rate $\scatrate^{\discrete}_{i}$ given in \cref{eqn: torsional heating general} in \cref{sec: interaction appendix} \cite{cirac_laser_1992,wilson-rae_cavity-assisted_2008}. There are two limiting cases that are of interest in \cref{sec: case study}: In the case where the atom-trap frequency is smaller than the lowest-frequency phonon mode $\phonindex_1$, $\itrapfreq < \phonfreq_{\phonindex_1}$, and detuned from resonance, $\phondecayrate_{\phonindex_1} \ll |\itrapfreq - \phonfreq_{\phonindex_1}|$, the ground-state  heating rate of the atom is
\begin{equation}\label{eqn: heating rate discrete phonon detuned}
  \scatrate_{i}^\discrete \simeq 2 \nbar \phondecayrate_{\phonindex_1} |\atphong_{\phonindex_1 i}|^2 \frac{\itrapfreq^2 + \phonfreq_{\phonindex_1}^2}{(\itrapfreq^2 - \phonfreq_{\phonindex_1}^2)^2}~.
\end{equation}
In the case where the atom trap is resonant with a single phonon mode $\phonindex$, $\phondecayrate_{\phonindex} \gg |\itrapfreq - \phonfreq_{\phonindex}|$, the rate is
\begin{equation}\label{eqn: heating rate discrete phonon}
  \scatrate_{i}^\discrete \simeq \frac{4 \nbar|\atphong_{\phonindex i}|^2}{\phondecayrate_{\phonindex}}~,
\end{equation}
where we assume $\nbar \gg 1$.

\medskip{}

The theory of atom-phonon interaction outlined in this section applies to any optical atom trap that relies on a photonic structure to shape light fields. The explicit calculation of atom-phonon coupling constants requires modeling of the dependence of the potential that the atom experiences on the displacement and the strain caused by the mechanical eigenmodes of the structure. Once the mechanical modes and corresponding atom-phonon coupling constants of a particular structure are known, \cref{eqn: Fermis golden rule,eqn: heating rate discrete phonon detuned,eqn: heating rate discrete phonon}, or more generally \cref{eqn: torsional heating general}, can be used to predict the phonon-induced heating of the atomic motion. In the next section, we apply this theory to explain heating rates observed in nanofiber-based atom traps.

\section{Case Study of a Nanofiber-based Trap}
\label{sec: case study}
\begin{table*}
  \begin{tabularx}{\textwidth}{c @{\qquad} l @{\quad}  l @{\qquad} l @{\quad} l @{\qquad} l @{\quad} l X}
    \toprule
    Trap & \multicolumn{2}{c}{$\Tmode_{01}$} & \multicolumn{2}{c}{$\Lmode_{01}$} & \multicolumn{2}{c}{$\Fmode_{11}$} & ~ \\
    \cmidrule{2-8}
    & $|\atphong^\radpress_{\phonindex i}|/2\pi$ (\si{\hertz}) & $|\atphong^\strainopt_{\phonindex i}|/2\pi$ (\si{\hertz})
    & $|\atphong^\radpress_{\phonindex i}|/2\pi$ ($\si{\hertz}\sqrt{\si{\meter}}$) & $|\atphong^\strainopt_{\phonindex i}|/2\pi$ ($\si{\hertz}\sqrt{\si{\meter}}$)
    & $|\atphong^\radpress_{\phonindex i}|/2\pi$ ($\si{\hertz}\sqrt{\si{\meter}}$) & $|\atphong^\strainopt_{\phonindex i}|/2\pi$ ($\si{\hertz}\sqrt{\si{\meter}}$) \\
    \midrule
    $\rpos$ & \num{0} & \num{5.47e-08} & \num{3.08e-09} & \num{1.56e-08} & \num{3.93e-4} & \num{2.18e-08} \\
    $\phipos$ & \num{0} & \num{7.81e-4} & \num{0} & \num{7.76e-11} & \num{2.28e-4} & \num{2.99e-10}\\
    $\zpos$ & \num{0} & \num{2.19e-12} & \num{0} & \num{1.05e-4} & \num{0} & \num{1.13e-10} \\
    \bottomrule
  \end{tabularx}
    \caption{Atom-phonon coupling constants. Listed are the contributions of displacement (\radpress) and strain (\strainopt) coupling to the coupling constants. The displacement coupling constants $\atphong^\radpress_{\phonindex i}$ are calculated according to \cref{eqn: 3d trap radiation pressure coupling constants}. The strain coupling constants  $\atphong^\strainopt_{\phonindex i}$ are obtained from \cref{eqn: definition coupling constants} with the coupling functions listed in \cref{tab: strain coupling functions} in \cref{sec: interaction appendix}. Coupling to modes on the continuous $\Lmode_{01}$ and $\Fmode_{11}$ bands is independent of the position of the trap site along the fiber axis. In contrast, the strain coupling constants to the discrete $\Tmode_{01}$ modes depend on the position since the torsional modes form standing waves; see \cref{sec: phonon appendix}. Listed here are the maximal coupling constants; for radial motion, the coupling is maximal at the end of the nanofiber ($\zpos = 0,\len$), while it is maximal at the center of the nanofiber ($\zpos = \len/2$) for the azimuthal and axial motion.}
    \label{tab: coupling constants}
\end{table*}

\begin{table}
\newcolumntype{C}[1]{>{\centering\arraybackslash}p{#1}}
\newcommand{\negligible}{$\ll$}
  \begin{center}
    \begin{tabular}{c C{2.1cm} C{2.1cm} C{2.1cm} }
      \toprule
      Trap & $\Tmode_{01}$ \qquad & $\Lmode_{01}$ & $\Fmode_{11}$ \\
       \midrule
      $\rpos$ & \negligible & \negligible & \SI{446}{\hertz} \\
      $\phipos$ & \negligible & \negligible & \SI{340}{\hertz} \\
      $\zpos$ & \negligible & \SI{8.36e-2}{\hertz} & \negligible \\
      \bottomrule
    \end{tabular}
    \caption{Atom heating rates. Listed are the contributions of the relevant phonon modes $\Tmode_{01}$, $\Lmode_{01}$, and $\Fmode_{11}$ to the heating rate $\scatrate_i^\thermal$ of a trapped atom in direction $i \in \{\rpos,\phipos,\zpos\}$ calculated according to \cref{eqn: Fermis golden rule,eqn: heating rate discrete phonon detuned}. Contributions below $10^{-4}\,\si{\hertz}$ are indicated by \lq{}\negligible.\rq{} The  rates are independent of the position of the trap site along the fiber. The fiber temperature is assumed to be $\temp = \SI{805}{\kelvin}$, and the remaining parameters are specified in \cref{sec: case study appendix}.}
    \label{tab: heating rates}
  \end{center}
\end{table}

Let us now use the framework sketched in \cref{sec: framework} to study the phonon-induced heating rates of the atomic motion in a nanofiber-based two-color atom trap. In particular, we consider a cesium atom trapped in the evanescent optical field surrounding a silica nanofiber \cite{dowling_evanescent_1996,le_kien_atom_2004}. The nanofiber is formed by the waist of an optical fiber which has been heated and pulled \cite{ward_optical_2014}. There have been several experimental realizations of this nanophotonic atom trap configuration \cite{vetsch_optical_2010,goban_demonstration_2012,kato_strong_2015,lee_inhomogeneous_2015,corzo_large_2016,ostfeldt_dipole_2017,meng_near-ground-state_2018,albrecht_fictitious_2016}. We calculate atom heating rates for the setup described in \cite{albrecht_fictitious_2016}, where a measured heating rate of $\scatrate_\phipos^\thermal = \SI{340(10)}{\hertz}$ in the azimuthal direction was reported. In order to explicitly calculate the phonon-induced heating rates, it is necessary to know the mechanical eigenmodes of the nanofiber close to resonance with the trap frequencies and to obtain the atom-phonon coupling constants. The latter calculation requires knowledge of the trap potential as well as the photonic eigenmodes of the nanofiber. \Cref{sec: photon appendix} summarizes the photonic eigenmodes of a nanofiber, and \cref{sec: atom appendix} provides details on the resulting trapping potential. \Cref{sec: phonon appendix} summarizes the phononic eigenmodes. In \cref{sec: interaction appendix}, we derive the resulting atom-phonon coupling constants for a nanofiber-based trap. The parameters of the particular setup considered in this \namecref{sec: case study} are listed in \cref{sec: case study appendix}.

Trapping of atoms is achieved by means of two lasers, one red and the other blue detuned with respect to the $D$ lines of cesium. The lasers are guided as photonic $\HEmode_{11}$ spatial modes in the nanofiber region; see \cref{sec: photon appendix}. \Cref{fig: trap} in \cref{sec: interaction appendix} shows the resulting trapping potential. The red-detuned laser is coupled into the fiber at both ends, leading to a standing wave that confines the atoms in the axial direction and creates a one-dimensional optical lattice. The laser beams are linearly polarized when coupled into the fiber, which leads to \emph{quasilinearly} polarized fields with intensity maxima at opposite poles of the fiber cross section in the nanofiber region \cite{le_kien_state-dependent_2013}. The corresponding electric field profiles are listed in \cref{sec: case study appendix}. The red- and blue-detuned field have orthogonal polarizations to obtain stronger azimuthal confinement \cite{vetsch_optical_2010}. There is an offset magnetic field oriented perpendicular to the fiber axis ($\zpos$ axis) along $\ezvecB = \cos(\phi) \exvec + \sin(\phi) \eyvec$, with $\phi = \SI{66}{\degree}$.  Atoms are initially prepared in the Zeeman substate $\atF=4$, $\atMF = -4$ of the hyperfine structure, where the offset magnetic field provides the quantization axis. The magnetic field causes a slight azimuthal shift of the trap sites. Nonetheless, the symmetry axes of the potential at the trap minimum are to a good approximation aligned with the radial, azimuthal, and axial unit vectors of a cylindrical coordinate system whose $\zpos$ axis coincides with the nanofiber axis. We can therefore use $i\in \{\rpos,\phipos,\zpos\}$ for the atom trap directions in the atom Hamiltonian \cref{eqn: atom Hamiltonian harmonic approximation}. The resulting frequencies of the atom trap are $(\rtrapfreq, \,\phitrapfreq, \,\ztrapfreq) = 2\pi \times(\num{123}, \,\num{71.8}, \,\num{193})\, \si{\kilo\hertz}$.

An infinitely long nanofiber supports three phonon bands which do not have a cutoff at low frequencies: the torsional $\Tmode_{01}$ band, longitudinal $\Lmode_{01}$ band, and flexural $\Fmode_{11}$ band; see \cref{sec: phonon appendix}.
\Cref{fig: phonon modes} shows the displacement of the nanofiber caused by phonon modes on each of these bands. The torsional band is linear and the longitudinal band asymptotically linear for low frequencies, with speeds of sound $\ctrans$ and $\cFL$ introduced in \cref{sec: phonon appendix}, respectively. The flexural band has a quadratic asymptote. The dispersion relations describing these bands as functions of the propagation constant $\phonk$ are
\begin{align}\label{eqn: fundamental phonon bands}
  \phonfreq_\Tmode &= \ctrans |\phonk| & \phonfreq_\Lmode &\simeq \cFL |\phonk| & \phonfreq_\Fmode &\simeq \frac{\cFL \rad}{2} \phonk^2~.
\end{align}
Here, $\rad$ is the radius of the nanofiber. These three fundamental bands are the only candidates for phonon-induced heating of the atomic motion since all other bands have frequencies much larger than the trap frequencies.

In experiments, the optical nanofibers used for atom trapping are typically realized as the waist of a tapered optical fiber \cite{vetsch_optical_2010}. The mechanical eigenmodes of this system -- including the nanofiber, the tapers, and the surrounding macroscopic fiber -- can be calculated either analytically or using finite-element methods \cite{wuttke_thermal_2013}. Since the fiber is finite in length, the eigenmodes are standing waves and the spectrum consists of discrete mechanical resonances. The system can in general support the same kinds of excitations as an infinite cylinder: torsional, longitudinal, and flexural. For some modes, the tapers act as reflectors and strongly localize them in the nanofiber region. Others are transmitted through the tapers and are delocalized over the entire fiber \cite{wuttke_thermal_2013}. In practice, all modes are damped. Dissipation occurs, among others, due to clamping losses \cite{pennetta_tapered_2016}, friction with the background gas \cite{wuttke_optically_2013}, material losses \cite{wiedersich_spectral_2000}, and surface losses \cite{penn_frequency_2006}. Depending on the magnitude of the damping $\kappa$ of each mode compared to the free spectral range (FSR), the actual spectrum ranges from  discrete ($\text{FSR} \gg \kappa$) to continuous ($\text{FSR} \ll \kappa$). In the case of a discrete spectrum, standing waves of finite lifetime $1/\kappa$ are a useful description of the mechanical dynamics of the fiber. In the limit of a continuous spectrum, the idealized eigenmodes of the system are no longer faithful representations, since the phonons interact too strongly with other degrees of freedom and are dissipated before they can form standing waves. Instead, it is more useful to represent the phonons as propagating modes of an infinite structure which interact with the atom once and then never return (analogous to an atom interacting with fiber-guided or free-space photons). Some of the damping mechanisms can be modeled theoretically \cite{penn_frequency_2006,wiedersich_spectral_2000}. However, more reliable results are obtained by measuring damping rates for the particular fiber in use. We perform measurements of the mechanical modes of the particular nanofiber setup considered here \cite{albrecht_fictitious_2016}, similar to \cite{wuttke_thermal_2013,fenton_spin-optomechanical_2018}. While torsional resonances are clearly visible, there is no indication of resonantly enhanced longitudinal or flexural nanofiber modes. The mode of lowest frequency is at $\phonfreq_\Tmode = 2\pi \times \SI{258}{\kilo\hertz}$ with a wavelength of $\SI{14.6}{\milli\meter}$ and a decay rate of $\kappa = 2\pi \times \SI{48(1)}{\hertz}$. The torsional modes can be modeled faithfully by imposing hard boundary conditions on an elastic cylinder; see \cite{wuttke_thermal_2013} and \cref{sec: phonon appendix}. The resulting spectrum is a discrete subset of the $\Tmode_{01}$ band of an infinite cylinder. In keeping with the absence of discrete resonances corresponding to longitudinal and flexural modes, we model these modes as the propagating modes of an infinite cylinder, with a continuous dispersion relation given by the longitudinal and flexural bands \cref{eqn: fundamental phonon bands}. The form of the longitudinal and flexural mechanical bands and the corresponding eigenmodes are then determined by the elastic mechanical properties of silica and the fiber radius alone. The wavelengths of the modes resonant with the azimuthal trap frequency, for instance, are $\SI{80.0}{\milli\meter}$ for the $\Lmode_{01}$ mode and $\SI{0.251}{\milli\meter}$ for the $\Fmode_{11}$ mode.

The theory derived in \cref{sec: framework} allows us to calculate atom heating rates based on these physical parameters. The only parameter not provided by ref.~\cite{albrecht_fictitious_2016} is the fiber temperature $\temp$. We choose the temperature such that the azimuthal heating rate $\scatrate_\phipos^\thermal$ observed in \cite{albrecht_fictitious_2016} is reproduced. Agreement with the measurement in ref.~\cite{albrecht_fictitious_2016} is achieved for $\temp = \SI{805}{\kelvin}$, which agrees well with the temperature of $\temp = \SI{850\pm150}{\kelvin}$ measured independently in \cite{wuttke_thermalization_2013} for a similar nanofiber at the given transmitted laser power. Heating in the azimuthal direction is dominantly caused by resonant flexural $\Fmode_{11}$ modes. To our knowledge, this is the first time that a theoretical prediction of the atom heating rate based on measured parameters and in quantitative agreement with measured heating rates has been obtained. We are then able to calculate the phonon-induced heating rates of the atomic motion in the radial, azimuthal, and axial direction, accounting for both displacement and strain coupling. The predicted atom-phonon coupling constants are listed in \cref{tab: coupling constants} and the resulting heating rates in \cref{tab: heating rates}.

The predicted heating rate for the radial degree of freedom is of a magnitude similar to the rate for the azimuthal degree of freedom. The calculated radial heating rate is $\scatrate_\rpos^\thermal = \SI{446}{\hertz}$, which agrees with the heating rate assumed in \cite{reitz_coherence_2013} to explain measured $T_2^\prime$ decoherence rates for nanofiber-trapped atoms. Heating along the radial axis, like heating in the azimuthal direction, is dominated by coupling to the resonant flexural $\Fmode_{11}$ modes. The coupling constants in \cref{tab: coupling constants} reveal that the coupling is due to displacement of the fiber surface, while coupling due to strain is lower by several orders of magnitude. \latin{A priori}, both longitudinal $\Lmode_{01}$ and flexural $\Fmode_{11}$ modes couple to the radial motion by displacement. However, the flexural modes lead to much higher heating rates for two reasons: First, flexural modes displace the fiber surface by a factor of $|\wmodecomp^\rpos_\Fmode/\wmodecomp^\rpos_\Lmode| \simeq \sqrt{\YoungE/(2 \dens)}/(\rtrapfreq \Poissonnu\rad) \simeq 10^5$ more than the longitudinal modes which leads to larger displacement coupling constants. Here, $\wmodecomp^\rpos_\Fmode$ and $\wmodecomp^\rpos_\Lmode$ are the radial components of the displacement eigenmode for the flexural and longitudinal modes, respectively. The quantity $\YoungE$ is Young's modulus and $\Poissonnu$ is the Poisson ratio; together, they describe the elastic properties of the nanofiber. The quantity $\dens$ is the mass density of the nanofiber and $\rtrapfreq$ the radial trap frequency. The second reason is that the density of states of the flexural modes is larger than the one of longitudinal modes by a factor of $\DOS_{\Fmode \rpos}/\DOS_\Lmode \simeq \sqrt{\cFL/(2 \rtrapfreq \rad)} \simeq 100$, and the heating rates are enhanced accordingly; see \cref{eqn: Fermis golden rule}.

Heating in the axial direction is predicted to be predominantly due to strain coupling to the resonant longitudinal $\Lmode_{01}$ mode, with a rate much smaller than the heating rates in the radial and azimuthal direction. To the best of our knowledge, the heating rate in the axial direction has not been measured so far.

One might expect heating by near-resonant torsional modes to be dominant because they are tightly confined to the nanofiber region, leading to Purcell enhancement of the coupling strength \cite{gerry_introductory_2005}. The strain induced by torsional modes causes a tilt of the quasilinear polarization of the light fields, see \cref{fig: discrete T modes blue and red scalar coupling r phi} in \cref{sec: interaction appendix}, which leads to coupling to the azimuthal motion of the atom in particular. In the present case, the contribution of torsional modes to the heating is negligible due to the large detuning between the torsional mode and trap frequencies compared to the phonon decay rate. However, we can use \cref{eqn: heating rate discrete phonon} with the coupling constants given in \cref{tab: coupling constants}  to obtain an estimate of the heating rates expected in the case when the torsional modes are resonant (e.g., in the case the nanofiber is longer). In this worst-case scenario, the predicted contribution to the heating rate in the azimuthal direction is $\scatrate^\discrete_\phipos = \SI{17.8}{\hertz}$, while heating in the other trap directions is still below $10^{-4}\,\si{\hertz}$, despite the Purcell enhancement. For the hypothetical case in which the torsional modes are not reflected at the ends of the nanofiber, our model predicts even lower heating rates. Hence, torsional modes are not a relevant source of heating in \cite{albrecht_fictitious_2016}, even if they are resonant with the trap frequencies.

In summary, the atom heating in the radial and azimuthal direction observed in experiments is well explained by the displacement coupling to the continuous $\Fmode_{11}$ band alone. In this case, \cref{eqn: Fermis golden rule} simplifies to the single equation
\begin{align}\label{eqn: final heating formula}
  \scatrate^\thermal_i &\simeq \frac{1}{2\sqrt{2}\pi}\frac{\boltzmann}{\hbar} \temp \atmass\sqrt{\frac{\itrapfreq}{\rad^5 \sqrt{\YoungE \dens^3}}} & i \in \{ \rpos,\phipos\}~,
\end{align}
where we use that $\hbar \itrapfreq \ll \boltzmann \temp$, such that the thermal occupation of the phonon modes is $\nbar_i \simeq \boltzmann \temp/\hbar \itrapfreq$. This simple formula agrees exceedingly well with calculations considering all phonon modes and both displacement and strain coupling.
\begin{figure*}
  \centering
  \setlength{\widthFigA}{180.9393pt}
  \setlength{\marginLeftFigA}{25.9393pt}
  \setlength{\marginRightFigA}{7pt}
  \setlength{\widthFigB}{162pt}
  \setlength{\marginLeftFigB}{7pt}
  \setlength{\marginRightFigB}{7pt}
  \setlength{\widthFigC}{160.6pt}
  \setlength{\marginLeftFigC}{5.6pt}
  \setlength{\marginRightFigC}{7pt}
  \parbox[t]{\widthFigA}{\vspace{0pt}\includegraphics[width = \widthFigA]{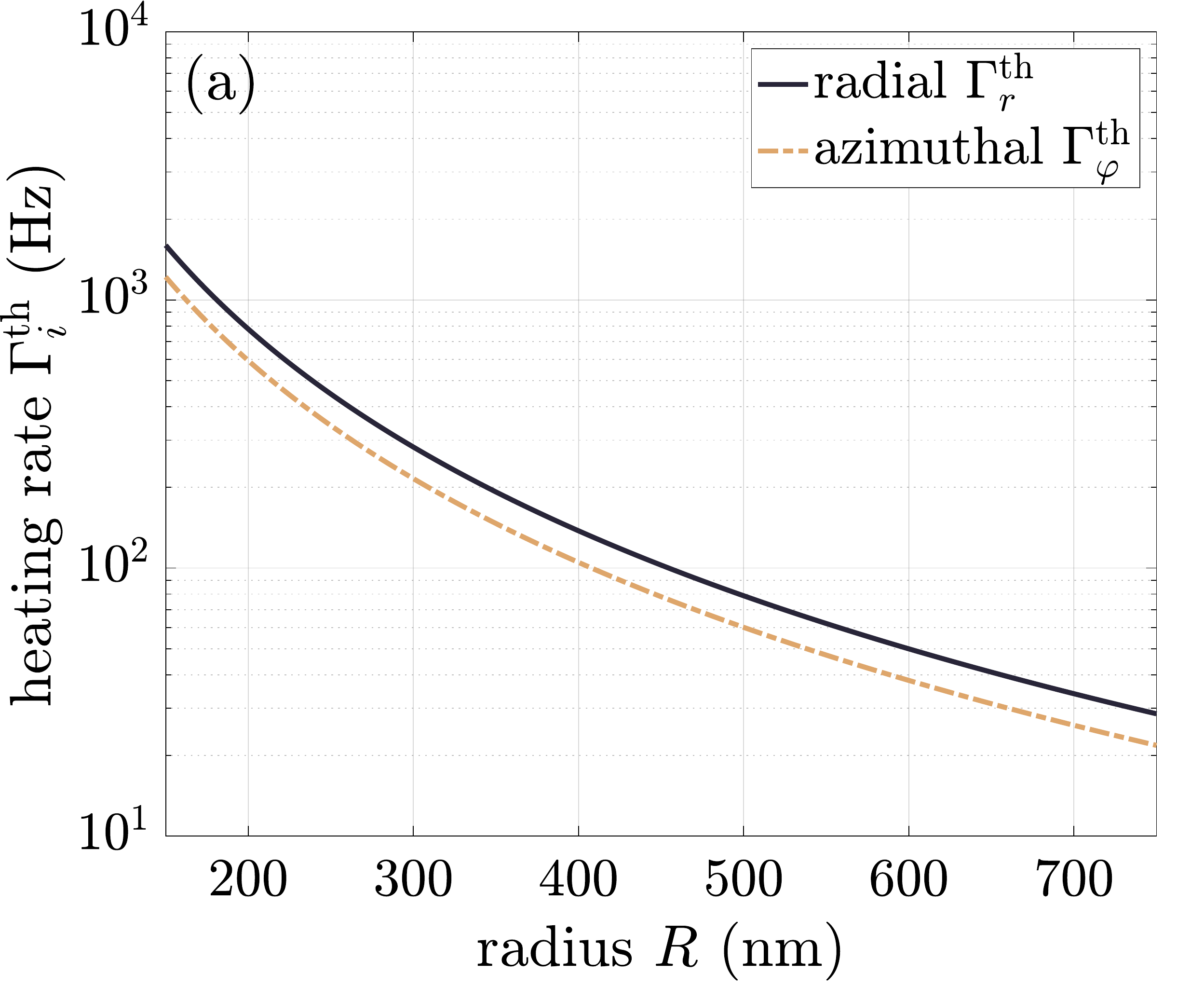}}
  \parbox[t]{\widthFigB}{\vspace{0pt}\includegraphics[width = \widthFigB]{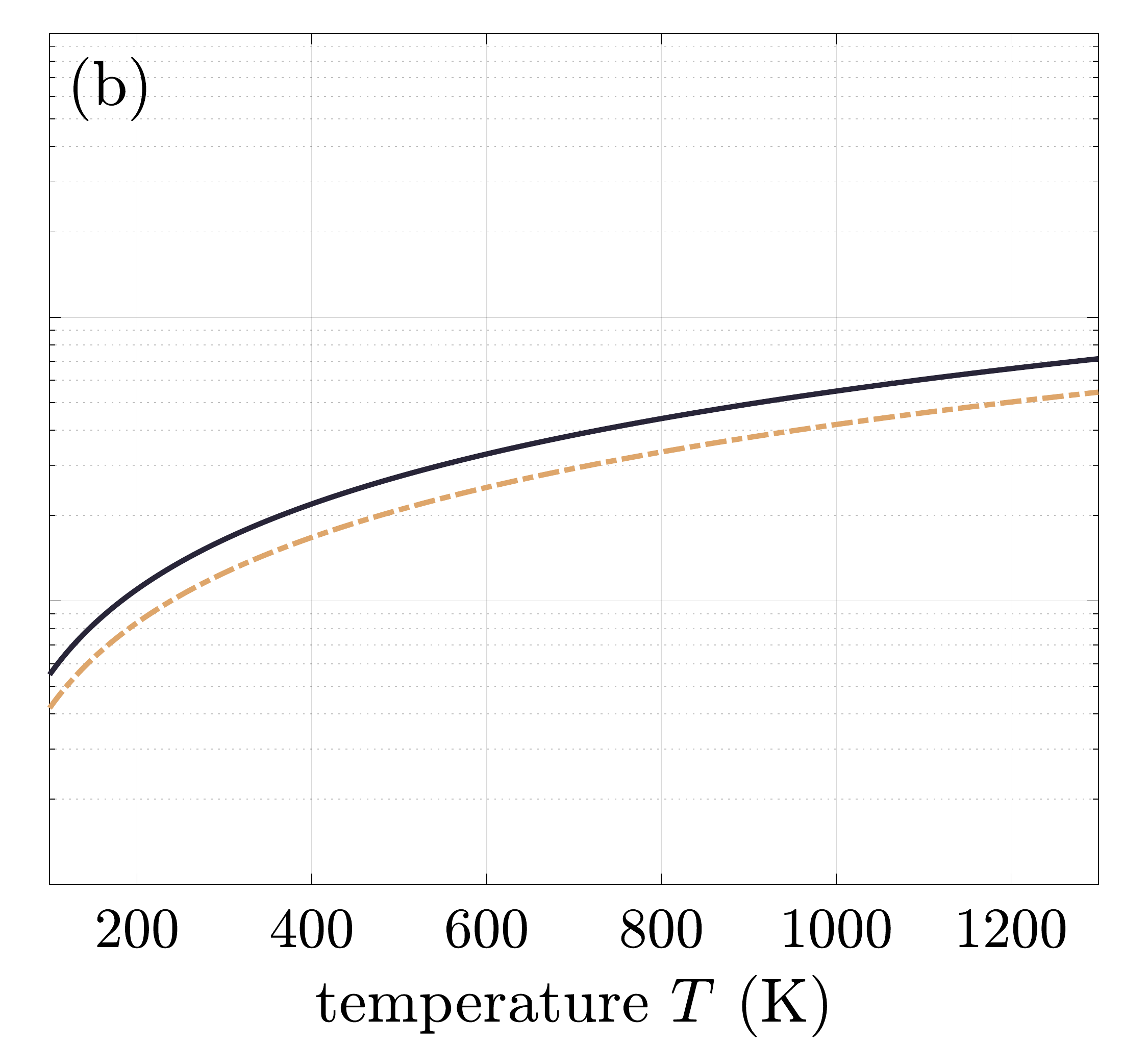}}
  \parbox[t]{\widthFigC}{\vspace{0pt}\includegraphics[width = \widthFigC]{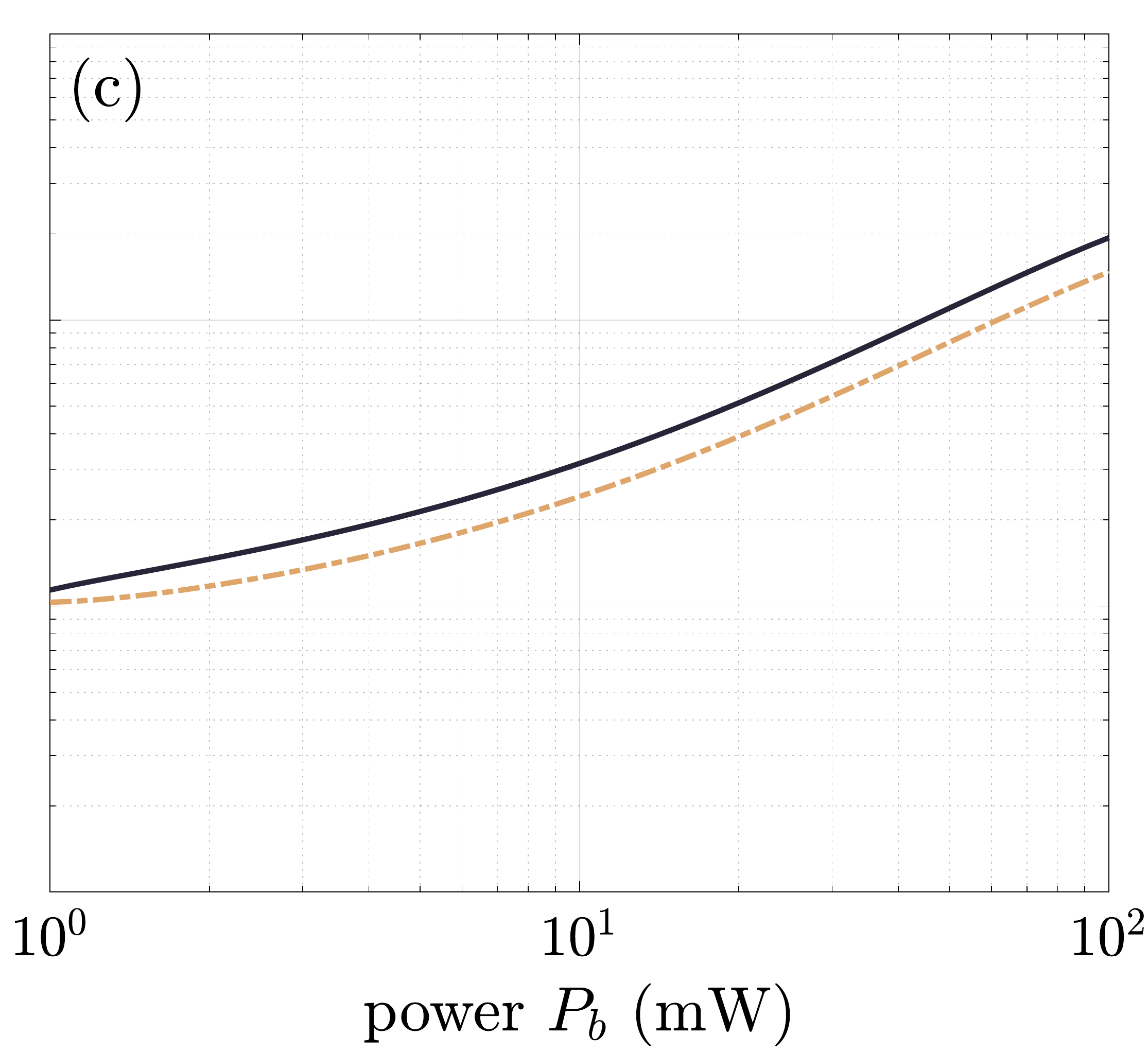}}\\
  \caption{Atom heating rate in the radial and azimuthal direction calculated using \cref{eqn: final heating formula} as function of (a) the nanofiber radius, (b) the temperature of the nanofiber, and (c) the power of the blue-detuned trapping laser. The difference between \cref{eqn: final heating formula} and the full theory \cref{eqn: heating rate contributions} is not discernible at the given scales. In \subcrefandb{fig: heating vs laser power}{a}{b}, all other parameters, in particular the trap frequencies, are unchanged. In \subcrefb{fig: heating vs laser power}{c}, the ratio between the power of the red- and blue-detuned laser is kept constant, $\power_\bluedetuned / \power_\reddetuned = 14.24$. The relation between the total laser power and temperature is modeled as $\temp(\power) = m_0 + m_1\power + m_2\power^2$, with $m_0 = \SI{400}{\kelvin}$, $m_1 = \SI{24}{\kelvin/\milli\watt}$, $m_2 = \SI{-0.062}{\kelvin/\milli\watt^2}$ based on the measurements in \cite{wuttke_thermalization_2013} for a nanofiber of radius $\rad = \SI{250}{\nano\meter}$ and length $\len = \SI{5}{\milli\meter}$. The temperature then varies from $\temp = \num{427}$ to $\SI{2298}{\kelvin}$ over the shown range of laser power. The trap frequencies simultaneously increase from $(\rtrapfreq,\,\phitrapfreq) = 2\pi \times (\num{29.1},\,\num{23.9})\,\si{\kilo\hertz}$ to $2\pi\times (\num{291},\,\num{168})\,\si{\kilo\hertz}$. The remaining parameters are specified in \cref{sec: case study appendix}.}
  \label{fig: heating vs laser power}
\end{figure*}
\Cref{fig: heating vs laser power} shows the dependence of the predicted heating rates in the radial and azimuthal direction on individual parameters, keeping the remaining parameters unchanged. Most pronounced is the scaling with the nanofiber radius as $\scatrate^\thermal_i \propto \rad^{-5/2}$, see \subcref{fig: heating vs laser power}{a}. The strong dependence on the radius is mostly due to the increased mechanical stability of larger nanofibers which leads to smaller vibrational amplitudes, see \cref{eqn: displacement F modes low-frequency limit}, in addition to a lower density of states. In contrast, the dependence on the fiber temperature is linear, see \subcref{fig: heating vs laser power}{b}, since the thermal occupation of the resonant phonon modes increases linearly with the temperature. Comparison of \subcref{fig: heating vs laser power}{a} and \subcref{fig: heating vs laser power}{b} shows that increasing the nanofiber radius by $\SI{150}{\nano\meter}$ to $\rad = \SI{400}{\nano\meter}$ at constant temperature has an effect comparable to cooling the fiber down to room temperature if all other parameters of the setup could be kept unchanged. \subCref{fig: heating vs laser power}{c} shows the dependence on the power of the blue-detuned laser, where the ratio of the power of the red- and blue-detuned lasers is kept constant. The temperature of the nanofiber increases with increased laser power since there is more absorption in the fiber \cite{wuttke_thermalization_2013}; see caption for details. Moreover, higher intensities lead to a tighter confinement of the atoms. The observed increase of the heating rate when raising the laser power is therefore caused by an increase of both the fiber temperature and the trap frequencies. While Young's modulus $\YoungE$ also slightly changes with $\temp$ \cite{spinner_elastic_1956}, the influence of this effect on the heating rate is negligible due to the weak dependence, $\scatrate^\thermal_i \propto \YoungE^{-1/4}$.

\medskip{}

Let us now discuss ways to reduce the atom heating caused by coupling to the continuous $\Fmode_{11}$ band. Lowering the overall fiber temperature in order to reduce the heating rates is difficult even in cryogenic environments because thermal coupling of the fiber to its surroundings is very weak \cite{wuttke_thermalization_2013}. However, based on the above analysis, different strategies to minimize the heating rates are conceivable. First of all, the fiber radius should be chosen as large as possible while maintaining the optical properties required for atom trapping. A second approach is to design the nanofiber such that it supports discrete, well-resolved resonances of flexural modes. While precise predictions of phonon linewidths are difficult, it may be possible to optimize the taper at both ends of the nanofiber and ensure that flexural modes are reflected and confined to $\zpos\in[0,\clamplen]$ with narrow linewidths, while the transmission of light is not reduced \cite{pennetta_tapered_2016}. Such a resonator of length $\clamplen$ for the flexural modes would effectively break the $\Fmode_{11}$ band into a discrete set of frequencies $\phonfreq_m$, and allow us to detune the atom trap from resonance with these mechanical modes.
\begin{figure*}[t]
  \raggedright
  \setlength{\widthFigA}{256.0448pt}
  \setlength{\marginLeftFigA}{29.0448pt}
  \setlength{\marginRightFigA}{7pt}
  \setlength{\widthFigB}{233pt}
  \setlength{\marginLeftFigB}{7pt}
  \setlength{\marginRightFigB}{6pt} %
  \parbox[t]{\widthFigA}{\vspace{0pt}\includegraphics[width = \widthFigA]{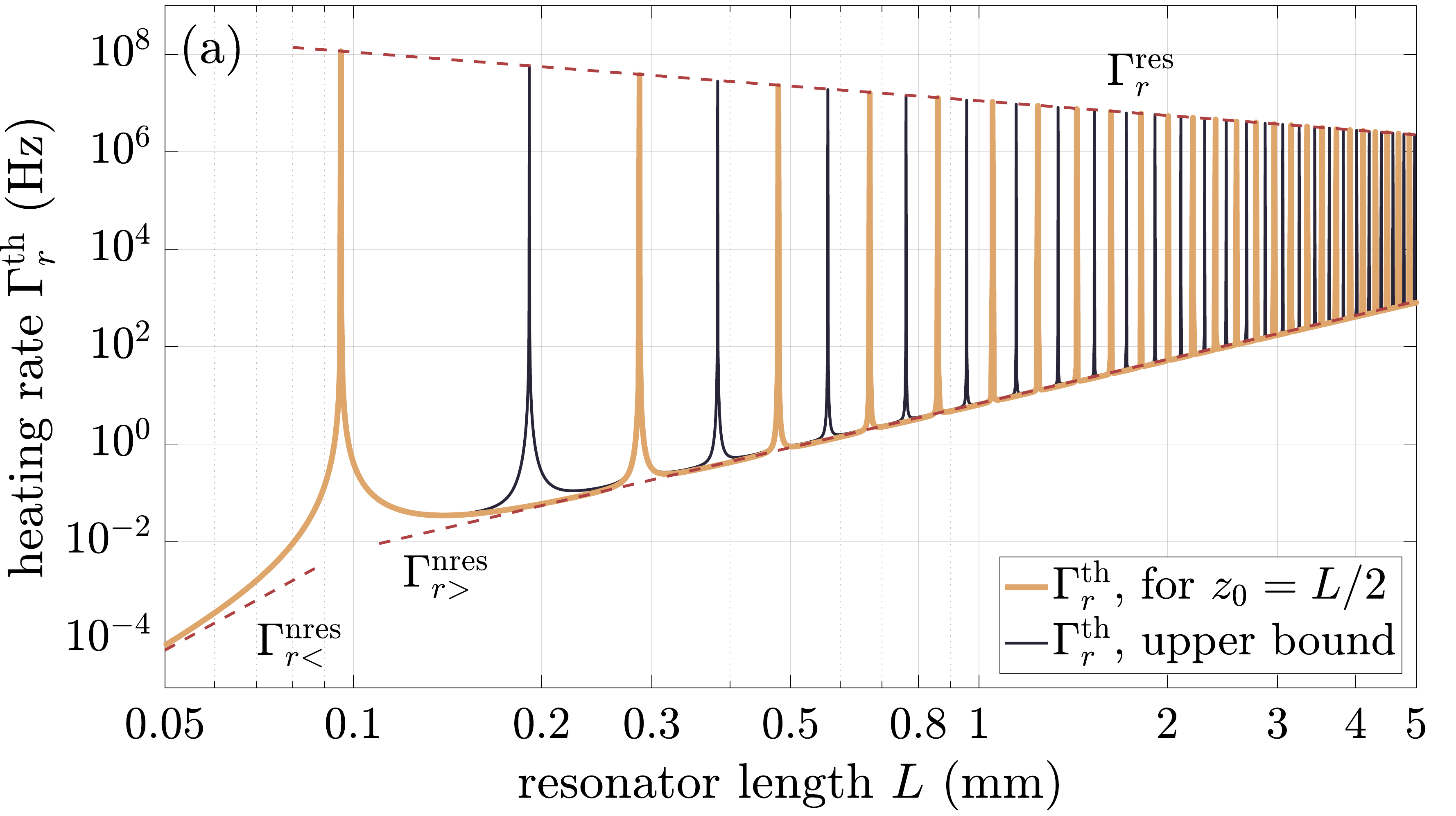}}
  \parbox[t]{\widthFigB}{\vspace{0pt}\includegraphics[width = \widthFigB]{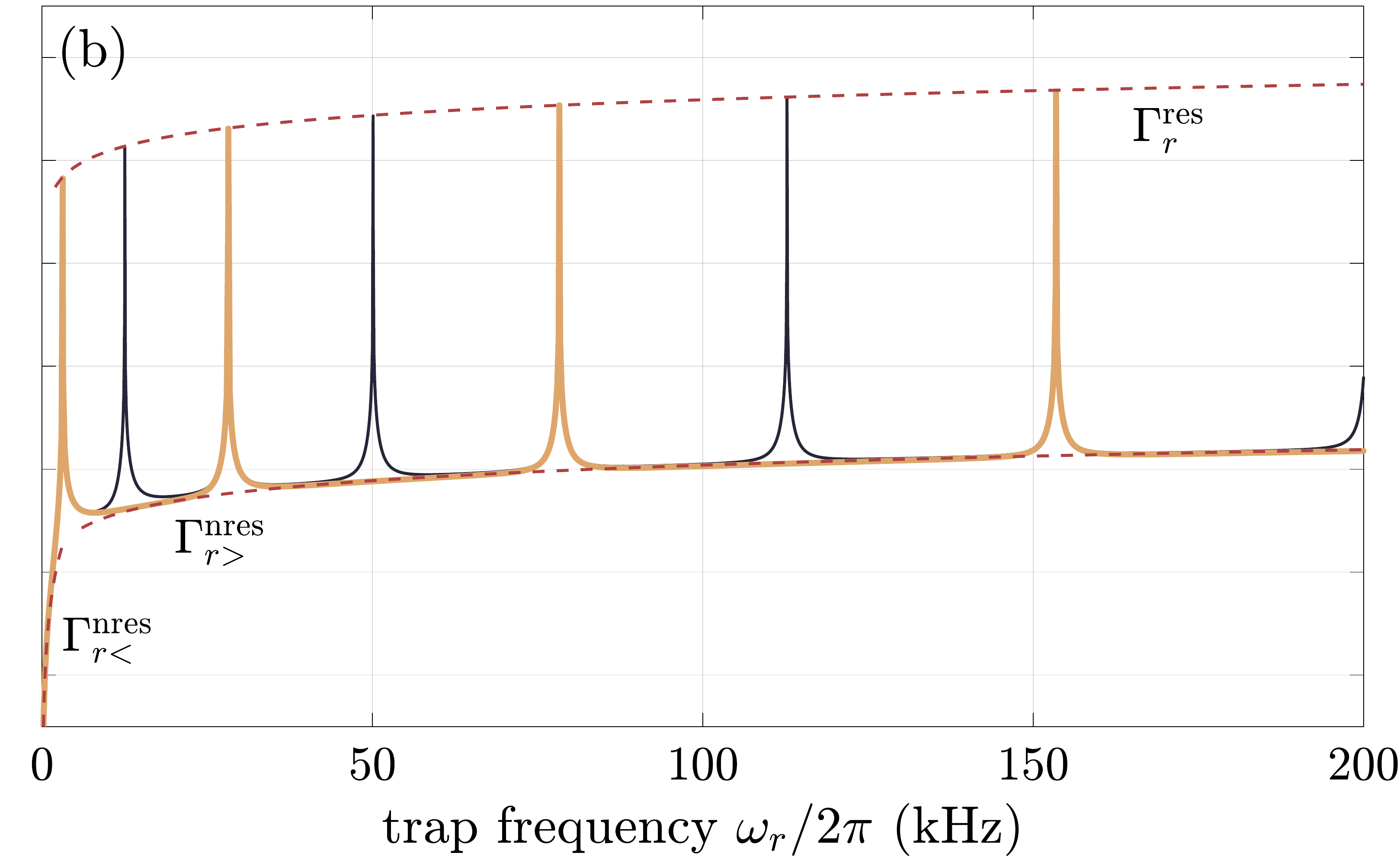}}\\
  \caption{Atom heating rate in the radial direction due to flexural resonator modes as a function of (a) the resonator length and (b) the trap frequency. In both cases, an exemplary decay rate of $\phondecayrate = 2\pi\times \SI{1.2}{\hertz}$ is assumed for all resonator modes. The bold yellow line corresponds to the heating rate experienced by an atom trapped at the center of the resonator, $\trapz = \clamplen/2$, calculated according to \cref{eqn: general heating discrete flexural modes}. The thin blue line represents a position-independent upper bound obtained by pretending that the atom sits at an antinode of each phonon mode simultaneously: In consequence, no resonance between the atom and resonator is masked by a vanishing position-dependent coupling rate. This approach is useful, since in experiments an entire ensemble of atoms is trapped at various positions along the fiber. The dashed red lines show the approximations \cref{eqn: heating off-resonant 1,eqn: heating off-resonant 2,eqn: resonant heating}. \subCrefb{fig: reduce heating}{a} assumes a trap frequency of $\rtrapfreq = 2\pi\times \SI{123}{\kilo\hertz}$, and \subcrefb{fig: reduce heating}{b} assumes a resonator length of $\clamplen = \SI{600}{\micro\meter}$.}
  \label{fig: reduce heating}
\end{figure*}
The flexural eigenmodes are then standing waves (see \cref{sec: phonon appendix}), with frequency spectrum
\begin{align}\label{eqn: flexural resonator spectrum}
  \phonfreq_m &\equiv m^2 \frac{\pi^2 \rad}{2\clamplen^2}\sqrt{\frac{\YoungE}{\dens}}~, & m &\in \N~.
\end{align}
The heating rate in the radial and azimuthal direction due to these flexural resonator modes then depends on the position $\zpos_0$ of the atom along the fiber axis; see \cref{sec: interaction appendix}. \Cref{fig: reduce heating} shows the dependence of the heating rate on the resonator length and trap frequency. Three regimes are clearly distinguishable: First, the trap is resonant with a flexural phonon mode. Second, the trap is off resonant and lies below the fundamental resonator frequency. Third, the trap is off resonant and lies above the fundamental resonator frequency. Assuming high thermal occupation of the phonon modes, $\nbar_m \gg 1$, simplified expressions for the heating rate can be obtained for each regime. If the trap frequency is below the fundamental phonon frequency but still much larger than the corresponding decay rate, $\phondecayrate_1 \ll \itrapfreq < \phonfreq_1$, as well as far detuned, $|\itrapfreq-\phonfreq_1|\gg \phondecayrate_1$, heating is dominated by off-resonant interaction with the fundamental phonon mode alone. In this case, the heating rate can be approximated as
\begin{equation}\label{eqn: heating off-resonant 1}
  \begin{split}
    \scatrate_i^\thermal &\simeq \scatrate_{i<}^\offresonant \sin^2(\pi \trapz/\clamplen)\\
    \scatrate_{i<}^\offresonant &\equiv \frac{16}{\pi^9}\frac{\boltzmann}{\hbar}\frac{\temp \atmass \dens \phondecayrate_1 \itrapfreq^3 \clamplen^7}{\YoungE^2 \rad^6} ~.
  \end{split}
\end{equation}

If the trap has a frequency larger than the fundamental resonator frequency, $\itrapfreq\gg \phonfreq_1$, while still being off resonant, $|\itrapfreq-\phonfreq_m|\gg \phondecayrate_m$, heating is mainly due to the low-frequency phonon modes below the trap frequency. Assuming in addition that the phonon decay rate is the same for all relevant modes, $\phondecayrate_m \simeq \phondecayrate$, an upper bound for the heating rate can be obtained:
\begin{equation}\label{eqn: heating off-resonant 2}
 \scatrate_i^\thermal \lesssim \scatrate_{i>}^\offresonant \equiv \frac{2}{45 \pi} \frac{\boltzmann}{\hbar}\frac{\temp\atmass \phondecayrate \itrapfreq \clamplen^3}{\YoungE \rad^4}.
\end{equation}
Here, we replace the sine in the coupling constant with $1$ for all modes, pretending the atom is located at an antinode of all modes simultaneously as a worst-case estimate. This approximation is useful because in experiments many atoms at different sites along the fiber axis are trapped at the same time.

If the trapped atom is resonant with a flexural phonon mode $m$, $|\itrapfreq-\phonfreq_m|\ll \phondecayrate_m$, and the contributions of the off-resonant modes can be neglected, the heating rate is
\begin{equation}\label{eqn: resonant heating}
  \begin{split}
  \scatrate_i^\thermal &\simeq \scatrate_{i}^\resonant \sin^2(\phonk_m \trapz)\\
  \scatrate_{i}^\resonant &\equiv \frac{2}{\pi}\frac{\boltzmann}{\hbar} \frac{\temp \atmass \itrapfreq}{\clamplen \dens \phondecayrate_m \rad^2}~.
  \end{split}
\end{equation}
The limiting expressions \cref{eqn: heating off-resonant 1,eqn: heating off-resonant 2,eqn: resonant heating} are shown as dashed black lines in \cref{fig: reduce heating}. Note that the dependence on decay rate and resonator length is inverted for off-resonant heating, \cref{eqn: heating off-resonant 1,eqn: heating off-resonant 2}, compared to resonant heating, \cref{eqn: resonant heating}. This inversion is expected, since large phonon linewidths $\phondecayrate_m$ assist off-resonant coupling, while small linewidths lead to a larger resonant enhancement. Small resonator lengths $\clamplen$ lead to higher coupling constants (Purcell enhancement), which increases resonant heating due to a single mode. In contrast, large resonator lengths result in a higher number of low-frequency modes and hence overcompensate the decrease in coupling strength and increase the heating due to off-resonant interaction.

In \cref{fig: reduce heating}, we exemplarily assume a decay rate of $\phondecayrate_m = 2\pi \times \SI{1.2}{\hertz}$ for all relevant flexural modes. This corresponds to a quality factor of $\phonfreq_\rpos/\phondecayrate_m = 10^5$ at the frequency of the radial trap. Quality factors of this magnitude have been achieved for silica microspikes by optimization of the shape of the taper \cite{pennetta_tapered_2016}. \subCref{fig: reduce heating}{a} shows that a decrease of the radial heating rate below the value expected without a resonator for flexural modes (see \cref{tab: heating rates}) is predicted for resonator lengths $\clamplen \lesssim \SI{3}{\milli\meter}$. A length of $\clamplen = \SI{50}{\micro\meter}$ to the very left of \subcref{fig: reduce heating}{a} can still be achieved for nanofibers, the calculated heating rate due to flexural phonon modes with the given decay rate is then as low as \SI{0.1}{\milli\hertz}. \subCref{fig: reduce heating}{b} assumes a resonator length of $\clamplen = \SI{600}{\micro\meter}$, achieving heating rates of around $\SI{1}{\hertz}$ and shows the dependence on the trap frequency. The spacing between resonances is on the order of $2\pi \times \SI{50}{\kilo\hertz}$, which would indeed render it possible to detune the radial and azimuthal trap from resonance.

These findings suggest that it may be possible to significantly reduce the heating rate of atomic motion in nanofiber-based traps by two orders of magnitude or more through optimization of the phononic properties of the fiber. Moreover, the scaling of the heating rate with the mass of the trapped particles as $\scatrate^\thermal_i \propto \atmass$ is highly relevant for optomechanical experiments. Setups with levitated nanoparticles, for instance, may feature comparable trap frequencies for particles that are orders of magnitude heavier than a single atom \cite{magrini_near-field_2018,diehl_optical_2018}. In order to stably trap heavier particles using nanophotonic structures and successfully cool their motion, it is imperative to carefully manage vibrations of the structure, for instance by improving the mechanical stability or by tuning mechanical modes out of resonance with the particle motion.

\section*{Conclusion}
In this article, we formulate a general theoretical framework for calculating the effect of phonons on guided optical modes and the resulting heating of atoms in nanophotonic traps. Our results are applicable to nanophotonic cold-atom systems~\cite{chang_colloquium_2018} and can readily be extended to the heating of dielectric nanoparticles trapped close to surfaces~\cite{magrini_near-field_2018,diehl_optical_2018}. In a case study for the example of cold cesium atoms in a two-color nanofiber-based optical trap, we predict heating rates of the atomic center-of-mass motion which are in excellent agreement with independently measured values~\cite{albrecht_fictitious_2016,reitz_coherence_2013}. In this system, the dominant contribution to heating stems from thermally occupied flexural modes of the nanofiber. We find that the heating rate scales with the fiber radius as $R^{-5/2}$. As a general design rule, this implies that structures of larger lateral dimensions are preferable regarding heating, albeit at the expense of smaller mode confinement and, hence, potentially lower atom-photon coupling strength. Given the fact that the heating rate is directly proportional to the temperature of the nanophotonic structure, reducing the absorption losses of the guided trapping light fields is advisable~\cite{ravets_intermodal_2013}. Moreover, heating is expected to decrease for smaller trap frequencies, $\Gamma \propto \sqrt{\omega}$. In general, our case study shows that careful design of the phononic properties of the nanophotonic system and, in particular, of its mechanical resonances is an effective strategy for reducing the heating. Finally, by providing a coherent theoretical framework in a single source, our work is instrumental in calculating, understanding, and managing heating in a plethora of nanophotonic traps.

\begin{acknowledgments}
We thank Y.~Meng for the experimental characterization of the torsional mode resonances of the tapered optical fiber in the nanofiber-based two-color trap setup. Financial support by the European Research Council (CoG NanoQuaNt) and the Austrian Academy of Sciences (ÖAW, ESQ Discovery Grant QuantSurf) is gratefully acknowledged. We acknowledge support by the Austrian Federal Ministry of Science, Research, and Economy (BMWFW).
\end{acknowledgments}

\appendix

\section{Photonic Eigenmodes}
\label{sec: photon appendix}
The potential experienced by an atom in a nanophotonic trap crucially depends the optical fields surrounding the photonic structure. The dynamics of optical fields in the presence of nonabsorbing matter is well described by the macroscopic Maxwell equations. In conjunction with linear response theory, they allow us to model materials using the relative permittivity and permeability tensors $\relpermitttens$ and $\relpermeabtens$, respectively \cite{jackson_classical_1999}. In this article, we consider dielectric materials that are not magnetizable ($\relpermeabtens = \id$). Motion and vibration of the dielectric can be modeled as a change of $\relpermitttens$ over time, provided this change happens on a timescale long compared to the frequency of electromagnetic radiation in the optical regime. With this in mind, we choose a description of the optical fields in terms of photonic eigenmodes \cite{joannopoulos_photonic_2011}, which lends itself well to a perturbative treatment of the effect of a modified permittivity on the optical fields \cite{snyder_optical_2012} as we discuss in \cref{sec: interaction appendix}.

After reviewing photonic eigenmodes in general, we describe the eigenmode structure of a nanofiber approximated as a homogeneous step-profile circular optical waveguide \cite{snyder_optical_2012,marcuse_light_1982,davis_lasers_1996}.

\subsection{Photonic Eigenmode Equation}

Consider a dielectric body in three dimensions. The body may be inhomogeneous and anisotropic, so its relative permittivity $\relpermitttens$ is a position-dependent tensor of second order. We assume that the permittivity is independent of frequency in the relevant interval, real valued, symmetric, and positive definite. In the vacuum outside the body $\relpermitttens=\id$. We are interested in the dynamics of the electromagnetic fields $\Efield$ and $\Bfield$ surrounding and permeating the dielectric. We express the electromagnetic fields through potentials and follow \cite{glauber_quantum_1991} in choosing the Coulomb gauge for the vector potential $\vecpot$,
\begin{equation}\label{eqn: Coulomb gauge condition}
  \div \spare{\relpermitttens(\pos) \vecpot(\post)} = 0~.
\end{equation}
Here, the juxtaposition of tensor and vector (or, more generally, of two tensors) indicates the maximal contraction $(\relpermitttens\vecpot)^i = \sum_j \relpermitttenscomp^{ij}\vecpotcomp^j$. In the absence of free charges, the electromagnetic fields can be represented solely through the vector potential, $\Efield = \ddelt\vecpot$ and $\Bfield = \curl \vecpot$. The macroscopic Maxwell equations reduce to
\begin{equation}\label{eqn: Maxwell equation for potentials}
  \frac{1}{\cvac^2} \relpermitttens(\pos) \ddeltt{\vecpot}(\post) = \Dphot \vecpot(\post)~,
\end{equation}
where $\Dphot \equiv - \curl \spare{ \curl \cdot~}$ is the double curl operator, each dot represents a time derivative $\ddelt\vecpot = \delt \vecpot$, and $\cvac$ is the vacuum light speed. In order to find solutions, one solves \cref{eqn: Maxwell equation for potentials} outside and inside the body separately and then uses continuity conditions to match the solutions at the interface: The magnetic field $\Bfield$ as well as the electric field component $\Efield\times \normvec$ orthogonal to the surface normal $\normvec$ are continuous across the surface. Normal to the surface, $(\relpermitttens \Efield) \cdot \normvec $ is continuous instead. \Cref{eqn: Maxwell equation for potentials}, together with the continuity conditions and the requirement that solutions be square integrable to ensure finite electromagnetic energy, has a unique solution given suitable initial conditions \cite{jackson_classical_1999}.

The above problem can be further reduced to an eigenvalue problem \cite{cohen-tannoudji_photons_2004,joannopoulos_photonic_2011,glauber_quantum_1991}, which is useful for describing phonon-induced perturbations of optical fields in \cref{sec: interaction appendix}. To this end, consider the generalized eigenvalue equation for photonic eigenmodes $\amode_\photindex$,
\begin{equation}\label{eqn: eigenvalue equation}
  \Dphot \amode_\photindex(\pos) = -\frac{\photfreq_\photindex^2}{\cvac^2}  \relpermitttens(\pos) \amode_\photindex(\pos)~,
\end{equation}
with the additional transversality constraint \cref{eqn: Coulomb gauge condition}. The eigenmodes are labeled by a suitable multi-index $\photindex$ which may contain both discrete and continuous indices. They span a subspace characterized by \cref{eqn: Coulomb gauge condition} of the space of square-integrable functions \cite{glauber_quantum_1991}. The eigenvalues $\photfreq_\photindex^2/\cvac^2$ are real and positive, since $(-\Dphot)$ acting on that space is a self-adjoint, positive semidefinite operator, and $\relpermitttens$ is a positive definite operator \cite{joannopoulos_photonic_2011}. For the same reason, different eigenmodes are orthogonal with respect to the measure $\relpermitttens(\pos)\dr$, and we assume that they are normalized according to
\begin{equation}\label{eqn: normalization photon a modes}
  \int \cconj\amode_\photindex(\pos) \cdot \spare{\relpermitttens(\pos)\amode_\photindexb(\pos)} \dd \pos = \kronecker_{\photindex\photindexb}~.
\end{equation}
For discrete indices, $\kronecker$ is the Kronecker symbol, while it is the delta distribution for continuous indices.

Any solution to Maxwell's equations can then be expanded in terms of eigenmodes of well-defined frequencies~%
\footnote{This equation can be derived by starting from the Lagrange density
$\Lagdens = \vacpermitt (\rootrelpermitttens \ddelt{\vecpot})^2/2 - ( \curl \vecpot )^2/2\vacpermeab$
adapted from \cite{glauber_quantum_1991} for anisotropic media, passing to the Hamilton formulation, and expanding all fields in terms of the eigenmodes \cite{cohen-tannoudji_photons_2004}. The tensor $\rootrelpermitttens$ here is the root of the permittivity, $\rootrelpermitttens^2 = \relpermitttens$, and the properties of the permittivity ensure that the root is unique, positive definite, and symmetric. The Hamilton functional is then of the form
$\Hamilfunc = \vacpermitt^{-1} \sum_\photindex \pare{ \photnormvar_\photindex\cconj\photnormvar_\photindex + \cconj\photnormvar_\photindex \photnormvar_\photindex}$.
},
\begin{equation}\label{eqn: mode expansion vector potential}
    \vecpot(\post) = \sum_\photindex \frac{1}{\photfreq_\photindex\vacpermitt} \spare{\photnormvar_\photindex e^{-\im \photfreq_\photindex \tm} \amode_\photindex(\pos) + \cc }~,
\end{equation}
where the coefficients $\photnormvar_\photindex \in \C$ are obtained from the initial conditions.
We define the modal fields of the electric and magnetic field as
\begin{align}\label{eqn: electric and magnetic modal fields}
    \emode_\photindex(\pos) &\equiv \frac{\im}{\vacpermitt} \amode_\photindex(\pos) &
    \bmode_\photindex(\pos) &\equiv \frac{1}{\photfreq_\photindex\vacpermitt} \curl \amode_\photindex(\pos)
\end{align}
for convenience, such that
\begin{equation}\label{eqn: modal field expansion electric and magnetic field}
  \begin{split}
  \Efield(\post) &= \sum_\photindex \spare{\photnormvar_\photindex \emode_\photindex(\pos)e^{-\im \photfreq_\photindex\tm} + \cc }\\
  \Bfield(\post) &= \sum_\photindex \spare{\photnormvar_\photindex \bmode_\photindex(\pos)e^{-\im \photfreq_\photindex\tm} + \cc }~.
  \end{split}
\end{equation}

The problem of solving Maxwell's equations in the presence of a dielectric body has therefore been reduced to finding the photonic eigenmodes $\amode_\photindex$ of that body.
Although it is sufficient to treat the optical fields classically for our purpose, note that such a description is also suitable for canonical quantization of the electromagnetic field in the presence of lossless media \cite{cohen-tannoudji_photons_2004,glauber_quantum_1991}.

\subsection{Photonic Fiber Eigenmodes}

\begin{figure*}
  \centering
  \setlength{\widthFigA}{177.2223pt}
  \setlength{\marginLeftFigA}{23.9223pt}
  \setlength{\marginRightFigA}{7pt}
  \setlength{\widthFigB}{160.3pt}
  \setlength{\marginLeftFigB}{7pt}
  \setlength{\marginRightFigB}{7pt}
  \setlength{\widthFigC}{164.3907pt}
  \setlength{\marginLeftFigC}{7pt}
  \setlength{\marginRightFigC}{11.0907pt}
  \parbox[t]{\widthFigA}{\vspace{0pt}\includegraphics[width = \widthFigA]{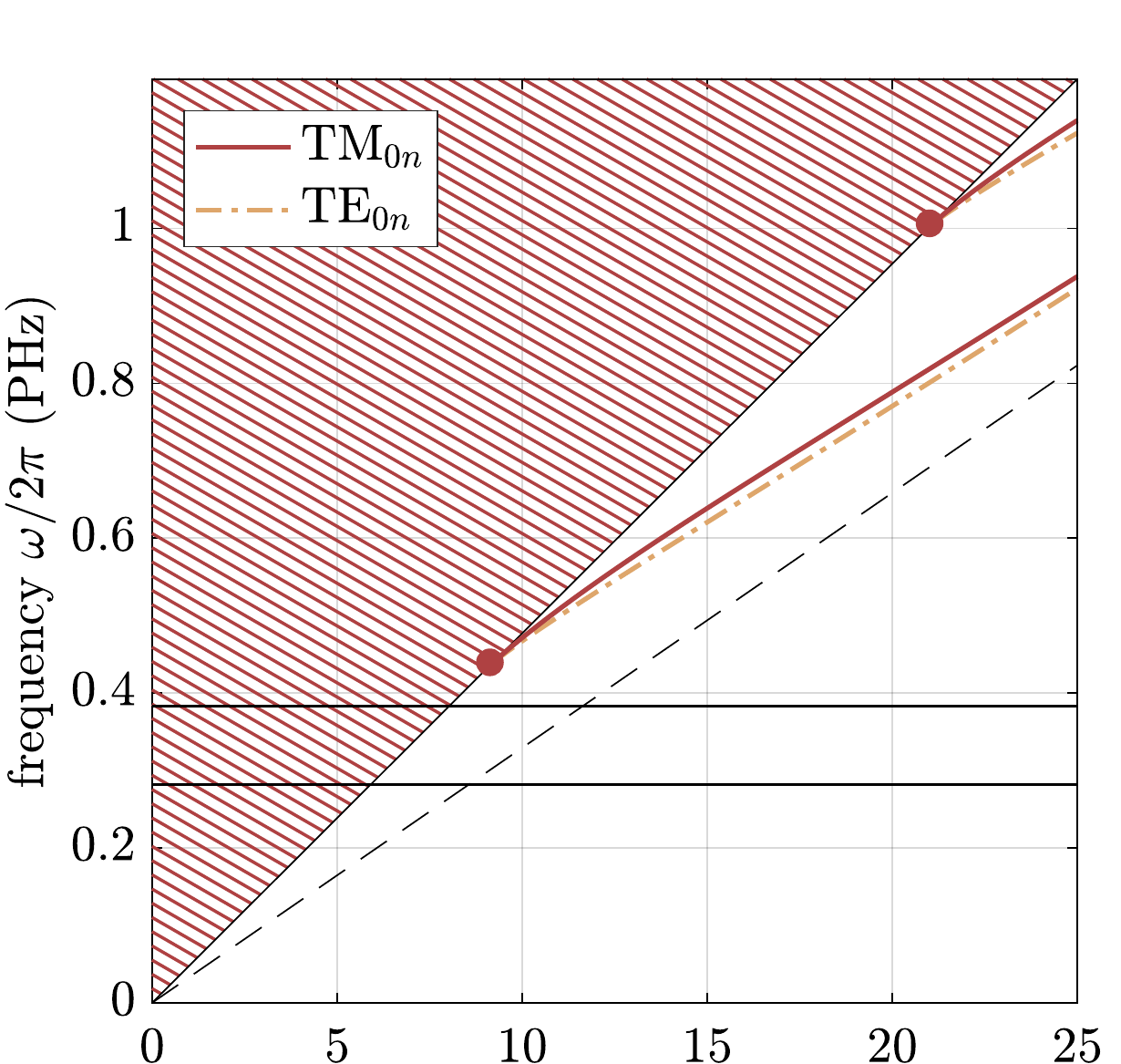}}%
  \parbox[t]{\widthFigB}{\vspace{0pt}\includegraphics[width = \widthFigB]{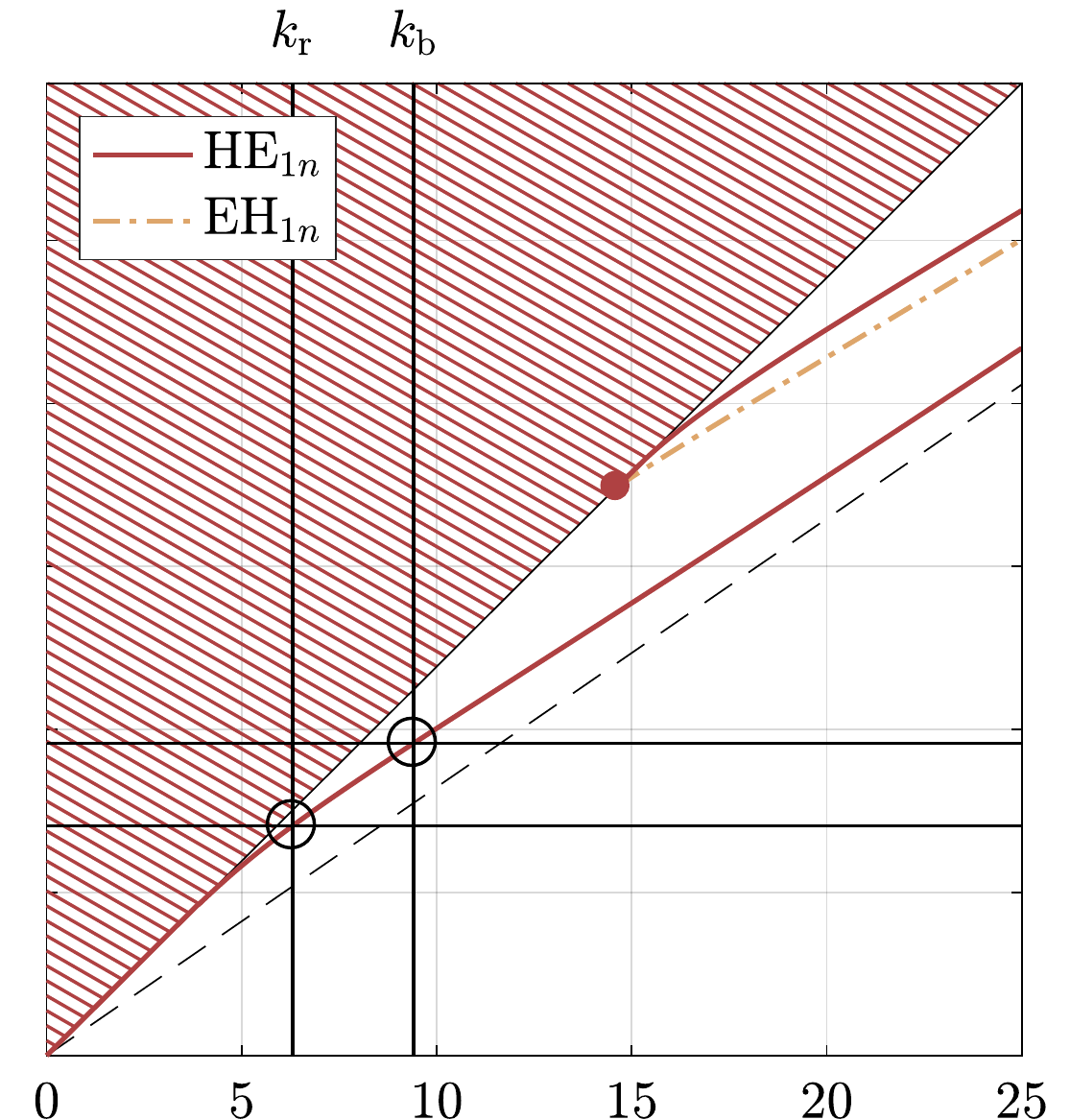}}%
  \parbox[t]{\widthFigC}{\vspace{0pt}\includegraphics[width = \widthFigC]{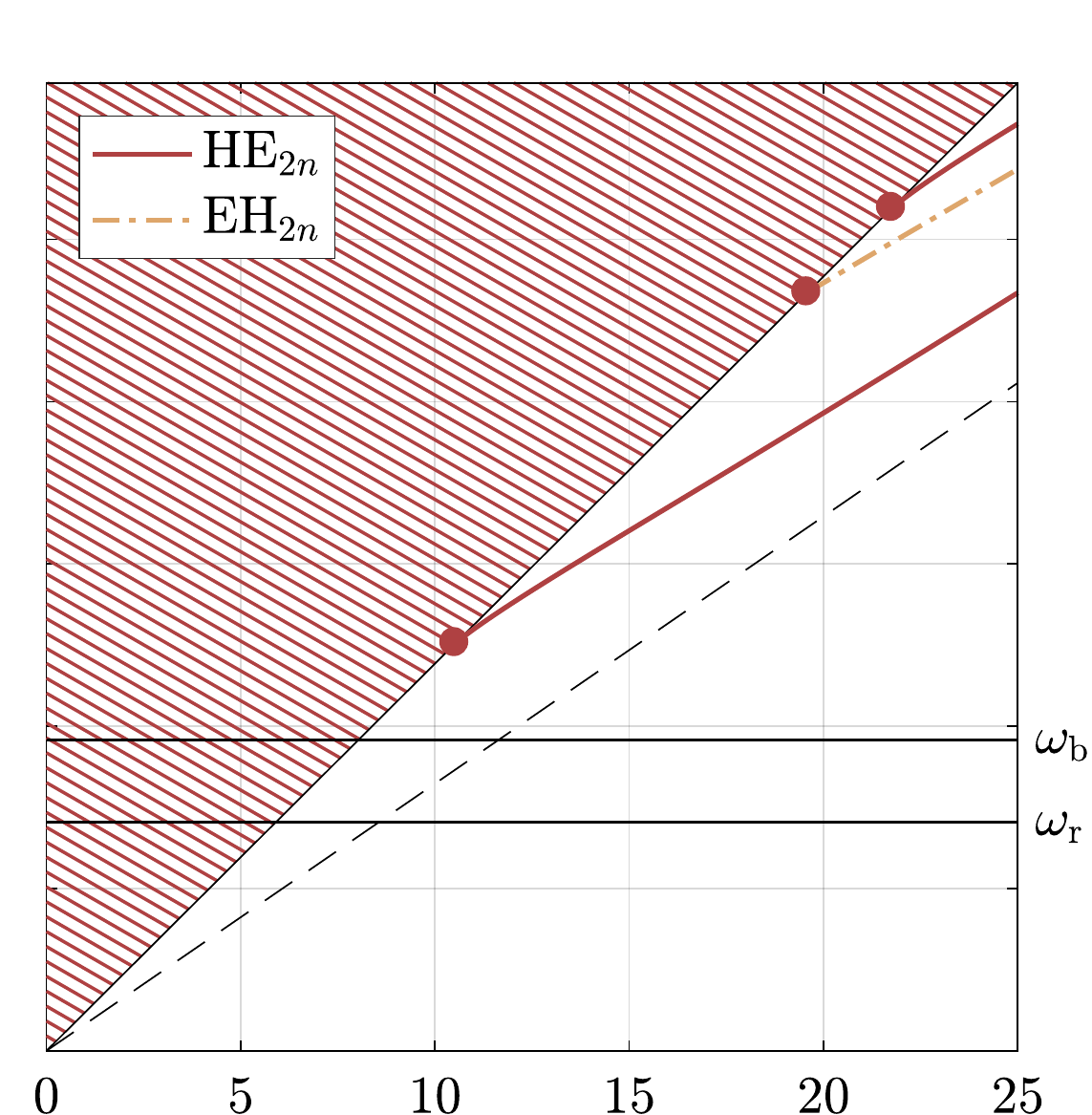}}\\
  \vspace{2pt}
  \parbox[t]{\widthFigA}{~}%
  \parbox[t]{\widthFigB}{\centering\footnotesize propagation constant $\photk$ $(1/\si{\micro\meter})$}%
  \parbox[t]{\widthFigC}{~}\\
  \vspace{7pt}
  \parbox[t]{\widthFigA}{\hspace*{\marginLeftFigA}(a) $\photl=0$\hspace*{\marginRightFigA}}%
  \parbox[t]{\widthFigB}{\hspace*{\marginLeftFigB}(b) $\photl=\pm1$\hspace*{\marginRightFigB}}%
  \parbox[t]{\widthFigC}{\hspace*{\marginLeftFigC}(c) $\photl=\pm2$\hspace*{\marginRightFigC}}%
  \caption{Band structures of nanofiber photon modes with azimuthal order $\photl=0$ in panel (a), $\photl=\pm 1$ in panel (b), and  $\photl=\pm 2$ in panel (c). The radius $\rad$ and the relative permittivity $\relpermitt$ are specified in \cref{sec: case study appendix}. The hatched area corresponds to radiative modes  delimited by the vacuum light line. The dashed line delineates the dielectric light line. The cutoff frequencies, where guided bands cross over into the radiative continuum, are indicated by solid points. The modes populated in the case study in \cref{sec: case study} are indicated by circles: Their frequencies $\photfreq_\reddetuned, \photfreq_\bluedetuned$ lie below all cutoff frequencies, so only modes on the fundamental $\HEmode_{11}$ band can be populated.}
  \label{fig: photon bands}
\end{figure*}

\begin{table}
  \setlength{\extrarowheight}{3pt}
  \newcolumntype{A}{>{\begin{math}}r<{\end{math}}}
  \newcolumntype{B}{>{\begin{math}}c<{\end{math}}}
  \newcolumntype{C}{>{\begin{math}}l<{\end{math}}}
  \begin{tabularx}{\columnwidth}[t]{ABC @{\quad} ABCX}
    \toprule
    \phota &=& \sqrt{\photfreq^2/\cbody^2 - \photk^2} & \photaa &=& -\im \phota & ~\\
    \photb &=& \sqrt{\photfreq^2/\cvac^2 - \photk^2} & \photbb &=& -\im \photb\\
    \photfreq_\cbody &=& \cbody |\photk| & \photfreq_\cvac &=& \cvac |\photk| \\
    \midrule
    \photaR &=& \phota\rad & \photaaR &=& \photaa \rad\\
    \photbR &=& \photb\rad & \photbbR &=& \photbb \rad\\
    \photkR &=& \photk \rad & \rposR &=& \rpos/ \rad\\
    \photw &=& \freq \rad/\cvac & \eta &=& \besselJ{\photl}(\photaR)/\besselK{\photl}(\photbbR)\\
    \gamma &=& \photkR \photw (\relpermitt - 1)\besselJ{\photl}(\photaR) \besselK{\photl}(\photbbR) \\
    && \multicolumn{5}{r}{$\times\cpare{\photaR \photbbR\spare{\photaR \besselJ{\photl}(\photaR) \besselK{\photl}'(\photbbR) + \photbbR \besselJ{\photl}'(\photaR) \besselK{\photl}(\photbbR) }}^{-1}$}  \\
    \signum_\photk &=& \photk/|\photk| & \signum_\photl &=& \photl/|\photl|\\
    \bottomrule
  \end{tabularx}
  \caption{Definitions of the radial constants $\phota$ and $\photb$, as well as the dimensionless quantities appearing in the photon modal fields. The definitions are given in terms of the azimuthal order $\photl$, propagation constant $\photk$, frequency $\photfreq$, radial position $\rpos$, fiber radius $\rad$, and the light speed $\cvac$ in vacuum and $\cbody \equiv \cvac/\sqrt{\relpermitt}$ in the fiber, respectively.}
  \label{tab: photon quantities}
\end{table}

We now consider an optical nanofiber in vacuum modeled as a cylinder of radius $\rad$, infinite length, and homogeneous and isotropic permittivity  $\relpermitttens = \relpermitt \id$. As the eigenmodes of such a fiber and their spectrum are well known \cite{snyder_continuous_1971,marcuse_light_1982,davis_lasers_1996,snyder_optical_2012}, we limit the discussion to their salient properties and list explicit expressions for the electric and magnetic modal fields as well as their dispersion relations.

We choose cylindrical coordinates $(\rpos,\phipos,\zpos)$, with \mbox{$\xpos = \rpos \cos\phipos$}, \mbox{$\ypos = \rpos \sin\phipos$}, and the $\zpos$ axis coinciding with the fiber axis. We do not solve the generalized eigenvalue equation (\ref{eqn: eigenvalue equation}) directly in order to obtain the eigenmodes of the vector potential. Instead, we remain on the level of electric and magnetic fields and solve the macroscopic Maxwell equations for constant permittivity $\relpermitt$ in the spectral domain,
\begin{equation}\label{eqn: Maxwell equations spectral domain}
  \begin{aligned}
        \div \emode(\pos) &= 0 &
        \div \bmode(\pos) &= 0 \\
        \curl \bmode(\pos) &= -\im\frac{\photfreq}{\cbody^2} \emode(\pos) &
        \curl \emode(\pos) &= \im \photfreq  \bmode(\pos)~.
  \end{aligned}
\end{equation}
Here, $\cbody\equiv\cvac/\sqrt{\relpermitt}$ inside the fiber, and $\cbody$ is replaced with $\cvac$ outside the fiber. \Cref{eqn: Maxwell equations spectral domain} can be solved in vacuum and in the dielectric separately. Both sets of solutions are then matched on the fiber surface according to the continuity conditions given above to find the modal fields $\emode_\photindex$ and $\bmode_\photindex$. The eigenmodes $\amode_\photindex$ can be obtained by inverting \cref{eqn: electric and magnetic modal fields}.

The first step is to solve \cref{eqn: Maxwell equations spectral domain} in the presence of an infinite isotropic medium of arbitrary homogeneous relative permittivity $\relpermitt>0$ (including vacuum $\relpermitt=1$ and dielectric $\relpermitt>1$). The solution space is spanned by electric and magnetic fields of the form
\begin{equation}\label{eqn: vector wave equation solutions}
  \begin{split}
    \emode(\pos) &= \frac{\emoder(\rpos)}{2\pi} e^{\im(\photl \phipos + \photk \zpos)} \\
    \bmode(\pos) &= \frac{\bmoder(\rpos)}{2\pi} e^{\im(\photl \phipos + \photk \zpos)}~.
  \end{split}
\end{equation}
The \emph{propagation constant} $\photk \in \R$ labels continuous excitations along the fiber axis, and the \emph{azimuthal order} $\photl \in \Z$ discrete excitations in the azimuthal direction. The radial partial waves $\emoder$ and $\bmoder$ depend on the magnitude of the frequency $\photfreq$ compared to the light line $\photfreq_\cbody \equiv \cbody |\photk|$ ($\cbody = \cvac$ for $\relpermitt=1$) and are given in \cref{tab: ME solutions cylindrical coordinates} in terms of the quantities defined in \cref{tab: photon quantities}.

In the second step, a first set of solutions $\emode,\bmode$ inside the fiber (with permittivity $\relpermitt>1$, light speed $\cbody\equiv\cvac/\sqrt{\relpermitt}$, and \emph{dielectric radial constant} $\phota$ defined in \cref{tab: photon quantities}) is matched to a second set $\tilde\emode,\tilde\bmode$ of solutions outside the fiber (with $\relpermitt=1$, light speed $\cvac$, and \emph{vacuum radial constant} $\photb$ defined in \cref{tab: photon quantities}). The continuity conditions require $\relpermitt \emodecomp^{\rpos} = \tilde\emodecomp^{\rpos}$ on the fiber surface, while the magnetic field and the remaining two components of the electric field need to be continuous. These conditions lead to electric and magnetic modal fields $\emode_\photindex,\bmode_\photindex$ together with frequency equations governing their eigenfrequency $\photfreq_\photindex$. Mode quadruplets $(\pm\photl,\pm\photk)$ are degenerate in frequency $\photfreq_\photindex$. Let us adopt the notation that solutions inside the fiber have radial partial waves with amplitudes $A,B,C,D$ (see \cref{tab: ME solutions cylindrical coordinates}), and solutions outside the fiber have primed amplitudes $A',B',C',D'$. At most two of these eight amplitudes are not fixed by the continuity conditions and the requirement that the modal fields be bounded, corresponding to one or two independent mode families.

The eigenmodes have markedly different properties depending on how their eigenfrequency $\photfreq_\photindex$ compares to the \emph{vacuum light line} $\photfreq_\cvac \equiv \cvac |\photk|$ and the \emph{dielectric light line} $\photfreq_\cbody \equiv \cbody |\photk|$. We distinguish three cases: Modes with frequencies above the vacuum light line are radiative modes, and modes with frequencies between the vacuum and dielectric light line are fiber-guided modes. Modes on the vacuum light line are weakly guided (they decay polynomially away from the fiber surface). On the dielectric light line and below, $\photfreq_\cbody \geq \photfreq_\photindex$, no modes can exist \cite{snyder_continuous_1971}.

\begin{table*}
  \setlength{\extrarowheight}{2pt}
  \newcolumntype{A}{>{\begin{math}}r<{\end{math}}}
  \newcolumntype{B}{>{\begin{math}}c<{\end{math}}}
  \newcolumntype{C}{>{\begin{math}}l<{\end{math}}}
  \begin{tabularx}{\textwidth}{l ABC @{\quad} ABC X}
    \toprule
    \multicolumn{4}{l}{Case} & \multicolumn{4}{l}{Solution} \\
    \midrule
    (1) & \photfreq &>& \photfreq_0 &
    \emodercomp^{\rpos} &=& \im\photaR^{-2} \cpare{ \photkR \photaR \spare{A \besselJ{\photl}'(\photaR \rposR) + B\besselY{\photl}'(\photaR \rposR)} + \im \photl \photw \spare{C \besselJ{\photl}(\photaR \rposR) + D\besselY{\photl}(\photaR \rposR)}/\rposR } & ~ \\ &
    \photl &\in& \Z &
    \emodercomp^{\phipos} &=&  \im \photaR^{-2} \cpare{ \im \photl \photkR \spare{A \besselJ{\photl}(\photaR \rposR) + B\besselY{\photl}(\photaR \rposR)}/\rposR - \photw\photaR \spare{C \besselJ{\photl}'(\photaR \rposR) + D\besselY{\photl}'(\photaR \rposR)} } \\ &&&&
    \emodercomp^{\zpos} &=&  A\besselJ{\photl}(\photaR \rposR) + B \besselY{\photl}(\photaR \rposR)\\ &&&&
    \bmodercomp^{\rpos} &=&  \im \photaR^{-2} \cpare{ \photkR \photaR \spare{C \besselJ{\photl}'(\photaR \rposR) + D\besselY{\photl}'(\photaR \rposR)} - \im \photl \relpermitt  \photw \spare{A \besselJ{\photl}(\photaR \rposR) + B\besselY{\photl}(\photaR \rposR)}/\rposR }/\cvac \\ &&&&
    \bmodercomp^{\phipos} &=&  \im \photaR^{-2} \cpare{ \im \photl \photkR \spare{C \besselJ{\photl}(\photaR \rposR) + D\besselY{\photl}(\photaR \rposR)}/\rposR + \relpermitt \photw\photaR \spare{A \besselJ{\photl}'(\photaR \rposR) + B\besselY{\photl}'(\photaR \rposR)} }/\cvac \\ &&&&

    \bmodercomp^{\zpos} &=&  \spare{ C\besselJ{\photl}(\photaR \rposR) + D \besselY{\photl}(\photaR \rposR)} / \cvac\\
    \midrule
    (2a) & \photfreq &=& \photfreq_0 &
    \emodercomp^{\rpos} &=& A \rposR^{-1} + B \rposR \\ &
    \photl&=&0 & %
    \emodercomp^{\phipos} &=& \im \pare{C \rposR^{-1} + D \rposR} \\ &&&&
    \emodercomp^{\zpos} &=& 2\im \photkR^{-1}B \\ &&&&
    \bmodercomp^{\rpos} &=& -\signum_\photk \im  \pare{C \rposR^{-1} + D \rposR}\sqrt{\relpermitt}/\cvac \\ &&&&
    \bmodercomp^{\phipos} &=& \signum_\photk  \pare{A \rposR^{-1} + B \rposR}\sqrt{\relpermitt}/\cvac\\ &&&&
    \bmodercomp^{\zpos} &=& \signum_\photk 2\photkR^{-1} D\sqrt{\relpermitt}/\cvac \\
    \midrule
    (2b) & \photfreq &=& \photfreq_0 &
    \emodercomp^{\rpos} &=& A \rposR^{-2} + B \rposR^{2} + C + D \ln(\rposR)\\ &
     |\photl|&=&1 & %
    \emodercomp^{\phipos} &=&  \signum_\photl \im \spare{- A \rposR^{-2} - B \rposR^{2}  + C + D \ln(\rposR)}\\ &&&&
    \emodercomp^{\zpos} &=& \im \photkR^{-1}\pare{ D \rposR^{-1} + 4B \rposR }\\ &&&&
    \bmodercomp^{\rpos} &=& \signum_\photk\signum_\photl \im \spare{ \pare{ A+ D/\photkR^2} \rposR^{-2} + B \rposR^{2} - \pare{C - 4B/\photkR^2 } - D \ln(\rposR) }\sqrt{\relpermitt}/\cvac \\ &&&&
    \bmodercomp^{\phipos} &=& \signum_\photk \spare{ \pare{ A+ D/\photkR^2} \rposR^{-2} + B \rposR^{2} + \pare{C - 4B/\photkR^2 } + D \ln(\rposR) } \sqrt{\relpermitt}/\cvac \\ &&&&
    \bmodercomp^{\zpos} &=& \signum_\photk\signum_\photl \photkR^{-1}  \pare{ D \rposR^{-1} -  4B \rposR } \sqrt{\relpermitt}/\cvac \\
    \midrule
    (2c) & \photfreq &=& \photfreq_0 &
    \emodercomp^{\rpos} &=& A \rposR^{-|\photl|-1} + B \rposR^{-|\photl|+1}
       + C \rposR^{|\photl|-1} + D \rposR^{|\photl|+1} \\ &
    |\photl|&\geq&2 & %
    \emodercomp^{\phipos} &=& \signum_\photl \im \pare{- A \rposR^{-|\photl|-1} + B \rposR^{-|\photl|+1}
       + C\rposR^{|\photl|-1} - D \rposR^{|\photl|+1}} \\ &&&&
    \emodercomp^{\zpos} &=& - 2\im \photkR^{-1} \spare{ (|\photl|-1)B \rposR^{-|\photl|} - (|\photl|+1)D \rposR^{|\photl|} }\\ &&&&
    \bmodercomp^{\rpos} &=& \signum_\photk \signum_\photl  \im \big\{ \spare{A - 2|\photl|(|\photl|-1) B/\photkR^2 } \rposR^{-|\photl|-1} - B \rposR^{-|\photl|+1} - \spare{C - 2|\photl|(|\photl|+1)D/\photkR^2} \rposR^{|\photl|-1} \\
    \multicolumn{8}{r}{$+ D \rposR^{|\photl|+1} \big\}\sqrt{\relpermitt}/\cvac$} \\ &&&&
    \bmodercomp^{\phipos} &=& \signum_\photk  \big\{ \spare{A - 2|\photl|(|\photl|-1) B/\photkR^2 } \rposR^{-|\photl|-1} + B \rposR^{-|\photl|+1} + \spare{C - 2|\photl|(|\photl|+1)D/\photkR^2} \rposR^{|\photl|-1} \\
    \multicolumn{8}{r}{$+ D \rposR^{|\photl|+1} \big\}\sqrt{\relpermitt}/\cvac$} \\ &&&&
    \bmodercomp^{\zpos} &=& - \signum_\photk \signum_\photl 2\photkR^{-1} \spare{ (|\photl|-1)B \rposR^{-|\photl|} + (|\photl|+1) D \rposR^{|\photl|} }\sqrt{\relpermitt}/\cvac\\
    \midrule
    (3) & \photfreq &<& \photfreq_0 &
    \emodercomp^{\rpos} &=& - \im \photaaR^{-2} \cpare{ \photkR \photaaR \spare{A \besselI{\photl}'(\photaaR \rposR) + B\besselK{\photl}'(\photaaR \rposR)}
    + \im \photl \photw \spare{C \besselI{\photl}(\photaaR \rposR) + D\besselK{\photl}(\photaaR \rposR)}/\rposR } \\ &
    \photl &\in& \Z &
    \emodercomp^{\phipos} &=& -\im \photaaR^{-2} \cpare{ \im \photl \photkR \spare{A \besselI{\photl}(\photaaR \rposR) + B\besselK{\photl}(\photaaR \rposR)}/\rposR
    - \photw\photaaR \spare{C \besselI{\photl}'(\photaaR \rposR) + D\besselK{\photl}'(\photaaR \rposR)} } \\ &&&&
    \emodercomp^{\zpos} &=&  A\besselI{\photl}(\photaaR \rposR) + B \besselK{\photl}(\photaaR \rposR)\\ &&&&
    \bmodercomp^{\rpos} &=& -\im\photaa^{-2} \cpare{ \photkR \photaaR \spare{C \besselI{\photl}'(\photaaR \rposR) + D\besselK{\photl}'(\photaaR \rposR)}
    - \im \photl \relpermitt  \photw \spare{A \besselI{\photl}(\photaa \rposR) + B\besselK{\photl}(\photaaR \rposR)}/\rposR }/\cvac \\ &&&&
    \bmodercomp^{\phipos} &=& -\im \photaaR^{-2} \cpare{ \im \photl \photkR \spare{C \besselI{\photl}(\photaaR \rposR) + D\besselK{\photl}(\photaaR \rposR)}/\rposR
    + \relpermitt \photw\photaaR \spare{A \besselI{\photl}'(\photaaR \rposR) + B\besselK{\photl}'(\photaaR \rposR)} }/\cvac\\ &&&&
    \bmodercomp^{\zpos} &=& \spare{ C\besselI{\photl}(\photaaR \rposR) + D \besselK{\photl}(\photaaR \rposR) }/\cvac\\
    \bottomrule
  \end{tabularx}
  \caption{Radial partial waves of the solutions \cref{eqn: vector wave equation solutions} to Maxwell's equations  using cylindrical coordinates in the spectral domain. Inside the fiber, $\photfreq_0=\photfreq_\cbody$. In vacuum outside the fiber, $\photfreq_0=\photfreq_\cvac$, and $\relpermitt=1$ such that $\photaR$ is replaced with $\photbR$. The quantities $A,B,C,D\in\C$ are amplitudes. The radial dependence is given by Bessel functions $\besselJ{\photl}, \besselY{\photl}$ above the light line, by modified Bessel functions $\besselI{\photl}, \besselK{\photl}$ below the light line, and by polynomials and the natural logarithm $\ln$ on the light line. The prime indicates the first derivative $\besselJ{\photl}'(\xpos) = \delx\besselJ{\photl}(\xpos)$. All other quantities used are defined in \cref{tab: photon quantities}.}
  \label{tab: ME solutions cylindrical coordinates}
\end{table*}

\begin{table*}
  \setlength{\extrarowheight}{4pt}
  \newcolumntype{A}{>{\begin{math}}r<{\end{math}}}
  \newcolumntype{B}{>{\begin{math}}c<{\end{math}}}
  \newcolumntype{C}{>{\begin{math}}l<{\end{math}}}
  \begin{tabular*}{\textwidth}{ABC}
    \toprule
    A' &=& \cpare{A\spare{\photbR \relpermitt \besselY{\photl}(\photbR) \besselJ{\photl}'(\photaR)/\photaR - \besselY{\photl}'(\photbR)\besselJ{\photl}(\photaR)}
    + \im C \photl \photkR \pare{\photbR^2/\photaR^2-1} \besselY{\photl}(\photbR)\besselJ{\photl}(\photaR)/\photbR \photw}/N\\
    B' &=& \spare{ A \cpare{ \besselJ{\photl}(\photaR)N -\besselJ{\photl}(\photbR) \spare{\photbR \relpermitt \besselY{\photl}(\photbR) \besselJ{\photl}'(\photaR)/\photaR - \besselY{\photl}'(\photbR)\besselJ{\photl}(\photaR)} }/\besselY{\photl}(\photbR)
     - \im C \photl \photkR \pare{\photbR^2/\photaR^2-1} \besselJ{\photl}(\photbR)\besselJ{\photl}(\photaR)/\photbR \photw }/N\\
    C' &=& \cpare{-\im A \photl \photkR \pare{\photbR^2/\photaR^2-1} \besselY{\photl}(\photbR)\besselJ{\photl}(\photaR)/\photbR \photw
    + C \spare{\photbR \besselY{\photl}(\photbR) \besselJ{\photl}'(\photaR)/\photaR - \besselY{\photl}'(\photbR)\besselJ{\photl}(\photaR)}}/N\\
    D' &=& \spare{\im A \photl \photkR \pare{\photbR^2/\photaR^2-1} \besselJ{\photl}(\photbR)\besselJ{\photl}(\photaR)/\photbR \photw
     + C \cpare{ \besselJ{\photl}(\photaR)N -\besselJ{\photl}(\photbR) \spare{\photbR \besselY{\photl}(\photbR) \besselJ{\photl}'(\photaR)/\photaR - \besselY{\photl}'(\photbR)\besselJ{\photl}(\photaR)} }/\besselY{\photl}(\photbR)}/N\\
    \bottomrule
  \end{tabular*}
  \caption{Radiative modes: The radial partial waves of modal fields \cref{eqn: vector wave equation solutions} are given by \mbox{case (1)} in \cref{tab: ME solutions cylindrical coordinates} both inside and outside the fiber, with amplitudes listed in this table. Unprimed amplitudes apply inside the fiber, primed amplitudes outside the fiber. The amplitudes $A,C\in\C$ are independent, and $B=D=0$. We define $N\equiv{\besselY{\photl}(\photbR)\besselJ{\photl}'(\photbR) - \besselY{\photl}'(\photbR)\besselJ{\photl}(\photbR)}$. All other quantities are defined in \cref{tab: photon quantities}.}
  \label{tab: radiative mode amplitudes}
\end{table*}

\begin{table*}
  \setlength{\extrarowheight}{5pt}
  \newcolumntype{A}{>{\begin{math}}r<{\end{math}}}
  \newcolumntype{B}{>{\begin{math}}c<{\end{math}}}
  \newcolumntype{C}{>{\begin{math}}l<{\end{math}}}
  \begin{tabular*}{\textwidth}{C @{\qquad} A B C @{\qquad} A B C }
    \toprule
    \multicolumn{1}{l}{Band} & \multicolumn{3}{l}{Fiber ($x<1$)} & \multicolumn{3}{l}{Vacuum ($x>1$)} \\
    \midrule
    \TEmode_{0\photn}, \TMmode_{0\photn} &
    \emodercomp^{\rpos}_{\photindex} &=& -\im \photkR \photaR^{-1} B \besselJ{1}(\photaR \rposR) &
    \emodercomp^{\rpos}_{\photindex} &=& \im  \eta\photkR \photbbR^{-1}  B \besselK{1}(\photbbR \rposR) \\&
    \emodercomp^{\phipos}_{\photindex} &=& - \photw \photaR^{-1} A \besselJ{1}(\photaR\rposR) & \emodercomp^{\phipos}_{\photindex} &=& \eta\photw \photbbR^{-1} A\besselK{1}(\photbbR\rposR) \\&
    \emodercomp^{\zpos}_{\photindex} &=& B\besselJ{0}(\photaR \rposR) &
    \emodercomp^{\zpos}_{\photindex} &=& \eta B \besselK{0}(\photbbR \rposR) \\&
    \bmodercomp^{\rpos}_{\photindex} &=& \photkR\photaR^{-1} A \besselJ{1}(\photaR\rposR)/\cvac &  \bmodercomp^{\rpos}_{\photindex} &=& -\eta\photkR \photbbR^{-1}  A\besselK{1}(\photbbR\rposR)/\cvac \\&
    \bmodercomp^{\phipos}_{\photindex} &=& -\im \relpermitt \photw \photaR^{-1} B \besselJ{1}(\photaR \rposR)/\cvac &
    \bmodercomp^{\phipos}_{\photindex} &=& \im \eta\photw \photbbR^{-1} B \besselK{1}(\photbbR \rposR)/\cvac \\&
    \bmodercomp^{\zpos}_{\photindex} &=& \im A \besselJ{0}(\photaR \rposR)/\cvac &
    \bmodercomp^{\zpos}_{\photindex} &=& \im\eta A\besselK{0}(\photbbR \rposR)/\cvac \\
    \midrule
    \HEmode_{|\photl|\photn}, \EHmode_{|\photl|\photn} &
    \emodercomp^{\rpos}_{\photindex} &=& \im \photaR^{-2} A\spare{ \photkR \photaR \besselJ{\photl}'(\photaR\rposR) -  \photl^2 \photw \gamma \besselJ{\photl}(\photaR \rposR)/\rposR } &
    \emodercomp^{\rpos}_{\photindex} &=& -\im \eta \photbbR^{-2} A\spare{ \photkR \photbbR \besselK{\photl}'(\photbbR\rposR) - \photl^2 \photw \gamma \besselK{\photl}(\photbbR \rposR)/\rposR } \\&
    \emodercomp^{\phipos}_{\photindex} &=& \photl\photaR^{-2} A\spare{ \photw \gamma \photaR \besselJ{\photl}'(\photaR\rposR) - \photkR \besselJ{\photl}(\photaR \rposR)/\rposR } &
    \emodercomp^{\phipos}_{\photindex} &=& - \photl\eta\photbbR^{-2}  A\spare{ \photw \gamma \photbbR \besselK{\photl}'(\photbbR\rposR) - \photkR \besselK{\photl}(\photbbR \rposR)/\rposR } \\&
    \emodercomp^{\zpos}_{\photindex} &=& A\besselJ{\photl}(\photaR \rposR) &
    \emodercomp^{\zpos}_{\photindex} &=& \eta A \besselK{\photl}(\photbbR \rposR) \\&
    \bmodercomp^{\rpos}_{\photindex} &=& -\photl\photaR^{-2} A\spare{ \photkR \gamma \photaR \besselJ{\photl}'(\photaR\rposR) - \relpermitt \photw \besselJ{\photl}(\photaR \rposR)/\rposR }/\cvac &
    \bmodercomp^{\rpos}_{\photindex} &=& \photl\eta \photbbR^{-2}  A\spare{ \photkR \gamma \photbbR \besselK{\photl}'(\photbbR\rposR) - \photw \besselK{\photl}(\photbbR \rposR)/\rposR }/\cvac \\&
    \bmodercomp^{\phipos}_{\photindex} &=& \im \photaR^{-2} A\spare{ \relpermitt \photw \photaR \besselJ{\photl}'(\photaR\rposR) - \photl^2 \photkR \gamma \besselJ{\photl}(\photaR \rposR)/\rposR }/\cvac &
    \bmodercomp^{\phipos}_{\photindex} &=& - \im \eta \photbbR^{-2} A\spare{ \photw \photbbR \besselK{\photl}'(\photbbR\rposR) - \photl^2 \photkR \gamma \besselK{\photl}(\photbbR \rposR)/\rposR }/\cvac \\&
    \bmodercomp^{\zpos}_{\photindex} &=& \im \photl\gamma  A\besselJ{\photl}(\photaR \rposR)/\cvac &
    \bmodercomp^{\zpos}_{\photindex} &=& \im \photl\eta\gamma A\besselK{\photl}(\photbbR \rposR)/\cvac \\
    \bottomrule
  \end{tabular*}
  \caption{Guided modes: Radial partial waves of the modal fields \cref{eqn: vector wave equation solutions} both inside the fiber and in the vacuum surrounding the fiber. The quantities $A,B\in\C$ are amplitudes. The $\TEmode_{0\photn}$ modes are characterized by $A=0$, and the $\TMmode_{0\photn}$ modes by $B=0$. The $\HEmode_{|\photl|\photn}$ and $\EHmode_{|\photl|\photn}$ modes have the same modal fields but distinct frequencies; see \cref{tab: guided modes frequency equations}. The remaining amplitude is determined from the normalization condition \cref{eqn: normalization radial partial waves of E field mode}. All other quantities are defined in \cref{tab: photon quantities}.}
  \label{tab: guided modes modal fields}
\end{table*}

\begin{table*}
  \setlength{\extrarowheight}{5pt}
  \newcolumntype{A}{>{\begin{math}}r<{\end{math}}}
  \newcolumntype{B}{>{\begin{math}}c<{\end{math}}}
  \newcolumntype{C}{>{\begin{math}}l<{\end{math}}}
  \begin{tabular*}{.25\textwidth}[t]{C @{\quad} ABC}
  \toprule
  \multicolumn{1}{l}{Band} & \multicolumn{3}{l}{Amplitudes}\\
  \midrule
  \TEmode_{0\photn} &
  C &\in& \C  \\&
  C' &=& \signum_{\photk}C \photkR \photaR^{-1} \besselJ{1}(\photaR) \\
  \TMmode_{0\photn} &
  A &\in& \C \\&
  A' &=& - \im A \relpermitt \photkR \photaR^{-1} \besselJ{1}(\photaR) \\
  \bottomrule
  \end{tabular*}
  \hfill
  \begin{tabular*}{.29\textwidth}[t]{C @{\quad} ABC}
  \toprule
  \multicolumn{1}{l}{Band} & \multicolumn{3}{l}{Amplitudes}\\
  \midrule
  \HEmode_{1\photn} &
  C &\in& \C \\&
  A' &=& \signum_\photk C \photl \photkR\photaR^{-1} \besselJ{\photl}'(\photaR)   /2 \\&
  C' &=& - \signum_\photk C \photl\photkR \photaR^{-1} \besselJ{\photl}'(\photaR)   /2\\
  \EHmode_{1\photn} &
  A &\in& \C \\&
  A' &=& \im A\relpermitt \photkR\photaR^{-2} \besselJ{\photl}'(\photaR)/2 \\&
  C' &=& \im A\relpermitt \photkR\photaR^{-2} \besselJ{\photl}'(\photaR)  /2 \\
  \bottomrule
  \end{tabular*}
  \hfill
  \begin{tabular*}{.43\textwidth}[t]{C @{\quad} ABC}
  \toprule
  \multicolumn{1}{l}{Band} & \multicolumn{3}{l}{Amplitudes}\\
  \midrule
  \HEmode_{|\photl|\photn} &
  A &\in& \C \\
  |\photl|\geq 2 &
  C &=& -\im A \signum_\photl \signum_\photk  \\&
  A' &=& \im A (\relpermitt - 1) \photkR \besselJ{\photl}(\photaR) / 2 (\relpermitt + 1) (|\photl|-1) \\&
  B' &=& \im A \photkR \besselJ{\photl}(\photaR)/ 2(|\photl|-1) \\
  \EHmode_{|\photl|\photn} &
  C &\in& \C \\
  |\photl|\geq 2 &
  A &=& -\im \signum_\photl \signum_\photk C /\relpermitt \\&
  A' &=& \signum_\photl \signum_\phonk C \photkR \photaR^{-1} \besselJ{\photl}'(\photaR)\\
  \bottomrule
  \end{tabular*}
  \caption{Weakly guided modes: The modal fields \cref{eqn: vector wave equation solutions} have radial partial waves given by \mbox{case (1)} in \cref{tab: ME solutions cylindrical coordinates} inside the fiber, and by \mbox{case (2)} outside the fiber, with nonzero amplitudes listed in this table. Unprimed amplitudes apply inside the fiber, primed amplitudes outside the fiber. Amplitudes that are not listed vanish. All other quantities are defined in \cref{tab: photon quantities}.}
  \label{tab: weakly guided mode amplitudes}
\end{table*}

\begin{table*}
  \setlength{\extrarowheight}{5pt}
  \newcolumntype{C}{>{\begin{math}}l<{\end{math}}}
  \begin{tabular*}{\textwidth}{C @{\qquad} C @{\qquad} C }
    \toprule
    \multicolumn{1}{l}{Band} & \text{Frequency equation} &  \text{Cutoff frequency} \\
    \midrule
    \TEmode_{0\photn} &
    \photaR \besselJ{0}(\photaR) \besselK{1}(\photbbR) + \photbbR \besselK{0}(\photbbR) \besselJ{1}(\photaR) = 0 &
    \besselJ{0}(\photaR) = 0 \\
    \TMmode_{0\photn} &
    \photaR \besselJ{0}(\photaR) \besselK{1}(\photbbR) + \relpermitt \photbbR \besselK{0}(\photbbR) \besselJ{1}(\photaR) = 0 &
    \besselJ{0}(\photaR) = 0 \\
    \midrule
    \HEmode_{|\photl|\photn} &
    \text{Odd roots of \cref{eqn: guided mode frequency equation}} &
    \spare{|\photl|(|\photl|-1) - \photaR^2 /(\relpermitt + 1) } \besselJ{\photl}(\photaR) + (|\photl|-1) \photaR \besselJ{\photl}'(\photaR) = 0 \\
    \EHmode_{|\photl|\photn} &
    \text{Even roots of \cref{eqn: guided mode frequency equation}}&
    \besselJ{\photl}(\photaR) = 0 \\
    \bottomrule
  \end{tabular*}
  \caption{Guided modes: Frequency equations and cutoff frequencies in terms of the quantities defined in \cref{tab: photon quantities}. The roots of these equation form discrete bands in the $(\photk,\photfreq)$ plane, as shown in \cref{fig: photon bands}.}
  \label{tab: guided modes frequency equations}
\end{table*}

\emph{Radiative modes} are distinguished by $\photfreq_\photindex > \photfreq_\cvac$. The modal fields $\emode_\photindex, \bmode_\photindex$ have radial partial waves given by \mbox{case (1)} in \cref{tab: ME solutions cylindrical coordinates} both inside and outside the fiber, with amplitudes listed in \cref{tab: radiative mode amplitudes}. There are two independent amplitudes, which implies two mode families can be distinguished. In defining these mode families, special care has to be taken to ensure they are orthogonal according to \cref{eqn: normalization photon a modes}; see \cite{snyder_continuous_1971} for details. Radiative modes are not confined to the fiber, but permeate all of space. In consequence, their excitation spectrum in the radial direction is continuous; that is, for each $(\photl,\photk)$ any eigenfrequency $\photfreq_\photindex > \photfreq_\cvac$ is admissible, and a continuous index is required to label them. A possible choice is the radial constant $\phota$, which is real valued and positive for radiative modes. The dispersion relation of radiative modes is hence
\begin{equation}\label{eqn: dispersion relation radiative modes}
  \photfreq_\photindex = \cbody \sqrt{ \photk^2 + \phota^2 }~,
\end{equation}
and they form a continuum above the vacuum light line in the $(\photk,\photfreq)$ plane, as shown in \cref{fig: photon bands}.

\emph{Guided modes} are characterized by $\photfreq_\cvac > \photfreq_\photindex > \photfreq_\cbody$. The radial partial waves of their modal fields are listed explicitly in \cref{tab: guided modes modal fields}, and the corresponding frequency equations in \cref{tab: guided modes frequency equations}. Guided modes can propagate inside the fiber but decay exponentially far outside the fiber. For guided modes with azimuthal order $\photl = 0$, either the electric field or the magnetic field may be transverse, while the other acquires longitudinal components. Hence, guided modes with $\photl=0$ fall into two mode families: transverse-electric ($\TEmode$) modes characterized by $\bmodecomp_\photindex^\zpos=0$ and transverse-magnetic ($\TMmode$) modes characterized by $\emodecomp^\zpos_\photindex = 0$. Only a discrete set of frequencies is admissible for each $(\photfam,\photl,\photk)$ because the fields are radially confined. These frequencies correspond to the roots of the frequency equations listed in \cref{tab: guided modes frequency equations}. The frequencies $\photfreq_\photindex(\photk)$ form discrete bands in the $(\photk,\photfreq)$ plane, see \subcref{fig: photon bands}{a}. The bands can be labeled by a band index $\photn \in \N$ starting from $\photn=1$ for the band of lowest frequency and increasing in frequency with $\photn$.
The modal fields of the $\TEmode$ and $\TMmode$ modes have the following symmetries with respect to the propagation constant:
\begin{equation}\label{eqn: guided modal field symmetries m=0}
  \begin{aligned}
    \emodecomp^{\rpos}(-\photk) &= -\emodecomp^{\rpos}(\photk) \qquad & \bmodecomp^{\rpos}(-\photk) &= - \bmodecomp^{\rpos}(\photk) \\
    \emodecomp^{\phipos}(-\photk) &= \emodecomp^{\phipos}(\photk) & \bmodecomp^{\phipos}(-\photk) &= \bmodecomp^{\phipos}(\photk) \\
    \emodecomp^{\zpos}(-\photk) &= \emodecomp^{\zpos}(\photk) & \bmodecomp^{\zpos}(-\photk) &= \bmodecomp^{\zpos}(\photk)~,
  \end{aligned}
\end{equation}
where we drop all constant mode indices.

For higher azimuthal excitations $|\photl| \geq 1$, both electric and magnetic field have longitudinal components and there is only a single hybrid mode family. The resulting bands shown in \subcrefand{fig: photon bands}{b}{c} derive from the frequency equation
\begin{multline}\label{eqn: guided mode frequency equation}
  \spare{ \photaR \besselJ{\photl}(\photaR) \besselK{\photl}'(\photbbR) + \photbbR \besselK{\photl}(\photbbR) \besselJ{\photl}'(\photaR) } \\
   \times \spare{ \photaR \besselJ{\photl}(\photaR) \besselK{\photl}'(\photbbR) + \relpermitt \photbbR \besselK{\photl}(\photbbR) \besselJ{\photl}'(\photaR) }\\
  = \spare{\photl \frac{\photkR \photw}{\photaR \photbbR}(\relpermitt-1)\besselJ{\photl}(\photaR) \besselK{\photl}(\photbbR) }^2~,
\end{multline}
with parameters defined in \cref{tab: photon quantities}. There are, however, two subclasses of modes called $\HEmode$ and $\EHmode$ distinguished by the asymptotes of their bands as $|\photk|\to\infty$. Each band of $\HEmode$ ($\EHmode$) modes asymptotically approaches a root of the Bessel function of the first kind $\besselJ{\photl-1}(\phota\rad)$ $\pare{\besselJ{\photl+1}(\phota\rad)}$, and the electric (magnetic) field has a longitudinal component of significant magnitude compared to its transverse components  \cite{snyder_optical_2012,davis_lasers_1996}. As the propagation constant $\photk$ is the only continuous index for guided modes, the orthonormality condition \cref{eqn: normalization photon a modes} in conjunction with \cref{eqn: electric and magnetic modal fields} reduces to a normalization condition for the electric radial partial waves:
\begin{equation}\label{eqn: normalization radial partial waves of E field mode}
  \int_0^\infty \rpos  |\emoder_\photindex(\rpos)|^2 \relpermitt(\rpos)\dd\rpos = \frac{1}{\vacpermitt^2}~.
\end{equation}
The modal fields of the $\HEmode$ and $\EHmode$ modes have the following symmetries with respect to the azimuthal order $\photl$ and propagation constant $\photk$:
\begin{align}\label{eqn: HE EH modes symmetries}
  \emodecomp^{\rpos}_{\signum\photl}(\signum' \photk) &= \signum^\photl \signum' \emodecomp^{\rpos}_{\photl}(\photk) & \bmodecomp^{\rpos}_{\signum\photl}(\signum' \photk) &= \signum^{\photl+1} \bmodecomp^{\rpos}_{\photl}(\photk) \nonumber \\
  \emodecomp^{\phipos}_{\signum\photl}(\signum' \photk) &= \signum^{\photl+1} \signum' \emodecomp^{\phipos}_{\photl}(\photk) & \bmodecomp^{\phipos}_{\signum\photl}(\signum' \photk) &= \signum^\photl \bmodecomp^{\phipos}_{\photl}(\photk) \nonumber\\
  \emodecomp^{\zpos}_{\signum\photl}(\signum' \photk) &= \signum^\photl  \emodecomp^{\zpos}_{\photl}(\photk) & \bmodecomp^{\zpos}_{\signum\photl}(\signum' \photk) &= \signum^{\photl+1} \signum' \bmodecomp^{\zpos}_{\photl}(\photk) ~,
\end{align}
where $\signum,\signum' = \pm1$.

\emph{Weakly guided modes} with frequencies on the vacuum light line, $\photfreq_\cvac = \photfreq_\photindex$, appear where bands of guided modes cross over into the radiative continuum. Inside the fiber, the radial partial waves of the modal fields are given by \mbox{case (1)} listed in \cref{tab: ME solutions cylindrical coordinates} (since $\photfreq_\photindex > \photfreq_\cbody$). Outside the fiber, they are given by \mbox{case (2)} with $\relpermitt=1$, and therefore replacing $\photaR$ with $\photbR$; see \cref{tab: photon quantities}. The corresponding amplitudes are listed in \cref{tab: weakly guided mode amplitudes}. The modal fields decay polynomially outside the fiber, as a limiting case between oscillatory behavior and exponential decay. The frequency equations on the vacuum light line listed in \cref{tab: guided modes frequency equations} mark the \emph{cutoff frequencies} below which a band of guided modes ceases to exist. The cutoff frequencies are indicated by solid dots in \cref{fig: photon bands}.

In summary, guided modes can be labeled by mode indices
\begin{equation}\label{eqn: fiber photon mode indices}
  \begin{aligned}
    \photl &\in \Z~, \quad & \photfam &\in \{\TEmode, \TMmode\}_{\photl=0} \text{ or } \{\HEmode,\EHmode\}_{\photl\neq0}~, \\
    \photk &\in \R~, & \photn &\in \N ~.
  \end{aligned}
\end{equation}
It is customary to name guided bands as $\photfam_{|\photl|\photn}$. At $\photl=0$, there are then $\TEmode_{0\photn}$ and $\TMmode_{0\photn}$ bands, and at $|\photl|\geq1$ there are $\HEmode_{|\photl|\photn}$ and $\EHmode_{|\photl|\photn}$ bands. For a given azimuthal order $\photl$, the $\HEmode$ and $\EHmode$ bands alternate, with the $\HEmode_{|\photl|1}$ band of lowest frequencies, see \subcrefand{fig: photon bands}{b}{c}. At sufficiently low frequencies, only the $\HEmode_{11}$ band is guided, while all other bands merge into the radiative continuum (single-mode regime); compare \subcref{fig: photon bands}{b} to \subcrefand{fig: photon bands}{a}{c}. These low-frequency $\HEmode_{11}$ modes are used as trapping fields in nanofiber-based cold-atom traps.

\section{Atom Trap and Motional States}
\label{sec: atom appendix}
In the first part of this \namecref{sec: atom appendix}, we summarize how to obtain the optical and surface potentials experienced by an atom trapped close to a photonic structure. In the second part, we discuss under which conditions the potential can be approximated as harmonic in the case of a nanofiber-based atom trap.

\subsection{Trapping Potential}

\begin{figure*}%
  \centering
  \setlength{\widthFigA}{148.0767pt}
  \setlength{\marginLeftFigA}{24.0767pt}
  \setlength{\marginRightFigA}{7pt}
  \setlength{\widthFigB}{160.2467pt}
  \setlength{\marginLeftFigB}{36.2467pt}
  \setlength{\marginRightFigB}{7pt}
  \setlength{\widthFigC}{184.172pt}
  \setlength{\marginLeftFigC}{7pt}
  \setlength{\marginRightFigC}{47.6323pt}
  \parbox[t]{\widthFigA}{\vspace{0pt}\includegraphics[width = \widthFigA]{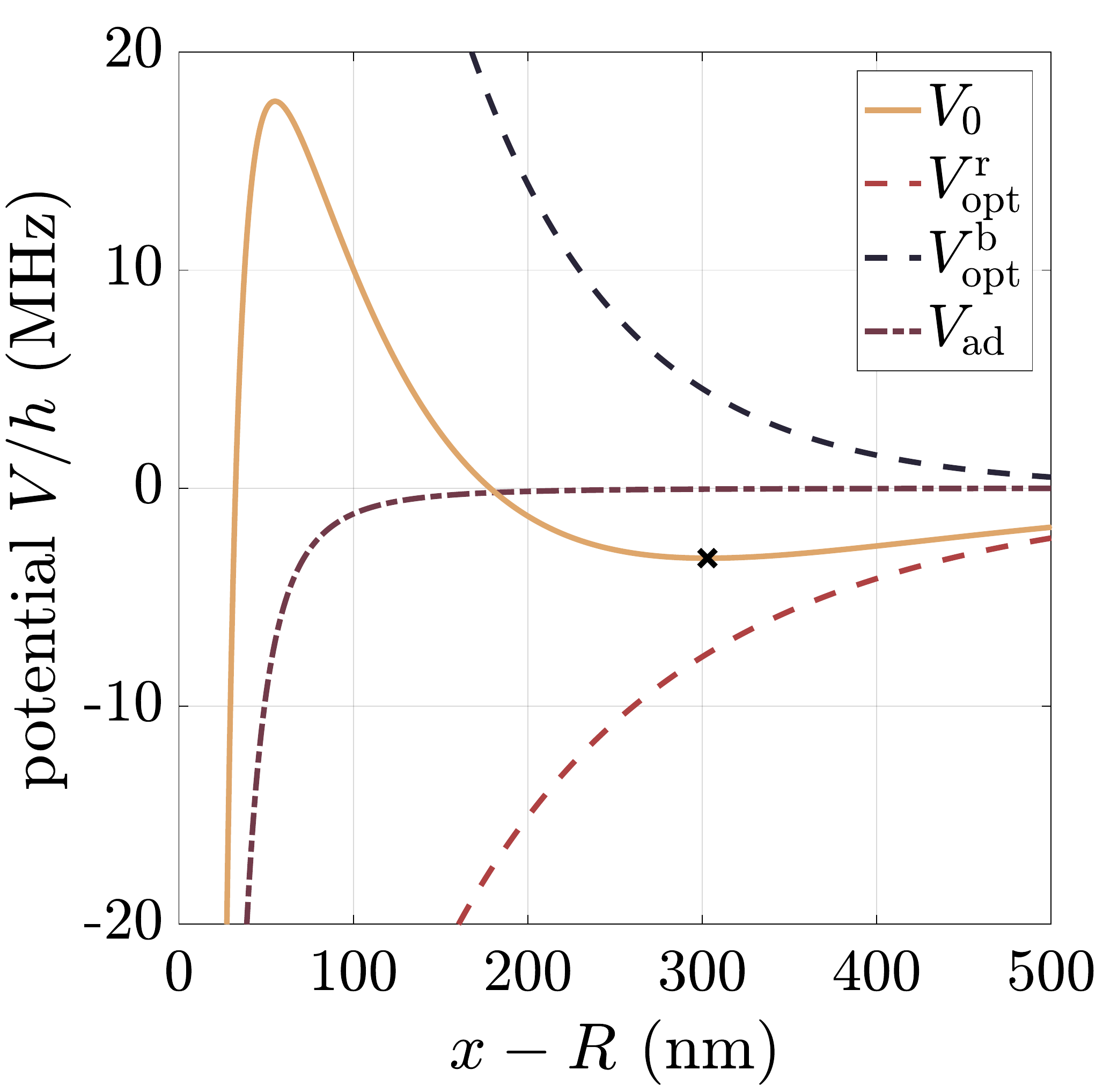}}%
  \parbox[t]{\widthFigB}{\vspace{0pt}\includegraphics[width = \widthFigB]{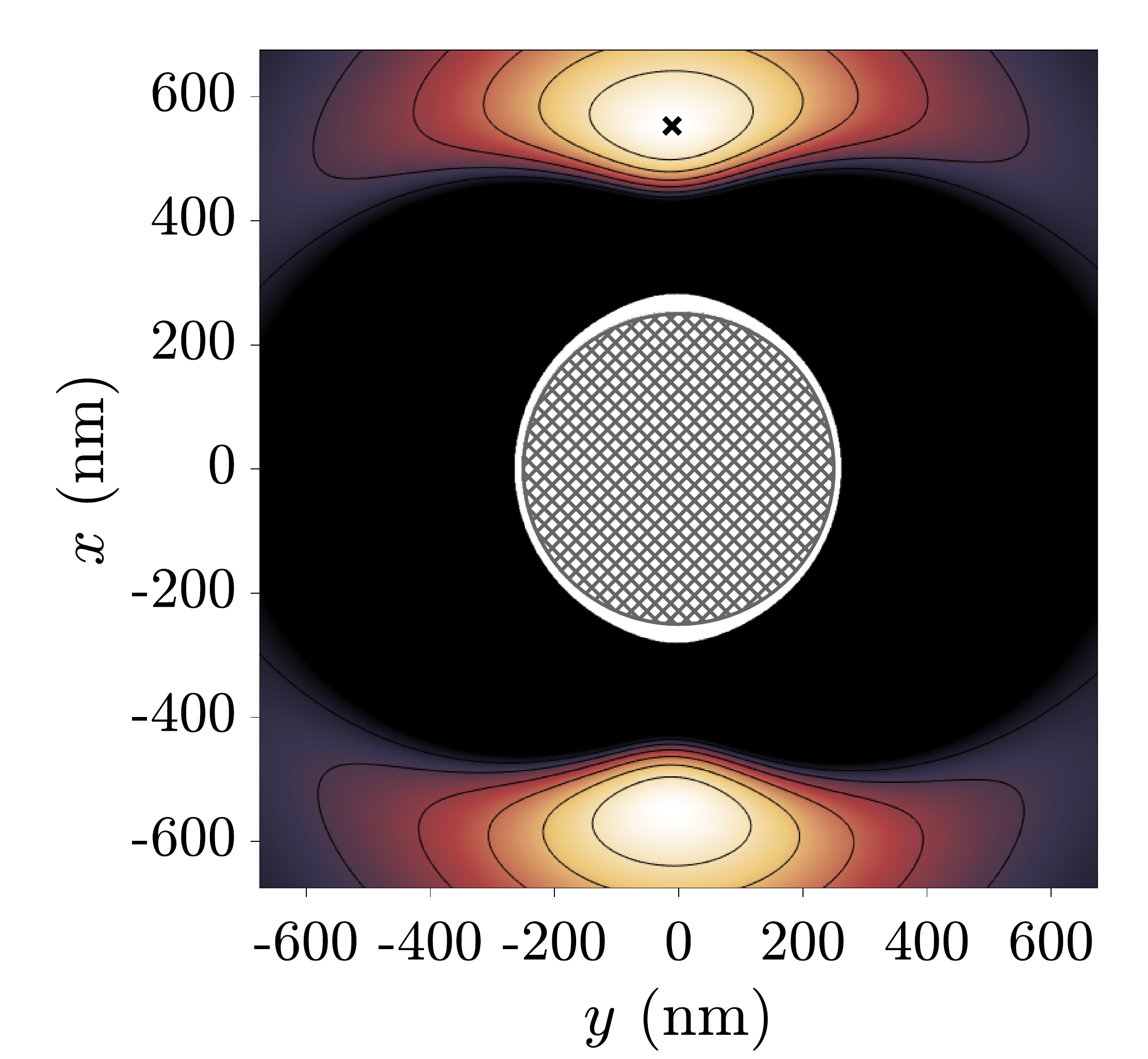}}%
  \parbox[t]{\widthFigC}{\vspace{0pt}\includegraphics[width = \widthFigC]{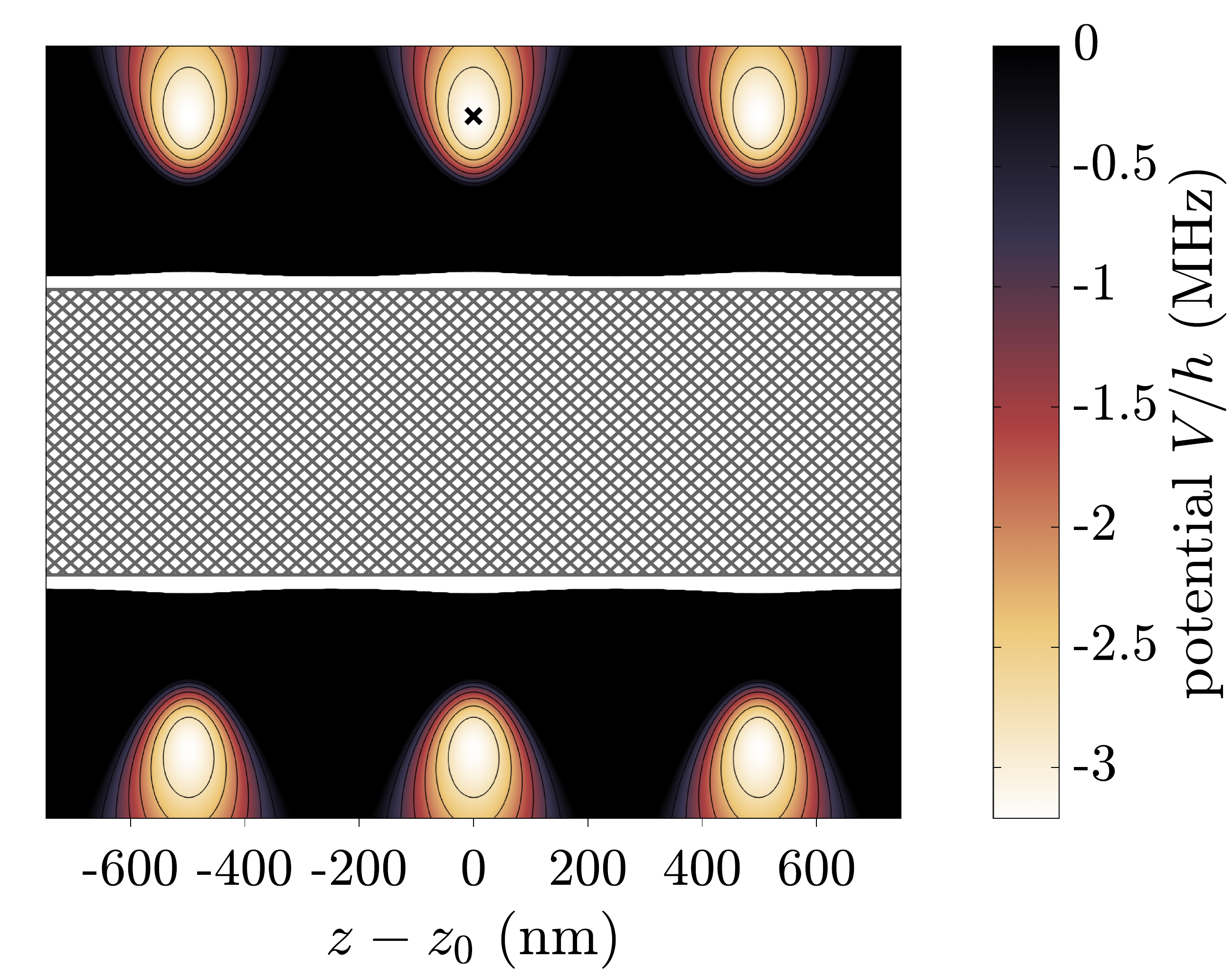}} \\
  \vspace{4pt}
  \parbox[t]{\widthFigA}{\hspace*{\marginLeftFigA}%
  \parbox[t]{\widthFigA-\marginLeftFigA-\marginRightFigA}{%
    (a) Contributions to potential at $\zpos = \trapz$, $\ypos = \trapy = 0$.%
  }%
  \hspace*{\marginRightFigA}}%
  \parbox[t]{\widthFigB}{\hspace*{\marginLeftFigB}%
  \parbox[t]{\widthFigB-\marginLeftFigB-\marginRightFigB}{%
    (b) Cross section at $\zpos = \trapz$.%
  }%
  \hspace*{\marginRightFigB}}%
  \parbox[t]{\widthFigC}{\hspace*{\marginLeftFigC}%
  \parbox[t]{\widthFigC-\marginLeftFigC-\marginRightFigC}{%
    (c) Longitudinal section at $\ypos = \trapy = 0$.%
  }%
  \hspace*{\marginRightFigC}}%
  \caption{Total potential experienced by the atom, with a trap minimum $\trappos$ indicated by the cross. \subCrefb{fig: trap}{a} shows the different contributions to the total potential $\pot_0$. \subCrefandb{fig: trap}{b}{c} show the total potential in a cross section and longitudinal section of the fiber, respectively. All positive values of the potential in \subcrefandb{fig: trap}{b}{c} are shown as black. The parameters used for this plot are listed in \cref{sec: case study appendix}.}
  \label{fig: trap}
\end{figure*}

The internal and external dynamics of an atom trapped close to a photonic structure is governed by the Hamiltonian
\begin{equation}\label{eqn: atom Hamiltonian general}
  \Hamilop_\atomic = \Hamilop_\internal + \frac{\atmomop^2}{2\atmass} + \potop_\optical + \potop_\adsorption
\end{equation}
in the absence of vibrations of the structure. Here, $\Hamilop_\internal$ describes the internal state of the atom, $\atmomop$ is the momentum operator of its center of mass, and the remaining terms are introduced in \cref{sec: framework}. The internal hyperfine-structure states of the atom can be labeled by the eigenvalues of a suitable set of commuting operators, for instance $\ket{\atHFSstate}\equiv\ket{\atintn \atS \atL \atJ \atI \atF}$, where $\atintn$ is the principal quantum number, $\atS$ is the electron spin, $\atL$ is the electron orbital angular momentum, $\atJ$ is the total electronic angular momentum, $\atI$ the nuclear spin, and $\atF$ the resulting total atom angular momentum \cite{le_kien_dynamical_2013}. The optical trapping fields are detuned from resonance with the atom, such that they do not excite the atom from the electronic ground state. Instead, both light and surface effects lead to a slight mixing (dressing) of the internal eigenstates $\ket{\atHFSstate}$ of the atom \cite{cohen-tannoudji_atom-photon_1998}. The new, dressed eigenstates have energies shifted by an amount typically much smaller than the splitting between hyperfine-structure levels of different $\atF$. The dressed eigenstates are therefore very similar to the bare eigenstates $\ket{\atHFSstate}$ and can be labeled using the same quantum numbers. Gradients in the light intensity then lead to position-dependent light shifts, which act as an optical potential for the center of mass and thus allow trapping of the atom \cite{grimm_optical_2000}. Both optical and surface potentials depend on the internal state $\ket{\atHFSstate}$ of the atom because the electric polarizability of the atom is state dependent \cite{le_kien_dynamical_2013,buhmann_dispersion_2012}. Moreover, the atom-light interaction can couple internal and motional states \cite{dareau_observation_2018,meng_near-ground-state_2018}.

We focus on scenarios without coupling of internal and motional states, such that the potential operators $\potop_\optical$ and $\potop_\adsorption$ are block diagonal in the dressed hyperfine-structure levels: $\potop_\optical + \potop_\adsorption = \sum_\atHFSstate \spare{\pot_{\optical, \atHFSstate}(\atposop) +\pot_{\adsorption,\atHFSstate}(\atposop)} \ket{\atHFSstate} \bra{\atHFSstate} $. The motion of the atom in the trap is thus governed by the Hamiltonian
\begin{equation}
  \Hamilop_\atomic = \frac{\atmomop^2}{2\atmass} + \pot_0(\atposop)~,
\end{equation}
in each subspace of the Hilbert space with fixed internal state $\ket{\atHFSstate}$.
Energies are measured relative to the energy $\braket{\atHFSstate|\Hamilop_\internal|\atHFSstate}$ of the internal state, and $\pot_0 \equiv \pot_{\optical,\atHFSstate} +\pot_{\adsorption,\atHFSstate}$ is the total state-dependent potential. In two-color traps in particular, two monochromatic light fields are used, tuned in opposite directions away from the resonance frequency of the atom \cite{le_kien_atom_2004}. The red-detuned field attracts the atom toward the surface of the nanophotonic structure. The repulsive blue-detuned field has a shorter decay length and dominates closer to the surface, keeping the atom at a distance. If the two light fields are sufficiently far detuned, interference between them is negligible, and the optical potential is the sum of the individual contributions $\pot_\optical^\reddetuned$ and $\pot_\optical^\bluedetuned$ of the red- and blue-detuned field, respectively. The total state-dependent potential is then
\begin{equation}\label{eqn: total potential}
  \pot_0 \equiv \pot_\optical^\reddetuned + \pot_\optical^\bluedetuned + \pot_\adsorption~,
\end{equation}
where we drop the index $\atHFSstate$ from notation. \Cref{fig: trap} shows an example of how these three contributions combine to a three-dimensional trapping potential close to an optical nanofiber.

The optical potential $\pot_\optical$ created by each monochromatic light field $\Efield(\post) = \Efield_0(\pos) e^{-\im\photfreq\tm} + \cc$ of frequency $\freq$ can be expressed in terms of a scalar, vector, and tensor light shift \cite{le_kien_dynamical_2013},
\begin{equation}\label{eqn: total light shift}
  \pot_\optical = \pot_\scalarshift + \pot_\vectorshift + \pot_\tensorshift~.
\end{equation}
We assume that there is a homogeneous magnetic offset field $\extBfield = \extBfieldcomp \ezvecB$ applied along the unit vector $\ezvecB$, which induces Zeeman splitting of the hyperfine structure. The internal dressed eigenstates of the atom are then the Zeeman substates $\ket{\atHFSstate} = \ket{\atFSstate \atF\atMF}$, provided the magnetic field is sufficiently strong to avoid mixing of the Zeeman substates by their relative light shifts. Here, $\ket{\atFSstate} = \ket{\atn \atS \atL \atJ \atI}$ is the fine-structure state, and $\atMF$ is the magnetic quantum number of the total atom angular momentum with respect to the quantization axis $\ezvecB$. The scalar light shift is
\begin{equation}\label{eqn: scalar light shift}
  \pot_\scalarshift(\atpos) = - \HFSscalarpolarizab |\Efield_0(\atpos)|^2~,
\end{equation}
with a scalar polarizability $\HFSscalarpolarizab$ that depends only on its fine-structure state $\ket{\atFSstate}$. The vector light shift is \cite{le_kien_dynamical_2013}
\begin{equation}\label{eqn: vector light shift}
  \pot_\vectorshift (\atpos) = - \frac{\HFSvectorpolarizab}{2\im}\frac{\atMF}{\atF}  \spare{\cconj\Efield_0(\atpos) \times \Efield_0(\atpos)} \cdot \ezvecB~.
\end{equation}
The vector polarizability $\HFSvectorpolarizab$ of the hyperfine-structure Zeeman substate $\ket{\atHFSstate}$ can be obtained from the vector polarizability $\FSvectorpolarizab$ of the fine-structure state $\ket{\atFSstate}$:
\begin{equation}
  \HFSvectorpolarizab = \frac{\atF(\atF+1) + \atJ(\atJ+1) - \atI (\atI+1)}{(\atF+1)2\atJ} \FSvectorpolarizab~.
\end{equation}
The tensor light shift is
\begin{equation}\label{eqn: tensor light shift}
  \pot_\tensorshift(\atpos) = - 3\HFStensorpolarizab \frac{3\atMF^2 - \atF(\atF+1)}{2\atF(2\atF-1)} \spare{|\Efieldcomp^\zposB_0(\atpos)|^2-\frac{1}{3}}~,
\end{equation}
where $\Efieldcomp_0^{\zposB} = \Efield_0 \cdot \ezvecB$, and the hyperfine-structure tensor polarizability $\HFStensorpolarizab$ is related to the fine-structure tensor polarizability $\FStensorpolarizab$ by
\begin{multline}
  \HFStensorpolarizab = (-1)^{(\atJ+\atI+\atF)} \sqrt{\frac{3(\atJ+1)(2\atJ+1)(2\atJ+3)}{2\atJ(2\atJ-1)}}\\
  \times \sqrt{\frac{2\atF(2\atF-1)(2\atF+1)}{3(\atF+1)(2\atF+3)} }  \begin{Bmatrix}\atF & 2 & \atF\\ \atJ & \atI & \atJ \end{Bmatrix} \FStensorpolarizab~.
\end{multline}
Here, $\{\cdot\}$ is the Wigner $6j$ symbol. The fine-structure polarizabilities $\FSscalarpolarizab = \HFSscalarpolarizab$, $\FSvectorpolarizab$, and $\FStensorpolarizab$ can be calculated from experimental data \cite{le_kien_dynamical_2013}.

At atom-surface separations realized in nanophotonic traps, surface effects are limited to dispersion forces~%
\footnote{%
At atomic distances from the surface, strong repulsion due to exchange energy can arise. Together with attractive dispersion forces, this can lead to adsorption of the atom on the surface \cite{kreuzer_physisorption_1986,zangwill_physics_1988}.
}.
The dispersion force between solids and atoms strongly depends on both geometry and material. Its effect can be modeled by the nonretarded Casimir-Polder potential for a two-level atom in the ground state, located in the vicinity of a dielectric object \cite{buhmann_dispersion_2012}. In the case of a nanofiber-based trap, it is sufficient to model the fiber as a dielectric half-space \cite{le_kien_atom_2004,le_kien_state-dependent_2013}, although
the potential for an atom close to a dielectric cylinder can in principle be calculated analytically \cite{schmeits_physical_1977,nabutovskii_interaction_1979,boustimi_van_2002,le_kien_atom_2004}. The potential is then
\begin{equation}\label{eqn: CP potential dielectric half-space}
  \pot_\adsorption(\rpos) = - \CPpotstrength (\rpos-\rad)^{-3}~.
\end{equation}
The parameter $\CPpotstrength>0$ can either be obtained experimentally \cite{stern_simulations_2011} or calculated from the electromagnetic properties of the material \cite{mclachlan_van_1964,schmeits_physical_1977,wylie_quantum_1984,buhmann_dispersion_2012}.

\subsection{Harmonic Trap Approximation}

A quadratic atom-phonon interaction Hamiltonian \cite{kustura_quadratic_2019} can be obtained by approximating the trapping potential as harmonic for atoms close in energy to the motional ground state. In the case of a nanofiber-trapped atom in particular, we choose cylindrical coordinates to describe the motion of the trapped atom. The Hamiltonian $\Hamilop_\atomic$ describing the motion of the atom in the trap in the absence of vibrations is then
\begin{equation}\label{eqn: Hamiltonian cylindrical coordinates}
  \Hamilop_\atomic = \frac{(\atrmomop)^2}{2\atmass}  - \frac{\hbar^2}{8\atmass\atrposop^2} + \frac{(\atphimomop)^2}{2\atmass\atrposop^2} + \frac{(\atzmomop)^2}{2\atmass} + \pot_0(\atposop)~,
\end{equation}
where $\atposop = (\atrposop, \atphiposop, \atzposop)$ is the position operator of the atom, and $(\atrmomop, \atphimomop, \atzmomop)$ are the components of the momentum operator. We expand the potential to second order around the local trap minimum $\trappos$. The corresponding motional frequencies of the atom with respect to the coordinates $(\atrposb, \atphiposb, \atzposb) \equiv (\rpos, \trapr\phipos,\zpos)$, are
\begin{align}\label{eqn: definition 3d trap frequencies}
  \itrapfreq &\equiv \sqrt{\frac{\partial_{x^i}^2 \pot(\trappos)}{\atmass}} &
  \ijtrapfreq &\equiv \sqrt{\frac{\partial_{x^i} \partial_{x^j} \pot(\trappos)}{\atmass}}~,
\end{align}
where $i,j \in \{\rpos,\phipos,\zpos\}$. The cross-derivatives of the potential may be nonzero, since the symmetry axes of the potential are in general not aligned with the coordinate axes (for instance when the magnetic offset field breaks the cylindrical symmetry of the setup). The Hamiltonian \cref{eqn: Hamiltonian cylindrical coordinates} then describes harmonic motion of the atom in each direction around the trap minimum, provided $\ijtrapfreq = 0$ and given that the trap is far from the fiber axis compared to its size: $\rtrapfreq \gg \hbar/(2 \atmass \trapr^2)$ and $\phitrapfreq \gg \hbar/(4\pi \atmass \trapr^2)$. Introducing ladder operators $\iaop$ and $\hconj\iaop$, the position operators $\atiposbop$ can then be expressed as
\begin{align}\label{eqn: zero-point fluctuations atom position 3d trap}
  \atiposbop & = \atiposbsd \pare{\iaop + \hconj\iaop} + \trapib
\end{align}
where $\trapib$ is the position of the trap minimum, and $(\atrposbsd,\atphiposbsd,\atzposbsd ) \equiv (\atrpossd, \trapr \atphipossd, \atzpossd)$ is the zero-point motion of the atom in the trap as defined in \cref{sec: framework}. If $\ijtrapfreq \neq 0$, there is additional cross-coupling between the motional modes of the atom, with coupling constants $\ijcrosscoupling = \ijtrapfreq^2/4\sqrt{\itrapfreq  \jtrapfreq }$.
The atom Hamiltonian is then in general
\begin{multline}\label{eqn: external Hamiltonian harmonic approximation}
  \Hamilop_\atomic = \sum_i \hbar \itrapfreq \pare{\hconj\ataop_i\ataop_i + \frac{1}{2}} \\
  + \sum_{i,j\neq i} \hbar \ijcrosscoupling ( \ataop_i + \hconj\ataop_i ) ( \ataop_j + \hconj\ataop_j )~,
\end{multline}
where we measure energies relative to the depth of the trapping potential, $\opticaltrapdepth = \pot(\trappos)$. The Hamiltonian can always be diagonalized~%
\footnote{%
The coupling between directions $i$ and $j\neq i$ can always be transformed away by selecting coordinates aligned with the symmetry axes of the potential in harmonic approximation.%
},
and hence simplifies to \cref{eqn: atom Hamiltonian harmonic approximation} after dropping the constant zero-point energy.

\section{Phononic Eigenmodes}
\label{sec: phonon appendix}
Vibrations of the photonic structure in a nanophotonic cold-atom trap alter the optical fields surrounding the structure. This variation leads to an interaction of the vibrations and the trapped atoms, as we discuss in detail in \cref{sec: interaction appendix}. Vibrations at frequencies relevant to nanophotonic traps can be modeled by linear elasticity theory, because the corresponding phonon wavelengths are sufficiently large not to resolve the microscopic structure of the solid. Linear elasticity theory describes the dynamics of elastic deformations of a continuous body around its equilibrium state \cite{achenbach_wave_1973,auld_acoustic_1973-1,gurtin_linear_1984}. The deformations are described by the \emph{displacement field} $\ufield$, a real valued vector field defined on the domain of the body. The displacement field indicates the magnitude and direction of the displacement of each point of the body from equilibrium at any given time.

Our objective is to provide a quantum description of the vibrations (in terms of a phonon field) and of the atom-phonon interaction. In this \namecref{sec: phonon appendix}, we review how a quantum formulation of linear elasticity can be obtained through canonical quantization, based on the concept of phononic eigenmodes. Subsequently, we discuss the phononic eigenmodes of a nanofiber.

\subsection{Quantum Elastodynamics}
\label{eqn: quantum elastodynamics}

Consider an elastic body in three dimensions. Within the framework of linear elasticity, its mechanical properties are described by the \emph{mass density} $\dens$ and the \emph{elasticity tensor} $\elastens$, both of which are in general position dependent. The elasticity tensor is of fourth order, with symmetries $\elastenscomp^{ijkl} = \elastenscomp^{jikl} = \elastenscomp^{ijlk} = \elastenscomp^{klij}$ \cite{gurtin_linear_1984}. In the case of a homogeneous elastic body, $\dens$ and $\elastens$ are constant. If the body is isotropic, the elasticity tensor has the form \cite{gurtin_linear_1984,achenbach_wave_1973}
\begin{equation}
  \elastenscomp^{ijkl} = \lamemu \spare{\kronecker^{ik}\kronecker^{jl} + \kronecker^{il}\kronecker^{jk}} + \lamelambda \kronecker^{ij}\kronecker^{kl}~.
\end{equation}
The two coefficients $\lamelambda$ and $\lamemu$ are called \emph{Lamé parameters}. The mechanical properties of a homogeneous and isotropic body are thus described by three real numbers: $\dens$, $\lamelambda$, and $\lamemu$. The density $\dens$ and Lamé's second parameter $\lamemu$ are positive, while Lamé's first parameter $\lamelambda$ may be negative \cite{gurtin_linear_1984}.
An alternative, widespread parametrization uses \emph{Young's modulus} $\YoungE$ and the \emph{Poisson ratio} $\Poissonnu$:
\begin{align}\label{eqn: Lame coefficients calculated}
    \lamelambda &= \frac{\Poissonnu \YoungE}{(1+\Poissonnu)(1-2\Poissonnu)} & \lamemu &= \frac{\YoungE}{2(1+\Poissonnu)}~.
\end{align}
The modulus is positive, as is the Poisson ratio for most materials \cite{gurtin_linear_1984}.

The dynamics of the displacement field is governed by the equation of motion \cite{gurtin_linear_1984,achenbach_wave_1973,auld_acoustic_1973-1}
\begin{equation}\label{eqn: phonon equation of motion}
  \dens  \uddotfield = \Dphon \ufield~,
\end{equation}
where we define the differential operator $\Dphon$ that acts on a vector field as $\spare{\Dphon\ufield}^i \equiv \sum_{jkl}\partial_j \elastenscomp^{ijkl}  \partial_k \ufieldcomp^l$. It is common to introduce the \emph{strain tensor} $\straintens$ describing deformations of the solid, and the \emph{stress tensor} $\stresstens$, which characterizes the forces needed to affect this strain:
\begin{equation}\label{eqn: definition strain and stress tensor}
  \begin{split}
    \straintenscomp^{ij} &\equiv \frac{1}{2}\pare{\partial_i \ufieldcomp^j + \partial_j \ufieldcomp^i} \\
    \stresstenscomp^{ij} &\equiv \sum_{kl}\elastenscomp^{ijkl}\straintenscomp^{kl}~.
  \end{split}
\end{equation}
Note that both strain and stress tensor are symmetric, $\straintenscomp^{ij} = \straintenscomp^{ji}$ and $\stresstenscomp^{ij} = \stresstenscomp^{ji}$. It is necessary to specify boundary conditions in order to obtain a unique solution given initial conditions \cite{gurtin_linear_1984}. For a body that is not subject to external forces, these boundary conditions are of Neumann type and state that on the surface of the body, the stress vanishes in direction $\normvec$ normal to the surface: $\stresstens \normvec = \zerovec$.

We are interested in quantizing the vibrations of a force-free, homogeneous, and isotropic elastic body. We start from the Lagrange density $\Lagdens = \dens\dot\ufield^2/2 - \sum_{ij}\straintenscomp^{ij}\stresstenscomp^{ij}/2$, which yields the correct equations of motion. The canonical conjugate momentum is then $\pifield = \dens\udotfield$, and the resulting classical Hamilton functional can be expressed as
\begin{equation}
  \Hamilfunc = \frac{1}{2}\int_\body \spare{\frac{\pifield^2}{\dens} - \ufield\cdot \Dphon\ufield}\dd\pos~,
\end{equation}
where $\body$ is the volume of the body~%
\footnote{%
The strain energy can be rewritten by partial integration as $\int_\body \sum_{ij}\straintenscomp^{ij} \stresstenscomp^{ij} \dd\pos =  \int_{\partial \body} \ufield\cdot (\stresstens\normvec) \dd \partial \body - \int_\body \ufield \cdot \Dphon\ufield \dd\pos$, where $\partial\body$ is the surface of the body. In the case of a force-free body, the surface term vanishes.%
}.

To proceed, we introduce phononic eigenmodes $\wmode_\phonindex$ with eigenfrequencies $\phonfreq_\phonindex$ as solutions of the eigenvalue equation
\begin{equation}\label{eqn: phonon eigenvalue equation}
  \Dphon \wmode_\phonindex(\pos) = -\rho \phonfreq^2_\phonindex \wmode_\phonindex(\pos)
\end{equation}
together with the boundary conditions for a force-free body. The eigenmodes are labeled using a multi-index $\phonindex$ which may contain both discrete and continuous indices. Since $(-\Dphon)$ is self-adjoint \cite{anghel_quantization_2007}, eigenmode solutions form an orthogonal basis for the space of admissible displacement fields and can be normalized according to
\begin{equation}\label{eqn: phonon orthonormality condition}
  \int_\body \cconj\wmode_\phonindex(\pos) \cdot \wmode_\phonindexb(\pos)\dd\pos = \kronecker_{\phonindex \phonindexb}~.
\end{equation}
Any solution to the equation of motion \cref{eqn: phonon equation of motion} can then be expressed as a linear combination of eigenmodes,
\begin{equation}\label{eqn: displacement mode expansion}
  \ufield(\post) = \sum_\phonindex \frac{1}{\dens \phonfreq_\phonindex} \spare{\phonnormvar_\phonindex e^{-\im \phonfreq_\phonindex \tm}\wmode_\phonindex(\pos) + \cc}~,
\end{equation}
with coefficients $\phonnormvar_\phonindex\in \C$ determined by the initial conditions.

Canonical quantization amounts to turning every eigenmode into a bosonic mode with ladder operators $\phonaop_\phonindex$ and $\hconj\phonaop_\phonindex$ that satisfy canonical commutation relations, $[\phonaop_\phonindex,\hconj\phonaop_\phonindexb] = \kronecker_{\phonindex\phonindexb}$.
The displacement field and its conjugate momentum are promoted to field operators with mode expansions
\begin{equation}\label{eqn: phonon field operator mode expansion}
  \begin{split}
    \ufieldop(\pos) &= \sum_\phonindex \ufieldmodedens_\phonindex \spare{ \phonaop_\phonindex \wmode_\phonindex(\pos)  + \hc}\\
    \pifieldop(\pos) &= -\im \sum_\phonindex \pifieldmodedens_\phonindex \spare{ \phonaop_\phonindex \wmode_\phonindex(\pos)  - \hc}~.
  \end{split}
\end{equation}
Here, $\ufieldmodedens_\phonindex \equiv \sqrt{\hbar/2\dens\phonfreq_\phonindex}$ and $\pifieldmodedens_\phonindex \equiv \sqrt{\hbar\dens\phonfreq_\phonindex/2}$ are the mode densities. The Hamiltonian takes the form
\begin{equation}\label{eqn: Hamiltonian phonons appendix}
  \Hamilop_\vibrational = \sum_\phonindex \hbar\phonfreq_\phonindex \hconj\phonaop_\phonindex\phonaop_\phonindex~,
\end{equation}
where we set the energy of the ground state to zero. Since strain plays an important role in the atom-phonon interaction discussed in \cref{sec: interaction appendix}, we introduce the tensorial strain modal fields $\strainmode_\phonindex$ with components
\begin{equation}\label{eqn: strain modal field}
  \strainmodecomp_\phonindex^{ij} \equiv \frac{1}{2} \spare{ \partial_i \wmodecomp_\phonindex^j + \partial_j \wmodecomp_\phonindex^i }
\end{equation}
such that the strain operator can be expressed as
\begin{equation}\label{eqn: strain operator mode expansion}
    \straintensop(\pos) = \sum_\phonindex \ufieldmodedens_\phonindex \spare{\phonaop_\phonindex \strainmode_\phonindex(\pos) + \hc}~.
\end{equation}

\subsection{Phononic Fiber Eigenmodes}
\label{sec: phonon fiber eigenmodes}

\begin{figure*}
  \centering
  \setlength{\widthFigA}{137.95pt}
  \setlength{\marginLeftFigA}{20.45pt}
  \setlength{\marginRightFigA}{7pt}
  \setlength{\widthFigB}{124.5pt}
  \setlength{\marginLeftFigB}{7pt}
  \setlength{\marginRightFigB}{7pt}
  \setlength{\widthFigC}{124.5pt}
  \setlength{\marginLeftFigC}{7pt}
  \setlength{\marginRightFigC}{7pt}
  \setlength{\widthFigD}{121.5pt}
  \setlength{\marginLeftFigD}{7pt}
  \setlength{\marginRightFigD}{7pt}
  \parbox[t]{\widthFigA}{\vspace{0pt}{\includegraphics[width = \widthFigA]{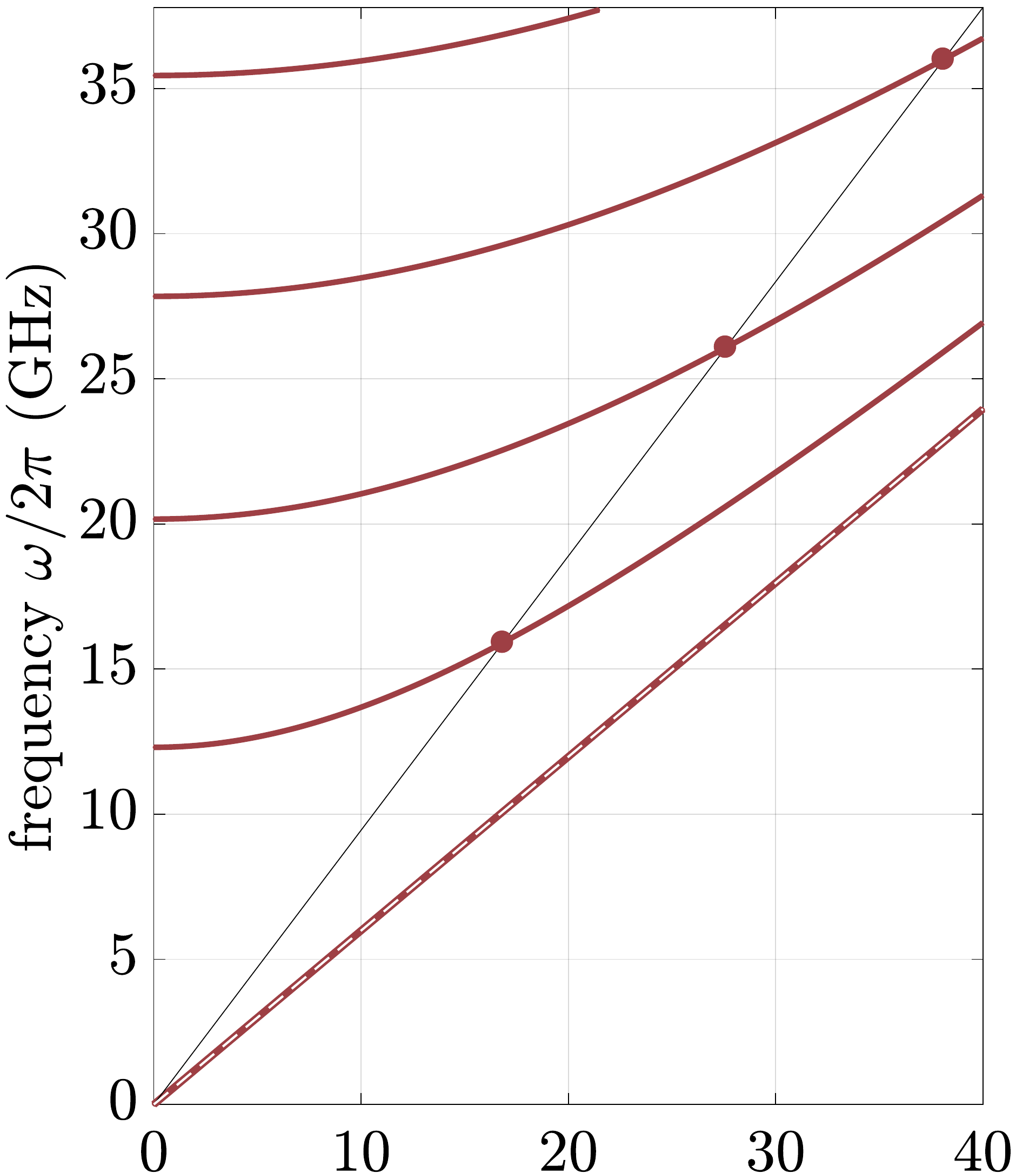}}}%
  \parbox[t]{\widthFigB}{\vspace{0pt}{\includegraphics[width = \widthFigB]{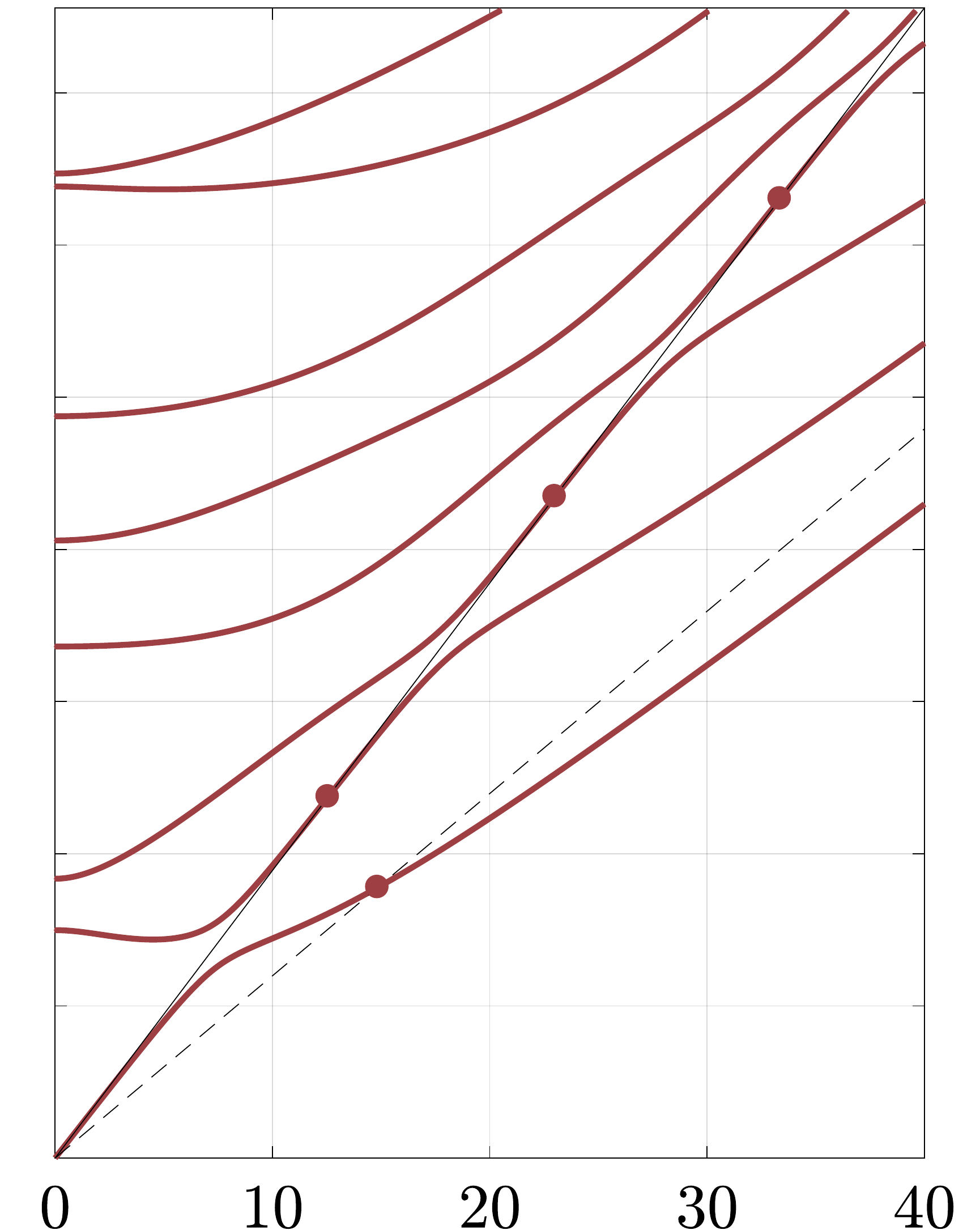}}}%
  \parbox[t]{\widthFigC}{\vspace{0pt}{\includegraphics[width = \widthFigC]{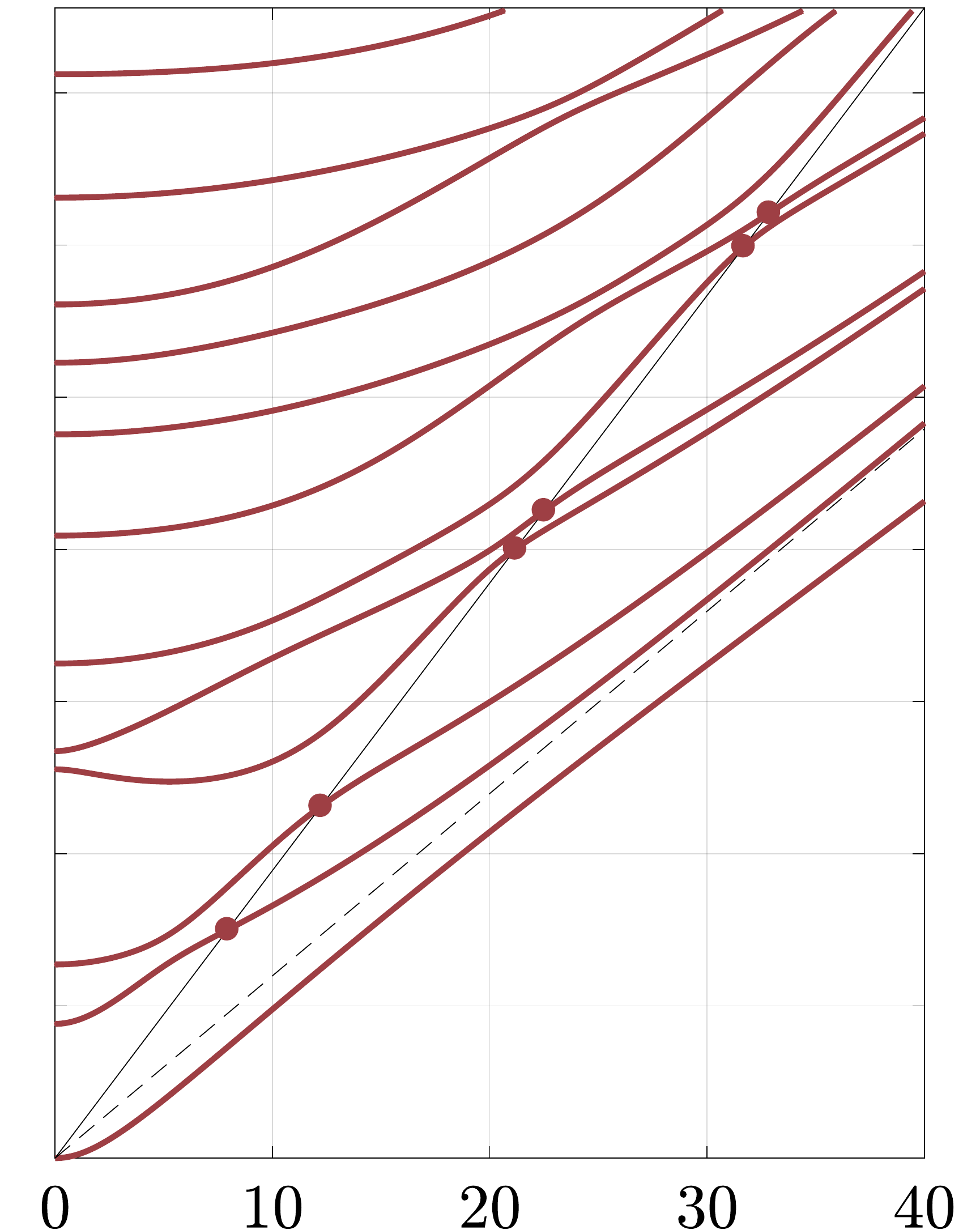}}}%
  \parbox[t]{\widthFigD}{\vspace{0pt}{\includegraphics[width = \widthFigD]{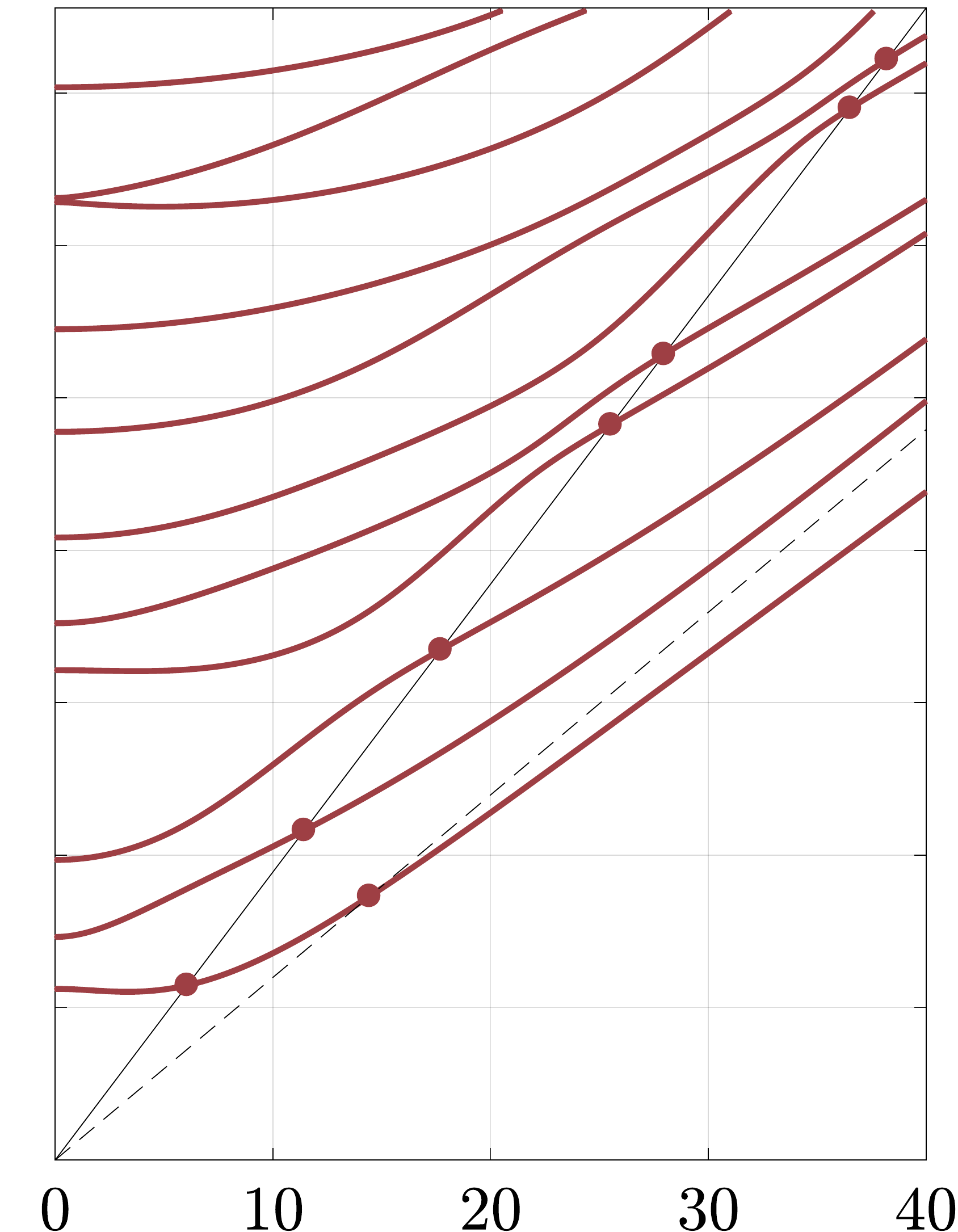}}}\\
  \vspace{2pt}
  \parbox[t]{\widthFigA}{~}%
  \parbox[t]{\widthFigB + \widthFigC}{\footnotesize propagation constant $\phonk$ $(1/\si{\micro\meter})$}%
  \parbox[t]{\widthFigD}{~}\\
  \vspace{7pt}
  \parbox[t]{\widthFigA}{\hspace*{\marginLeftFigA}%
  \parbox[t]{\widthFigA-\marginLeftFigA-\marginRightFigA}{%
    (a) $\Tmode_{0\phonn}$ bands at $\phonl=0$%
  }\hspace*{\marginRightFigA}}%
  \parbox[t]{\widthFigB}{\hspace*{\marginLeftFigB}%
  \parbox[t]{\widthFigB-\marginLeftFigB-\marginRightFigB}{%
    (b) $\Lmode_{0\phonn}$ bands at $\phonl=0$%
  }\hspace*{\marginRightFigB}}%
  \parbox[t]{\widthFigC}{\hspace*{\marginLeftFigC}%
  \parbox[t]{\widthFigC-\marginLeftFigC-\marginRightFigC}{%
    (c) $\Fmode_{1\phonn}$ bands at $\phonl=\pm1$%
  }\hspace*{\marginRightFigC}}%
  \parbox[t]{\widthFigD}{\hspace*{\marginLeftFigD}%
  \parbox[t]{\widthFigD-\marginLeftFigD-\marginRightFigD}{%
    (d) $\Fmode_{2\phonn}$ bands at $\phonl=\pm2$%
  }\hspace*{\marginRightFigD}}%
  \caption{Band structures of nanofiber phonon modes with different azimuthal order $\phonl$, for radius $\rad$ and mechanical properties specified in \cref{sec: case study appendix}. Panel (a) shows torsional bands, panel (b) longitudinal bands, and panel (c) flexural bands of azimuthal order $\phonl=\pm 1$ and $\phonl=\pm 2$, respectively. The solid black line delineates the longitudinal sound line, the dashed line the transverse sound line [which coincides with the lowest band in panel (a)]. Solid dots mark intersections of phonon bands with either sound line.}
  \label{fig: phonon bands}
\end{figure*}

\begin{table}
  \newcolumntype{A}{>{\begin{math}}r<{\end{math}}}
  \newcolumntype{C}{>{\begin{math}}l<{\end{math}}}
  \begin{tabularx}{\columnwidth}[t]{A @{${}={}$} C @{\qquad} A @{${}={}$} C}
    \toprule
    \ctrans & \sqrt{\YoungE/2\dens(1+\Poissonnu)}  & \clong & \sqrt{\YoungE(\Poissonnu - 1)/\dens(\Poissonnu + 2 \Poissonnu^2-1)} \\
    \phona & \sqrt{\phonfreq^2/\ctrans^2 - \phonk^2} & \phonaa & -\im \phona\\
    \phonb & \sqrt{\phonfreq^2/\clong^2 - \phonk^2} & \phonbb & -\im \phonb\\
    \phonfreq_\trans & \ctrans |\phonk| & \phonfreq_\longitud & \clong |\phonk| \\
    \midrule
    \phonaR & \phona\rad & \phonaaR & \phonaa \rad\\
    \phonbR & \phonb\rad & \phonbbR & \phonbb \rad\\
    \phonkR & \phonk \rad & \rposR & \rpos/ \rad\\
    \eta & \besselJ{1}(\phonbR)/\besselJ{1}(\phonaR) & \tilde{\eta} & \besselI{1}(\phonbbR)/\besselJ{1}(\phonaR) \\
    \breve{\eta} & \besselI{1}(\phonbbR)/\besselI{1}(\phonaaR) \\
    \signum_\phonk & \phonk/|\phonk| & \signum_\phonl & \phonl/|\phonl|\\
    \bottomrule
  \end{tabularx}
  \caption{Definitions of the longitudinal and the transverse sound velocity $\clong$ and $\ctrans$, as well as the radial constants $\phona$ and $\phonb$, and the dimensionless quantities appearing in the phonon modal fields. The definitions are given in terms of the density $\dens$, Young's modulus $E$, Poisson ratio $\nu$, fiber radius $\rad$, azimuthal order $\phonl$, propagation constant $\phonk$, and radial position $\rpos$.}
  \label{tab: phonon quantities}
\end{table}

We now consider vibrations of a nanofiber modeled as a homogeneous and isotropic cylinder of radius $\rad$ and of infinite length along the $\zpos$ axis. In the following, we summarize the resulting phononic eigenmodes and frequency equations. Details on the derivation can be found in \cite{achenbach_wave_1973,auld_acoustic_1973-1,armenakas_free_1969,meeker_guided_1964}.

Solving the eigenvalue equation \cref{eqn: phonon eigenvalue equation} in cylindrical coordinates leads to phononic eigenmodes and corresponding strain modal fields of the form
\begin{equation}\label{eqn: phonon cylinder eigenmode partial wave decomposition}
  \begin{split}
    \wmode_\phonindex(\pos) &=  \frac{\wmoder_\phonindex(\rpos)}{2\pi}  e^{\im (\phonl\phipos + \phonk \zpos)}\\
    \strainmode_\phonindex(\pos) &=  \frac{\strainmoder_\phonindex(\rpos)}{2\pi}  e^{\im (\phonl\phipos + \phonk \zpos) }~,
  \end{split}
\end{equation}
in close analogy to the photon modes \cref{eqn: vector wave equation solutions}. Here, $\phonk \in \R$ is the propagation constant and $\phonl \in \Z$ the azimuthal order. Mode quadruplets $(\pm\phonl,\pm\phonk)$ are degenerate in eigenfrequency $\phonfreq_\phonindex$. Since phonon modes are radially confined by the finite fiber radius, there is only a discrete set of frequencies $\phonfreq_\phonindex$ admissible for each $(\phonl,\phonk)$, analogous to the case of guided photonic fiber modes. The eigenfrequencies $\phonfreq_\phonindex(\phonk)$ form discrete bands in the $(\phonk,\photfreq)$ plane; see \cref{fig: phonon bands}. We can therefore count radial excitations using a discrete band index $\phonn\in \N$ starting from $\phonn = 1$ for the band of lowest frequency and increasing in frequency with $\phonn$. Since the propagation constant is the only continuous mode index, the orthonormality relation \cref{eqn: phonon orthonormality condition} reduces to a normalization condition for the radial partial waves:
\begin{equation}\label{eqn: normalization phonon radial partial wave}
  \int_0^\rad \rpos |\wmoder_{\phonindex}(\rpos)|^2 \dd \rpos = 1~.
\end{equation}

Elastodynamics, unlike electrodynamics, allows for longitudinal in addition to transverse polarizations even in the absence of surfaces. In the nanofiber the presence of a surface forces these excitations to hybridize, forming eigenmodes that can have both transverse and longitudinal contributions. Transverse waves propagate with the \emph{transverse sound velocity} $\ctrans$, while longitudinal waves propagate with the \emph{longitudinal sound velocity} $\clong$ which is typically larger than the transverse sound velocity:
\begin{equation}\label{eqn: definition sound speeds}
  \begin{split}
    \ctrans &\equiv \sqrt{\frac{\YoungE}{2\dens(1+\Poissonnu)}} \\
    \clong &\equiv \sqrt{\frac{\YoungE(\Poissonnu - 1)}{\dens(\Poissonnu + 2 \Poissonnu^2-1)}}~.
  \end{split}
\end{equation}
The radial partial waves $\wmoder_\phonindex$ have three contributions,
\begin{equation}\label{eqn: radial partial waves flexural eigenmodes}
  \wmoder_\phonindex(\rpos) = A \wmoderA(\rpos) + B \wmoderB(\rpos) + C \wmoderC(\rpos)~,
\end{equation}
where $A,B,C\in \C$ are amplitudes. The components of the three vectorial terms $\wmoderA$, $\wmoderB$, and $\wmoderC$ are listed in \cref{tab: flexural eigenmodes}. The form of the eigenmodes depends on the magnitude of the eigenfrequency $\phonfreq_\phonindex$ compared to the \emph{longitudinal sound line} $\phonfreq_\longitud \equiv \clong |\phonk|$ and the \emph{transverse sound line} $\phonfreq_\trans \equiv \ctrans |\phonk|$. We assume that $\clong > \ctrans$ in the following discussion, as is the case for most materials. If this is not the case, the radial partial waves of eigenmodes are different from the ones given in this \namecref{sec: phonon appendix}.

The eigenmodes have to meet the boundary conditions discussed above to ensure a stress-free surface. In terms of the stress modal field
\begin{equation}
  \stressmodecomp_\phonindex^{kl}(\pos) \equiv \sum_{mn}\elastenscomp^{klmn}\strainmodecomp_\phonindex^{mn}(\pos)~,
\end{equation}
defined here using the strain modal field \cref{eqn: strain modal field}, the boundary conditions are $\stressmode_\phonindex \normvec = \zerovec$ at $\rpos=\rad$. For the nanofiber, the exterior surface normal vector is $\normvec = \ervec$. The stress modal field can be decomposed into partial waves
\begin{equation}
  \stressmode_\phonindex(\pos) =  \frac{\stressmoder_\phonindex(\rpos)}{2\pi}  e^{\im (\phonl\phipos +\im \phonk \zpos)}~,
\end{equation}
and the radial partial waves of the relevant components brought into the form
\begin{equation}\label{eqn: stress vector field cylindrical coordinates}
  \begin{split}
    \stressmodercomp_\phonindex^{\rpos\rpos}(\rpos) &= -\frac{2\lamemu\rad}{\rpos^2} \spare{A \phononBCmatrixcomp^{\rpos a}(\rpos) + B \phononBCmatrixcomp^{\rpos b}(\rpos) + C \phononBCmatrixcomp^{\rpos c}(\rpos) }\\
    \stressmodercomp_\phonindex^{\phipos\rpos}(\rpos) &= \im\frac{\lamemu\rad}{\rpos^2} \spare{A \phononBCmatrixcomp^{\phipos a}(\rpos) + B \phononBCmatrixcomp^{\phipos b}(\rpos) + C \phononBCmatrixcomp^{\phipos c}(\rpos) }\\
    \stressmodercomp_\phonindex^{\zpos\rpos}(\rpos) &= \im\frac{\lamemu\rad}{\rpos^2} \spare{A \phononBCmatrixcomp^{\zpos a}(\rpos) + B \phononBCmatrixcomp^{\zpos b}(\rpos) + C \phononBCmatrixcomp^{\zpos c}(\rpos) }~.
  \end{split}
\end{equation}
The nine components $\phononBCmatrixcomp^{kl}$, $k=\rpos,\phipos,\zpos$, $l=a,b,c$ evaluated on the fiber surface $\rpos=\rad$ are listed in \cref{tab: stress tensor components}. Let $\tens{\phononBCmatrixcomp}$ be the matrix with coefficients $\phononBCmatrixcomp^{kl}(\rpos=\rad)$. The boundary conditions can then be written as
\begin{equation}\label{eqn: free body boundary conditions matrix notation}
    \tens{\phononBCmatrixcomp} \begin{pmatrix}A \\B\\C\end{pmatrix} = \zerovec
\end{equation}
and yield relations between the three amplitudes $A,B,C$. For each mode family, one independent amplitude remains which can subsequently be determined from the normalization condition \cref{eqn: normalization phonon radial partial wave}. The trivial solution $A=B=C=0$ corresponds to zero displacement and is of no interest to us. The boundary conditions \cref{eqn: free body boundary conditions matrix notation} can be met with nontrivial amplitudes if and only if
\begin{equation}\label{eqn: phonon frequency equation}
  \det \tens{\phononBCmatrixcomp} = 0~.
\end{equation}
This relation is the frequency equation for the eigenmodes, as it constrains the admissible eigenfrequencies $\phonfreq_\phonindex$ for a given propagation constant $\phonk$ and azimuthal order $\phonl$. It is therefore an implicit equation for the dispersion relation $\phonfreq_\phonindex(\phonk)$.

We distinguish the cases $\phonl=0$ and $|\phonl|\geq1$. For azimuthal order $\phonl = 0$, $\phononBCmatrixcomp^{\rpos c} = \phononBCmatrixcomp^{\phipos a} = \phononBCmatrixcomp^{\zpos c} = 0$, and the frequency equation \cref{eqn: phonon frequency equation} factorizes,
\begin{equation}\label{eqn: phonon frequency equation 2}
\phononBCmatrixcomp^{\phipos c} \det \begin{pmatrix}\phononBCmatrixcomp^{\rpos a} & \phononBCmatrixcomp^{\rpos b}\\\phononBCmatrixcomp^{\zpos a} & \phononBCmatrixcomp^{\zpos b}\end{pmatrix} = 0~,
\end{equation}
giving rise to two independent mode families: \emph{torsional} ($\Tmode$) modes with $\phononBCmatrixcomp^{\phipos c}=0$ and \emph{longitudinal} ($\Lmode$) modes for which the determinant in \cref{eqn: phonon frequency equation 2} vanishes. For higher azimuthal excitations $|\phonl|\geq1$, this is not the case, and there is only one family of \emph{flexural} ($\Fmode$) modes.

Torsional modes can exist only above and on the transverse sound line $\phonfreq_\trans$, see \subcref{fig: phonon bands}{a}. Longitudinal and flexural modes, on the other hand, are hybrids of longitudinal and transverse waves, and exhibit different behavior depending on the magnitude of their frequency $\phonfreq_\phonindex$ compared to the transverse sound line $\phonfreq_\trans$ and the longitudinal sound line $\phonfreq_\longitud$. In the case when the frequency lies above the longitudinal sound line $\phonfreq_\phonindex > \phonfreq_\longitud > \phonfreq_\trans$, both transverse and longitudinal excitations can propagate in the bulk of the cylinder, and the eigenmode shows oscillatory behavior in $\rpos$. We refer to these modes as \emph{bulk modes}. In the case when the frequency lies between the two sound lines $\phonfreq_\longitud > \phonfreq_\phonindex > \phonfreq_\trans$, we call the modes \emph{mixed modes}: Only the transverse contributions to the eigenmodes oscillate in $\rpos$, while the longitudinal contributions decay as modified Bessel functions of the first kind $\besselI{\phonn}$ away from the surface towards the center of the cylinder. \emph{Surface modes} are characterized by $\phonfreq_\longitud > \phonfreq_\trans > \phonfreq_\phonindex$. Neither transverse nor longitudinal excitations can propagate in the bulk of the fiber, and eigenmodes are confined to the fiber surface. Such modes are often called surface acoustic waves \cite{achenbach_wave_1973,eringen_elastodynamics_1975}. On each sound line, the respective contributions to the eigenmodes are polynomial in $\rpos$.

\begin{table}
  \setlength{\extrarowheight}{3pt}
  \newcolumntype{A}{>{\begin{math}}r<{\end{math}}}
  \newcolumntype{B}{>{\begin{math}}c<{\end{math}}}
  \newcolumntype{C}{>{\begin{math}}l<{\end{math}}}
  \begin{tabular*}{\columnwidth}[t]{ABC @{\quad} ABC  ABC}
    \toprule
    \multicolumn{3}{l}{Case} & \multicolumn{3}{l}{Eigenmode component \hspace*{1ex}} & \multicolumn{3}{l}{Frequency equation} \\
    \midrule
    \phonfreq_\phonindex &>& \phonfreq_\trans &
    \wmodercomp^\phipos_\phonindex(\rpos) &=& C \besselJ{1}(\phonaR\rposR) &
    \besselJ{2}(\phonaR) &=& 0 \\
    \phonfreq_\phonindex &=& \phonfreq_\trans &
    \wmodercomp^\phipos_\phonindex(\rpos) &=& C \rposR &
    \phonfreq_\phonindex(\phonk) &=& \ctrans |\phonk| \\
    \bottomrule
  \end{tabular*}
  \caption{Torsional ($\Tmode$) fiber eigenmodes: Nonzero component of the displacement eigenmode, and frequency equations. The amplitude $C\in\C$ is fixed by the normalization condition \cref{eqn: normalization phonon radial partial wave}. All quantities are defined in \cref{tab: phonon quantities}.}
  \label{tab: torsional modes}
\end{table}

\begin{figure*}
  \setlength{\widthFigA}{117.3389pt}
  \setlength{\marginLeftFigA}{25.3389pt}
  \setlength{\marginRightFigA}{7pt}
  \setlength{\widthFigB}{99pt}
  \setlength{\marginLeftFigB}{7pt}
  \setlength{\marginRightFigB}{7pt}
  \setlength{\widthFigC}{93pt}
  \setlength{\marginLeftFigC}{7pt}
  \setlength{\marginRightFigC}{1pt}
  \parbox[t]{\widthFigA}{\vspace{0pt}{\includegraphics[width = \widthFigA]{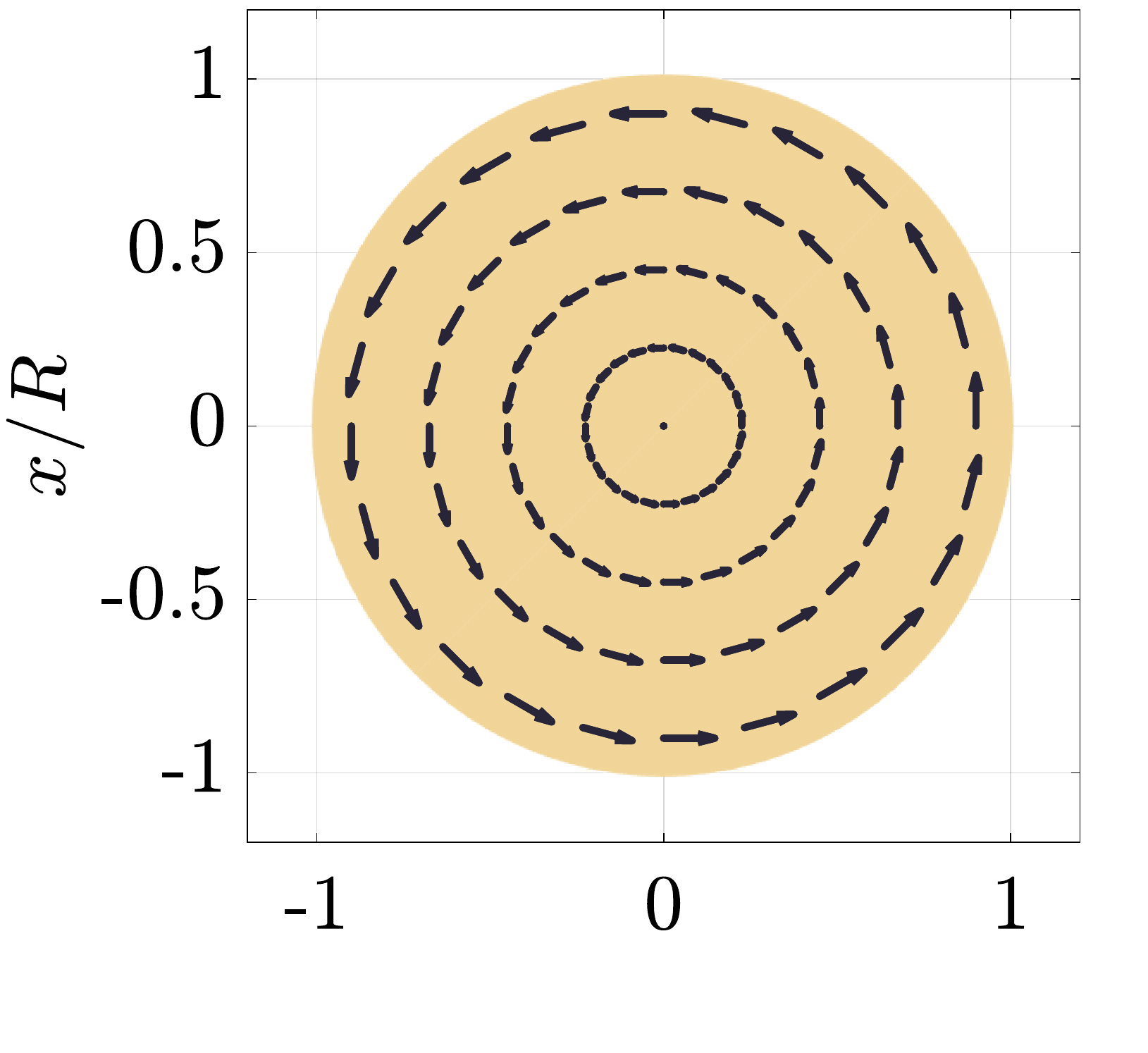}}}%
  \parbox[t]{\widthFigB}{\vspace{0pt}{\includegraphics[width = \widthFigB]{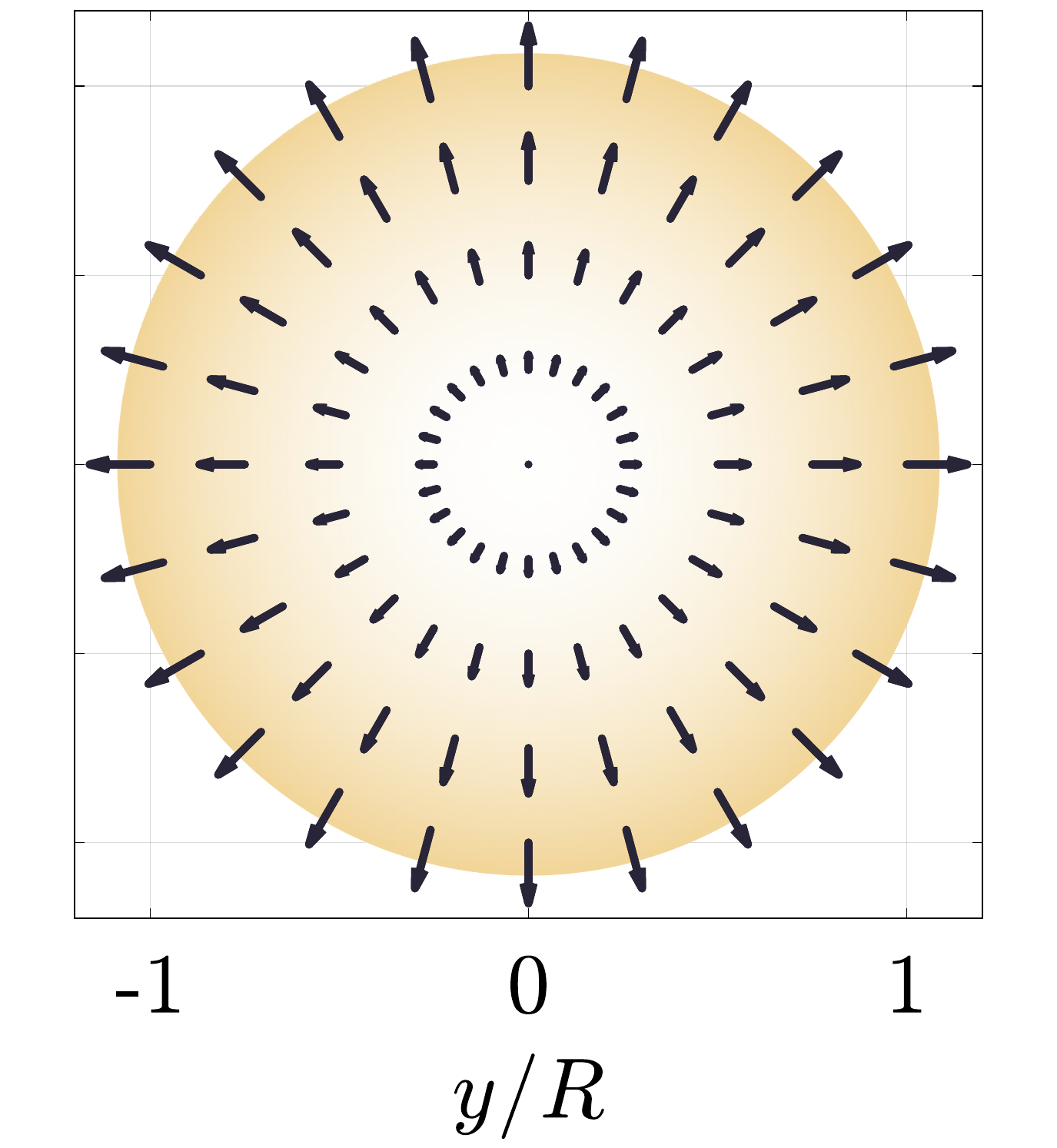}}}%
  \parbox[t]{\widthFigC}{\vspace{0pt}{\includegraphics[width = \widthFigC]{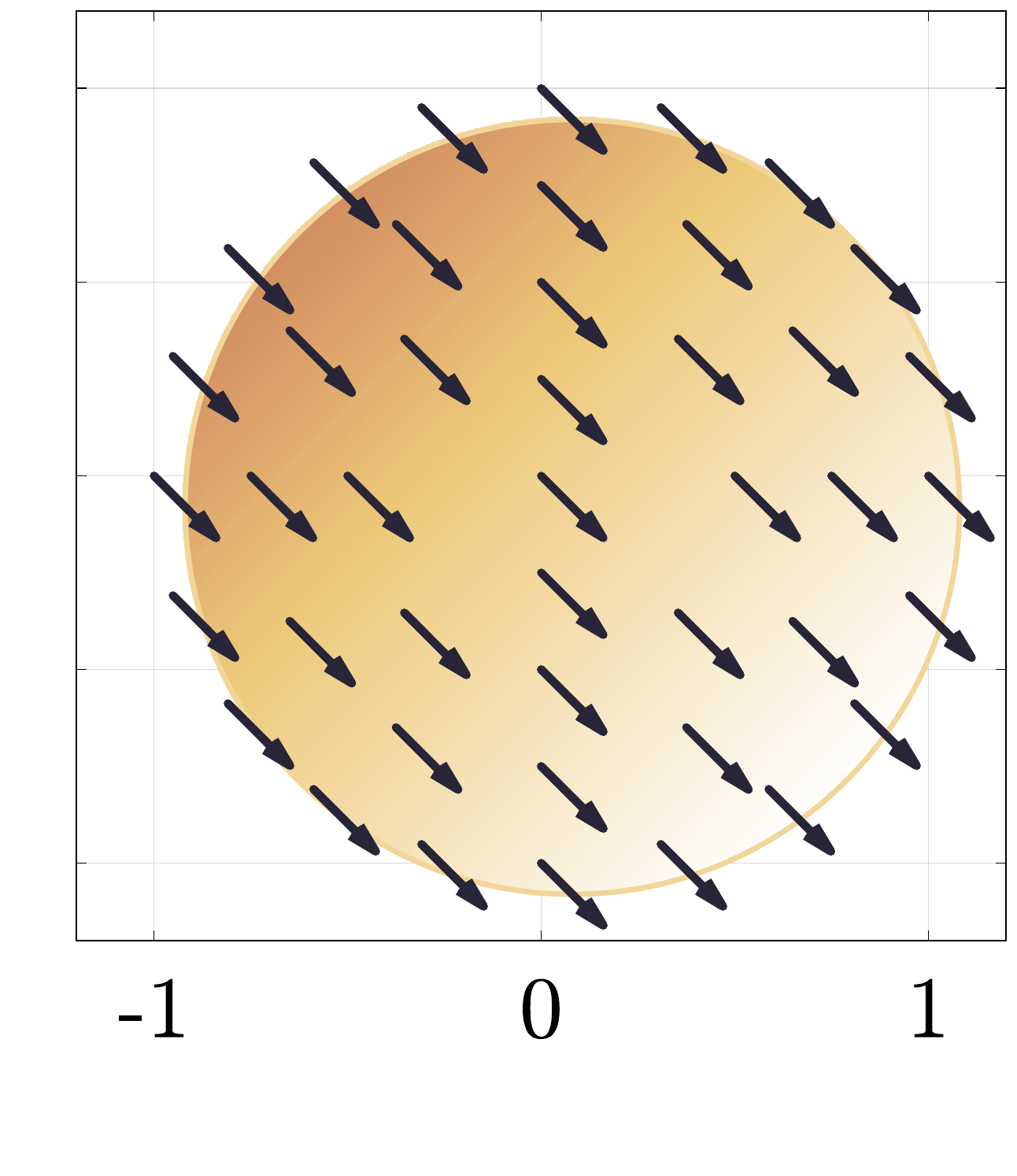}}}\\
  \vspace{4pt}
  \parbox[t]{\widthFigA}{\hspace*{\marginLeftFigA}%
  \parbox[t]{\widthFigA-\marginLeftFigA-\marginRightFigA}{%
    (a) $\Tmode_{01}$ at $\zpos = \lambda/4$%
  }\hspace*{\marginRightFigA}}%
  \parbox[t]{\widthFigB}{\hspace*{\marginLeftFigB}%
  \parbox[t]{\widthFigB-\marginLeftFigB-\marginRightFigB}{%
    (b) $\Lmode_{01}$ at $\zpos = 0$%
  }\hspace*{\marginRightFigB}}
  \parbox[t]{\widthFigC}{\hspace*{\marginLeftFigC}%
  \parbox[t]{\widthFigC-\marginLeftFigC-\marginRightFigC}{%
    (c) $\Fmode_{11}$ at $\zpos = \lambda/8$%
  }\hspace*{\marginRightFigC}}\\
  \vspace{5pt}
  \setlength{\widthFigA}{202.3389pt}
  \setlength{\marginLeftFigA}{25.3389pt}
  \setlength{\marginRightFigA}{7pt}
  \setlength{\widthFigB}{180pt}
  \setlength{\marginLeftFigB}{7pt}
  \setlength{\marginRightFigB}{3pt}
  \parbox[t]{\widthFigA}{\vspace{0pt}{\includegraphics[width = \widthFigA]{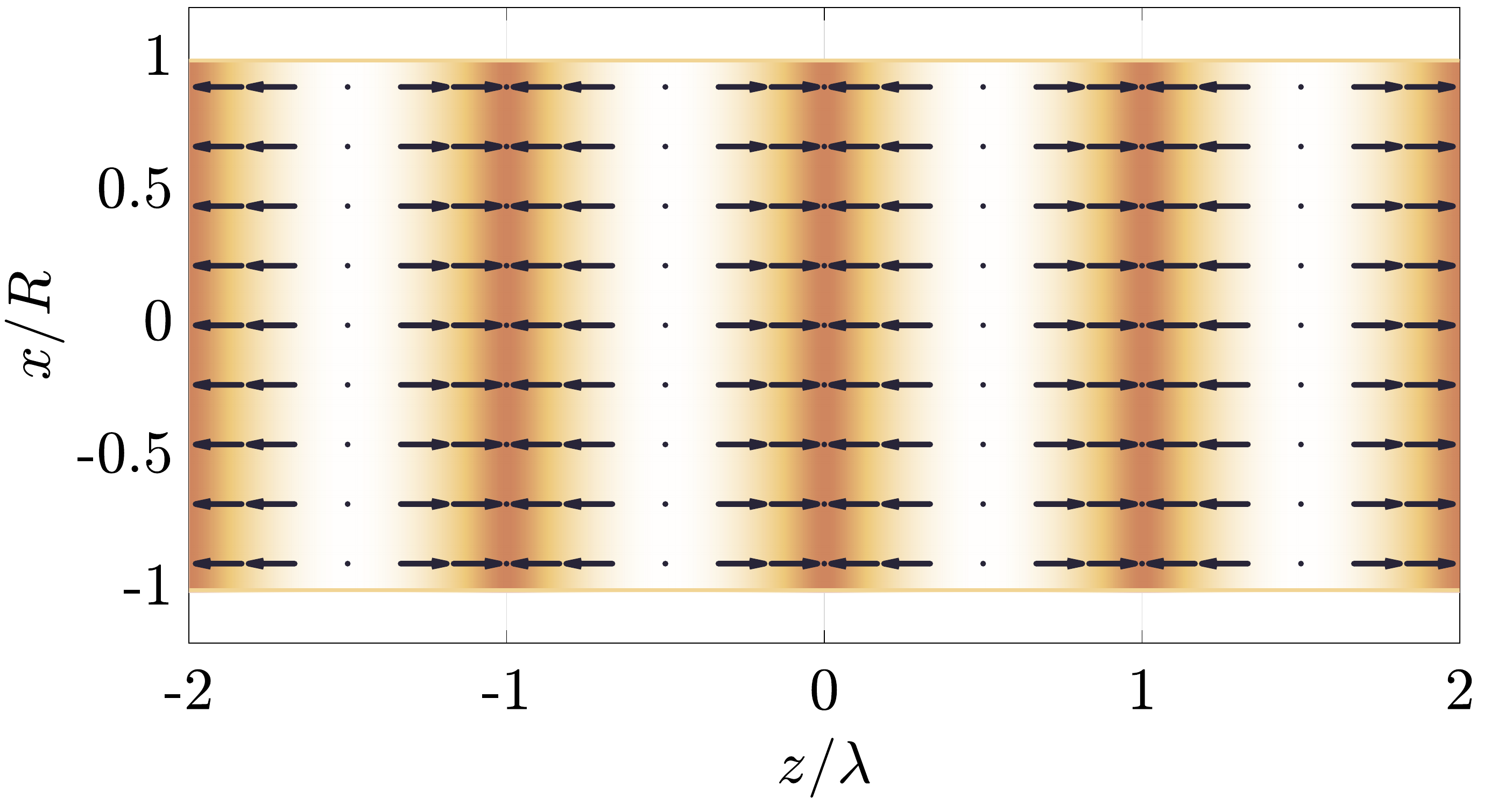}}}%
  \parbox[t]{\widthFigB}{\vspace{0pt}{\includegraphics[width = \widthFigB]{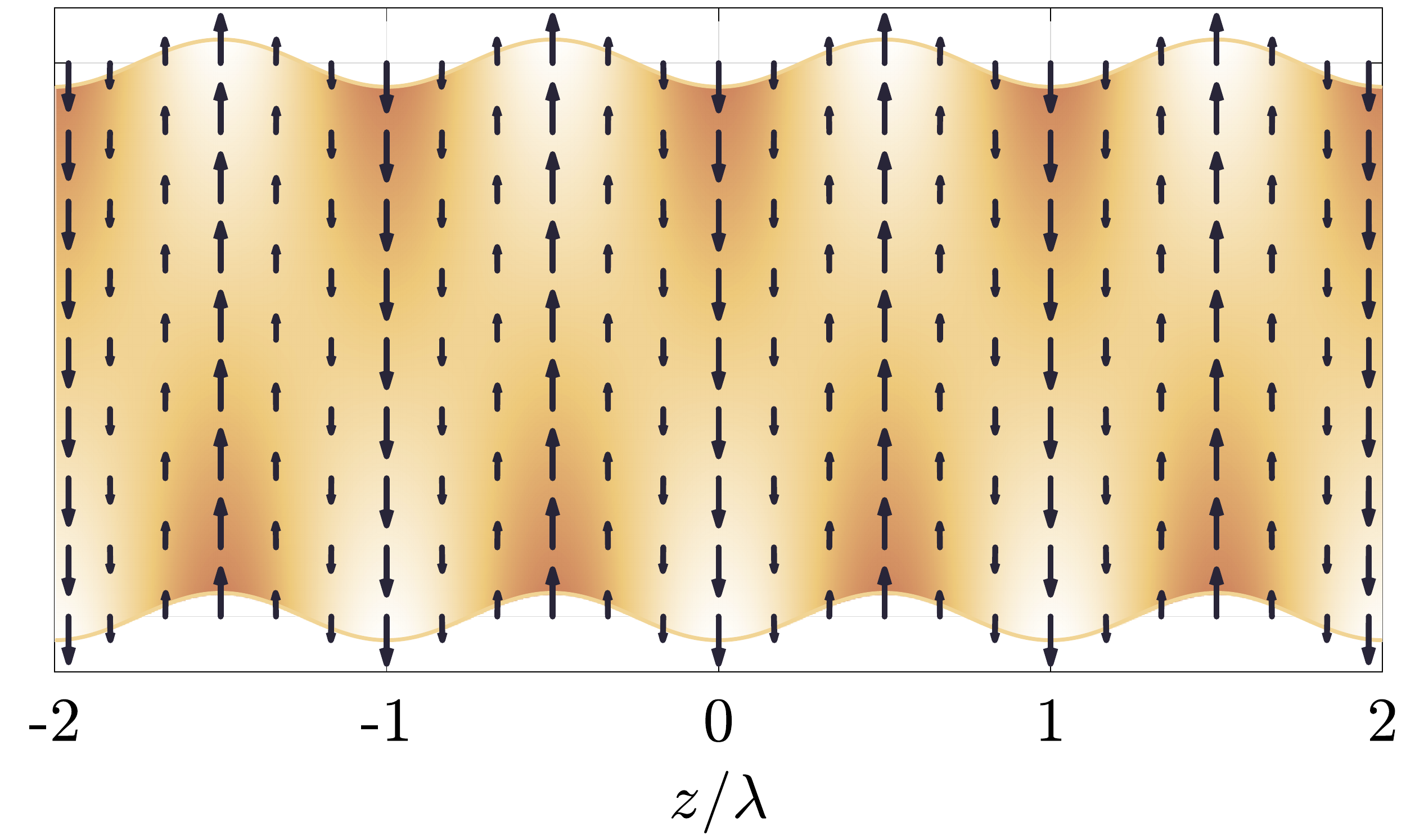}}}\\
  \vspace{4pt}
  \parbox[t]{\widthFigA}{\hspace*{\marginLeftFigA}%
  \parbox[t]{\widthFigA-\marginLeftFigA-\marginRightFigA}{%
    (d) $\Lmode_{01}$ at $\ypos = 0$%
  }\hspace*{\marginRightFigA}}%
  \parbox[t]{\widthFigB}{\hspace*{\marginLeftFigB}%
  \parbox[t]{\widthFigB-\marginLeftFigB-\marginRightFigB}{%
    (e) $\Fmode_{01}$ at $\ypos = 0$%
  }\hspace*{\marginRightFigB}}%
  \caption{Mechanical nanofiber eigenmodes on the three fundamental bands. \subCreftob{fig: phonon modes}{a}{c} show a cross section of the nanofiber, \subcrefandb{fig: phonon modes}{d}{e} a longitudinal section. The arrows correspond to the displacement field of a single mode with arbitrary amplitude $\phonnormvar_\phonindex$; see \cref{eqn: displacement mode expansion}. The color gradients show density variations $\div \ufield$ displayed at $\pos+\ufield(\pos)$ to simultaneously indicate the displacement of the fiber section. Darker areas are of higher density than lighter areas. Displacements in the vector plot are exaggerated by a factor of two compared to the density plot. The mechanical properties of the fiber are given in \cref{sec: case study appendix}. All three phonon modes have frequency $\phonfreq = 2\pi \times  \SI{123}{\kilo\hertz}$ resonant with the radial atom trap; see \cref{sec: case study}. The wavelengths $\wavelen=2\pi/\phonk$ are much larger than the nanofiber radius and vary between modes. The longitudinal sections therefore do not preserve the aspect ratio of the fiber: The $\zpos$ axis is compressed by a factor of $\wavelen/\rad \simeq \num{4e5}$ relative to the $\xpos$ axis for the $\Lmode$ mode in \subcrefb{fig: phonon modes}{d} and by $\wavelen/\rad \simeq\num{2e3}$ for the $\Fmode$ mode in \subcrefb{fig: phonon modes}{e}, while both components of the displacement field are drawn to the same scale. The $\Tmode_{01}$ mode in \subcrefb{fig: phonon modes}{a} leads only to azimuthal displacement and does not induce density variations. The $\Lmode_{01}$ has a dominant longitudinal component resulting in density waves along the fiber axis, see \subcrefb{fig: phonon modes}{d}. Its radial component causes breathing of the fiber radius and is smaller by a factor $\cFL/(\phonfreq\rad\Poissonnu) \simeq \num{1.8e5}$. The radial component in \subcrefb{fig: phonon modes}{b} is exaggerated by a corresponding factor compared to the axial component in \subcrefb{fig: phonon modes}{d} in order to be visible. The $\Fmode_{11}$ mode displaces the entire fiber cross section orthogonal to the fiber axis in a circular motion, see \subcrefb{fig: phonon modes}{c}. The density variations in \subcrefandb{fig: phonon modes}{c}{e} result from longitudinal displacement that is smaller than the transverse displacement visible in the vector plot by a factor of $\sqrt{\cFL/(2 \phonfreq \rad)}\simeq \num{120}$.}
  \label{fig: phonon modes}
\end{figure*}

\begin{table*}
  \setlength{\extrarowheight}{3pt}
  \newcolumntype{A}{>{\begin{math}}r<{\end{math}}}
  \newcolumntype{B}{>{\begin{math}}c<{\end{math}}}
  \newcolumntype{C}{>{\begin{math}}l<{\end{math}}}
  \begin{tabular}{C @{\quad} ABC @{\quad} C @{\quad} C @{\quad} C}
    \toprule
    \multicolumn{1}{l}{Term} & \multicolumn{3}{l}{Case} & \multicolumn{3}{l}{Eigenmode component} \\
    \cmidrule{5-7}
    & & & & \multicolumn{1}{l}{$k=\rpos$} & \multicolumn{1}{l}{$k=\phipos$} & \multicolumn{1}{l}{$k=\zpos$} \\
    \midrule
    \wmoderAicomp &
    \phonfreq_\phonindex &>&\phonfreq_\longitud &
    \phonbR \besselJ{\phonl}'(\phonbR \rposR) &
    \im \phonl \besselJ{\phonl}(\phonbR \rposR)/\rposR &
    \im \phonkR \besselJ{\phonl}(\phonbR \rposR) \\&
    \phonfreq_\phonindex &=&\phonfreq_\longitud &
    |\phonl| \rposR^{|\phonl|-1} &
    \im \phonl \rposR^{|\phonl|-1} &
    \im \phonkR \rposR^{|\phonl|} \\&
    \phonfreq_\phonindex &<&\phonfreq_\longitud &
    \phonbbR \besselI{\phonl}'(\phonbbR\rposR) &
    \im \phonl \besselI{\phonl}(\phonbbR \rposR)/\rposR &
    \im \phonkR \besselI{\phonl}(\phonbbR\rposR) \\
    \midrule
    \wmoderBicomp &
    \phonfreq_\phonindex &>&\phonfreq_\trans &
    \phonkR \besselJ{\phonl+1}(\phonaR \rposR) &
    -\im \phonkR \besselJ{\phonl+1}(\phonaR \rposR) &
    \im \phonaR \besselJ{\phonl}(\phonaR \rposR) \\&
    \phonfreq_\phonindex &=&\phonfreq_\trans &
    \phonkR \rposR^{|\phonl|+1} &
    - \signum_\phonl \im \phonkR \rposR^{|\phonl|+1} &
    2\im (|\phonl|+1) \rposR^{|\phonl|} \\&
    \phonfreq_\phonindex &<&\phonfreq_\trans &
    \phonkR \besselI{\phonl+1}(\phonaaR \rposR)  &
    -\im \phonkR \besselI{\phonl+1}(\phonaaR\rposR) &
    \im \phonaaR \besselI{\phonl}(\phonaaR\rposR) \\
    \midrule
    \wmoderCicomp &
    \phonfreq_\phonindex &>&\phonfreq_\trans &
    \phonl \besselJ{\phonl}(\phonaR \rposR)/\rposR &
    \im\phonaR \besselJ{\phonl}'(\phonaR\rposR) &
    0 \\&
    \phonfreq_\phonindex &=&\phonfreq_\trans &
    \phonkR \rposR^{|\phonl|-1} &
    \signum_\phonl \im \phonkR \rposR^{|\phonl|-1} &
    0 \\&
    \phonfreq_\phonindex &<&\phonfreq_\trans &
    \phonl \besselI{\phonl}(\phonaaR \rposR)/\rposR &
    \im \phonaaR \besselI{\phonl}'(\phonaaR \rposR) &
    0 \\
    \bottomrule
  \end{tabular}
  \caption{Fiber eigenmodes: Terms in the radial partial wave $\wmoder_\phonindex$ of the displacement eigenmode in \cref{eqn: radial partial waves flexural eigenmodes}. The radial dependence is given by Bessel functions $\besselJ{\phonl}$ and modified Bessel functions $\besselI{\phonl}$ of the first kind. The prime indicates the first derivative, $\besselJ{\phonl}'(\xpos) = \delx\besselJ{\phonl}(\xpos)$. All remaining quantities are defined in \cref{tab: phonon quantities}.}
  \label{tab: flexural eigenmodes}
\end{table*}

\begin{table*}
  \setlength{\extrarowheight}{3pt}
  \newcolumntype{A}{>{\begin{math}}r<{\end{math}}}
  \newcolumntype{B}{>{\begin{math}}c<{\end{math}}}
  \newcolumntype{C}{>{\begin{math}}l<{\end{math}}}
  \newcolumntype{F}{>{\begin{math}\thinmuskip=.75mu\medmuskip=1mu}l<{\end{math}}}
  \begin{tabularx}{\textwidth}[t]{X @{~} ABBBC F @{~} F @{~} F}
    \toprule
    \multicolumn{1}{l}{Term} & \multicolumn{5}{l}{Case} & \multicolumn{3}{l}{Component} \\
    \cmidrule{7-9}
    & & & & & & \multicolumn{1}{l}{$k=\rpos$} & \multicolumn{1}{l}{$k=\phipos$} & \multicolumn{1}{l}{$k=\zpos$} \\
    \midrule
    $\phononBCmatrixcomp^{ka}$ &
    \phonfreq_\phonindex &>& \phonfreq_\longitud &>& \phonfreq_\trans &
    \spare{\pare{\phonaR^2 - \phonkR^2}/2 - (\phonl^2-\phonl)}\besselJ{\phonl}(\phonbR )- \phonbR\besselJ{\phonl+1}(\phonbR ) &
    2(\phonl^2-\phonl)J_{\phonl}(\phonbR ) - 2 \phonl \phonbR\besselJ{\phonl+1}(\phonbR ) &
    2\phonl\phonkR\besselI{\phonl}(\phonbbR) + 2\phonkR\phonbbR \besselI{\phonl+1}(\phonbbR) \\&
    \phonfreq_\phonindex &=& \phonfreq_\longitud &>& \phonfreq_\trans &
    \spare{\pare{\phonaR ^2 - \phonkR^2}/2 - (\phonl^2-|\phonl|)} &
    2\phonl(|\phonl|-1)  &
    2 |\phonl| \phonkR \\&
    \phonfreq_\longitud &>& \phonfreq_\phonindex &>& \phonfreq_\trans &
    \spare{\pare{\phonaR^2 - \phonkR^2}/2 - (\phonl^2-\phonl)}\besselI{\phonl}(\phonbbR ) + \phonbbR\besselI{\phonl+1}(\phonbbR) &
    2(\phonl^2-\phonl)\besselI{\phonl}(\phonbbR) + 2 \phonl \phonbbR\besselI{\phonl+1}(\phonbbR) &
    2\phonl\phonkR \besselI{\phonl}(\phonbbR) + 2\phonkR\phonbbR \besselI{\phonl+1}(\phonbbR) \\&
    \phonfreq_\longitud &>& \phonfreq_\phonindex &=& \phonfreq_\trans &
    \spare{ -\phonkR^2/2 - (\phonl^2-\phonl)}\besselI{\phonl}(\phonbbR) + \phonbbR\besselI{\phonl+1}(\phonbbR) &
    2(\phonl^2-\phonl)\besselI{\phonl}(\phonbbR) + 2 \phonl\phonbbR\besselI{\phonl+1}(\phonbbR) &
    2\phonl\phonkR\besselI{\phonl}(\phonbbR) + 2\phonkR \phonbbR \besselI{\phonl+1}(\phonbbR) \\&
    \phonfreq_\longitud &>& \phonfreq_\trans &>& \phonfreq_\phonindex &
    \spare{-\pare{\phonaaR^2 + \phonkR^2}/2 - (\phonl^2-\phonl)}\besselI{\phonl}(\phonbbR) + \phonbbR\besselI{\phonl+1}(\phonbbR) &
    2(\phonl^2-\phonl)\besselI{\phonl}(\phonbbR) + 2 \phonl \phonbbR\besselI{\phonl+1}(\phonbbR) &
    2\phonl\phonkR \besselI{\phonl}(\phonbbR) + 2\phonkR \phonbbR \besselI{\phonl+1}(\phonbbR) \\
    \midrule
    $\phononBCmatrixcomp^{kb}$ &
    \phonfreq_\phonindex &>& \phonfreq_\trans &&&
    - \phonkR\phonaR\besselJ{\phonl}(\phonaR ) + (\phonl+1)\phonkR\besselJ{\phonl+1}(\phonaR ) &
    -\phonkR\phonaR\besselJ{\phonl}(\phonaR ) + 2(\phonl+1)\phonkR\besselJ{\phonl+1}(\phonaR ) &
    \phonl\phonaR\besselJ{\phonl}(\phonaR ) + \pare{\phonkR^2-\phonaR^2}\besselJ{\phonl+1}(\phonaR) \\&
    \phonfreq_\phonindex &=& \phonfreq_\trans &&&
    -(|\phonl|+1) \phonkR &
    0 &
    2 (\phonl^2+|\phonl|) + \phonkR ^2  \\&
    \phonfreq_\phonindex &<& \phonfreq_\trans &&&
    - \phonkR \phonaaR\besselI{\phonl}(\phonaaR) + (\phonl+1)\phonkR\besselI{\phonl+1}(\phonaaR) &
    -\phonkR\phonaaR\besselI{\phonl}(\phonaaR) + 2(\phonl+1)\phonkR \besselI{\phonl+1}(\phonaaR) &
    \phonl\phonaaR\besselI{\phonl}(\phonaaR) + \pare{\phonkR^2+\phonaaR^2}\besselI{\phonl+1}(\phonaaR) \\
    \midrule
    $\phononBCmatrixcomp^{kc}$ &
    \phonfreq_\phonindex &>& \phonfreq_\trans &&&
    (\phonl-\phonl^2)\besselJ{\phonl}(\phonaR ) + \phonl\phonaR\besselJ{\phonl+1}(\phonaR ) &
    \spare{2(\phonl^2-\phonl) - \phonaR^2}\besselJ{\phonl}(\phonaR ) + 2\phonaR\besselJ{\phonl+1}(\phonaR ) &
    \phonl\phonkR\besselJ{\phonl}(\phonaR ) \\&
    \phonfreq_\phonindex &=& \phonfreq_\trans &&&
    -(1-\kronecker_{\phonl 0})(|\phonl|-1) \phonkR &
    2 (1-\kronecker_{\phonl 0}) (\phonl - \signum_\phonl) \phonkR  &
    (1-\kronecker_{\phonl 0})\phonkR ^2 \\&
    \phonfreq_\phonindex &<& \phonfreq_\trans &&&
    (\phonl-\phonl^2)\besselI{\phonl}(\phonaaR) - \phonl \phonaaR\besselI{\phonl+1}(\phonaaR) &
    \spare{2(\phonl^2-\phonl) + \phonaaR^2}\besselI{\phonl}(\phonaaR) - 2\phonaaR\besselI{\phonl+1}(\phonaaR) &
    \phonl\phonkR\besselI{\phonl}(\phonaaR) \\
    \bottomrule
  \end{tabularx}
  \caption{Fiber eigenmodes: Terms in the radial partial wave $\stressmoder_\phonindex$ of the stress modal field in \cref{eqn: stress vector field cylindrical coordinates}. The terms are evaluated on the fiber surface ($\rpos=\rad$) and correspond to the matrix elements of $\phononBCmatrix$ in the boundary condition \cref{eqn: free body boundary conditions matrix notation} and frequency equation \cref{eqn: phonon frequency equation}. All remaining quantities are defined in \cref{tab: phonon quantities}.}
  \label{tab: stress tensor components}
\end{table*}

\begin{table*}
  \setlength{\extrarowheight}{3pt}
  \newcolumntype{A}{>{\begin{math}}r<{\end{math}}}
  \newcolumntype{B}{>{\begin{math}}c<{\end{math}}}
  \newcolumntype{C}{>{\begin{math}}l<{\end{math}}}
  \begin{tabularx}{\textwidth}[t]{CBCX}
    \toprule
    \strainmodercomp^{\rpos\rpos}_\phonindex &=& \big\{B \phonkR \phonaaR \besselI{\phonl}(\phonaaR \rposR) + C\phonl(\phonl-1) \besselI{\phonl}(\phonaaR \rposR)/\rposR^2
    - \spare{ B(\phonl+1)\phonkR - C\phonl \phonaaR } \besselI{\phonl+1}(\phonaaR \rposR)/\rposR + A \phonbbR^2 \besselI{\phonl}(\phonbbR \rposR)
    + A \phonl(\phonl-1) \besselI{\phonl}(\phonbbR \rposR)/\rposR^2 & ~ \\
    \multicolumn{4}{r}{$- A \phonbbR \besselI{\phonl+1}(\phonbbR \rposR)/\rposR \big\}/\rad$}\\
    \strainmodercomp^{\phipos\phipos}_\phonindex &=& \cpare{- C \phonl(\phonl-1) \besselI{\phonl}(\phonaaR \rposR)/\rposR^2
    + \spare{ B (\phonl+1) \phonkR - C \phonl \phonaaR } \besselI{\phonl+1}(\phonaaR \rposR)/\rposR - A \phonl(\phonl-1) \besselI{\phonl}(\phonbbR \rposR)/\rposR^2
    + A \phonbbR \besselI{\phonl+1}(\phonbbR \rposR)/\rposR}/\rad\\
    \strainmodercomp^{\zpos\zpos}_\phonindex &=& \spare{- B \phonkR \phonaaR \besselI{\phonl}(\phonaaR \rposR) - A \phonkR^2 \besselI{\phonl}(\phonbbR \rposR)}/\rad\\
    \strainmodercomp^{\rpos\phipos}_\phonindex &=& \big\{ - \im \pare{B \phonkR \phonaaR - C \phonaaR^2} \besselI{\phonl}(\phonaaR \rposR)/2
    + \im C \phonl(\phonl-1) \besselI{\phonl}(\phonaaR \rposR)/\rposR^2 + \im \spare{ B (\phonl+1) \phonkR - C \phonaaR } \besselI{\phonl+1}(\phonaaR \rposR)/\rposR
    + \im A \phonl(\phonl-1) \besselI{\phonl}(\phonbbR \rposR)/\rposR^2 & ~ \\
    \multicolumn{4}{r}{ $+ \im A \phonl \phonbbR \besselI{\phonl+1}(\phonbbR \rposR)/\rposR \big\}/\rad$ }\\
    \strainmodercomp^{\rpos\zpos}_\phonindex &=& \spare{\im \phonl \pare{B \phonaaR + C \phonkR} \besselI{\phonl}(\phonaaR \rposR)/2\rposR
    + \im B \pare{\phonaaR^2 + \phonkR^2} \besselI{\phonl+1}(\phonaaR \rposR)/2 +  \im \phonl A \phonkR \besselI{\phonl}(\phonbbR \rposR)/\rposR
    + \im A \phonkR \phonbbR \besselI{\phonl+1}(\phonbbR \rposR)}/\rad\\
    \strainmodercomp^{\phipos\zpos}_\phonindex &=& \spare{-\phonl \pare{B \phonaaR + C\phonkR} \besselI{\phonl}(\phonaaR \rposR)/2\rposR
    + \pare{B\phonkR^2 - C\phonkR \phonaaR} \besselI{\phonl+1}(\phonaaR \rposR)/2 - A \phonl \phonkR \besselI{\phonl}(\phonbbR \rposR)/\rposR}/\rad\\
    \bottomrule
  \end{tabularx}
  \caption{Fiber eigenmodes: Components of the radial partial wave $\strainmoder_\phonindex$ of the strain modal field for the surface mode sector, $\phonfreq < \phonfreq_\trans$. The amplitudes $A,B,C$ are determined by the boundary conditions \cref{eqn: free body boundary conditions matrix notation} and the normalization condition \cref{eqn: normalization phonon radial partial wave}. All remaining quantities are defined in \cref{tab: phonon quantities}.}
  \label{tab: flexural strain tensor components}
\end{table*}

\begin{table*}
  \setlength{\extrarowheight}{3pt}
  \newcolumntype{A}{>{\begin{math}}r<{\end{math}}}
  \newcolumntype{B}{>{\begin{math}}c<{\end{math}}}
  \newcolumntype{C}{>{\begin{math}}l<{\end{math}}}
  \begin{tabular}{ABBBC @{\quad} C @{\quad} C}
    \toprule
    \multicolumn{5}{l}{Case} & \multicolumn{2}{l}{Eigenmode component}\\
    \cmidrule{6-7}
    & & & & & \multicolumn{1}{l}{$k=\rpos$} & \multicolumn{1}{l}{$k=\zpos$} \\
    \midrule
    \phonfreq_\phonindex &>& \phonfreq_\longitud &>& \phonfreq_\trans &
    \phonbR  A \spare{  2 {\phonkR^2\eta} \besselJ{1}(\phonaR \rposR)/\pare{\phonkR^2 - \phonaR^2} -\besselJ{1}(\phonbR\rposR) } &
    \im \phonkR A\spare{ 2 {\phonaR\phonbR\eta} \besselJ{0}(\phonaR \rposR)/\pare{\phonkR^2 - \phonaR^2} + \besselJ{0}(\phonbR \rposR) }\\
    \phonfreq_\phonindex &=& \phonfreq_\longitud &>& \phonfreq_\trans &
    A \pare{\phonaR^2 - \phonkR^2} \besselJ{1}(\phonaR \rposR)/\spare{2 \phonaR \besselJ{0}(\phonaR)} &
    \im \phonkR A \cpare{ 1+ \pare{\phonaR^2 - \phonkR^2} \besselJ{0}(\phonaR \rposR)/\spare{2 \phonkR^2 \besselJ{0}(\phonaR)} } \\
    \phonfreq_\longitud &>& \phonfreq_\phonindex &>& \phonfreq_\trans &
     \phonbbR A \spare{ - 2 {\phonkR^2\tilde{\eta}}/\pare{\phonkR^2 - \phonaR^2} \besselJ{1}(\phonaR \rposR) +\besselI{1}(\phonbbR\rposR) } &
     \im \phonkR A \spare{ - 2 {\phonaR \phonbbR \tilde{\eta} }/\pare{\phonkR^2 - \phonaR^2}  \besselJ{0}(\phonaR \rposR) + \besselI{0}(\phonbbR\rposR) } \\
    \phonfreq_\longitud &>& \phonfreq_\phonindex &=& \phonfreq_\trans &
    \phonbbR A \spare{ \besselI{1}(\phonbbR \rposR) - 2 \besselI{1}(\phonbbR) \rposR} &
    \im A \spare{ \phonkR \besselI{0}(\phonbbR \rposR) -4  {\phonbbR}\besselI{1}(\phonbbR)/\phonkR } \\
    \phonfreq_\longitud &>& \phonfreq_\trans &>& \phonfreq_\phonindex &
    \phonbbR A \spare{ - 2 {\phonkR^2 \breve{\eta}}\besselI{1}(\phonaaR \rposR)/\pare{\phonkR^2 + \phonaaR^2}   +\besselI{1}(\phonbbR\rposR) } &
     \im \phonkR A { - 2 {\phonaaR \phonbbR \breve{\eta} } \besselI{0}(\phonaaR \rposR) /\pare{\phonkR^2 + \phonaaR^2}+ \besselI{0}(\phonbbR \rposR) } \\
    \bottomrule
  \end{tabular}
  \caption{Longitudinal ($\Lmode$) fiber eigenmodes: Nonzero components of the radial partial wave of the displacement eigenmode. The amplitude $A\in\C$ can be obtained from the normalization condition \cref{eqn: normalization phonon radial partial wave}, all other quantities are defined in \cref{tab: phonon quantities}. These expressions are valid as long as $\phononBCmatrixcomp^{\phipos c}\neq0$, that is, away from intersections with torsional bands. Otherwise, the displacement eigenmode can be obtained from the general expressions in \cref{tab: flexural eigenmodes} in conjunction with \cref{eqn: free body boundary conditions matrix notation}.}
  \label{tab: longitudinal eigenmodes}
\end{table*}

\begin{table*}
  \setlength{\extrarowheight}{3pt}
  \newcolumntype{A}{>{\begin{math}}r<{\end{math}}}
  \newcolumntype{B}{>{\begin{math}}c<{\end{math}}}
  \newcolumntype{C}{>{\begin{math}}l<{\end{math}}}
  \begin{tabular}{ABBBC @{\quad} ABC}
    \toprule
    \multicolumn{5}{l}{Case} & \multicolumn{3}{l}{Frequency equation} \\
    \midrule
    \phonfreq_\phonindex &>& \phonfreq_\longitud &>& \phonfreq_\trans &
    0 &=& 2 \phonbR \pare{\phonaR^2 + \phonkR^2 } \besselJ{1}(\phonaR) \besselJ{1}(\phonbR)  - \pare{\phonaR^2-\phonkR^2}^2 \besselJ{1}(\phonaR) \besselJ{0}(\phonbR) - 4 \phonaR \phonbR \phonkR^2 \besselJ{0}(\phonaR) \besselJ{1}(\phonbR) \\
    \phonfreq_\phonindex &=& \phonfreq_\longitud &>& \phonfreq_\trans &
    0 &=& \besselJ{1}(\phonaR) \\
    \phonfreq_\longitud &>&\phonfreq_\phonindex &>& \phonfreq_\trans &
    0 &=& -2 \phonbbR \pare{\phonaR^2 + \phonkR^2} \besselJ{1}(\phonaR) \besselI{1}(\phonbbR) - \pare{\phonaR^2-\phonkR^2}^2 \besselJ{1}(\phonaR) \besselI{0}(\phonbbR) + 4 \phonaR \phonbbR  \phonkR^2 \besselJ{0}(\phonaR) \besselI{1}(\phonbbR) \\
    \phonfreq_\longitud &>& \phonfreq_\phonindex &=& \phonfreq_\trans &
    0 &=& \phonkR^2 \besselI{0}(\phonbbR) - 6 \phonbbR \besselI{1}(\phonbbR) \\
    \phonfreq_\longitud &>&\phonfreq_\trans &>& \phonfreq_\phonindex &
    0 &=& 2 \phonbbR \pare{\phonaaR^2 - \phonkR^2}  \besselI{1}(\phonaaR) \besselI{1}(\phonbbR) - \pare{\phonaaR^2+\phonkR^2}^2 \besselI{1}(\phonaaR) \besselI{0}(\phonbbR) + 4 \phonaaR \phonbbR \phonkR^2 \besselI{0}(\phonaaR) \besselI{1}(\phonbbR) \\
    \bottomrule
  \end{tabular}
  \caption{Longitudinal ($\Lmode$) fiber eigenmodes: Frequency equations. All quantities are defined in \cref{tab: phonon quantities}.}
  \label{tab: longitudinal modes frequency equations}
\end{table*}

\emph{Torsional modes} are characterized by $\phonl=0$ and the frequency equation $\phononBCmatrixcomp^{\phipos c}=0$. They have zero longitudinal and radial displacement $\wmodecomp_\phonindex^\rpos = \wmodecomp_\phonindex^\zpos = 0$ and are therefore purely transverse excitations; see \subcref{fig: phonon modes}{a}. The radial partial waves of the eigenmode and the frequency equations are listed in \cref{tab: torsional modes}. The roots of the frequency equation form bands $\Tmode_{0\phonn}$ in the $(\phonk,\phonfreq)$ plane plotted in \subcref{fig: phonon bands}{a}.

On the transverse sound line ($\phonfreq_\phonindex=\phonfreq_\trans$), the frequency equation is always satisfied, while it has no solution below the transverse sound line ($\phonfreq_\phonindex<\phonfreq_\trans$). The fundamental (i.e., lowest frequency) torsional band $\Tmode_{01}$ thus coincides with the transverse sound line. The radial partial wave of the strain modal field on the $\Tmode_{01}$ band has two nonzero components:
\begin{equation}\label{eqn: fundamental torsional strain modal field}
  \begin{split}
    \strainmodercomp^{\phipos\zpos}_\phonindex(\rpos) &= \strainmodercomp^{\zpos\phipos}_\phonindex(\rpos) = \frac{\im}{2}\frac{C}{\rad}\phonkR\rposR~;
  \end{split}
\end{equation}
see \cref{tab: phonon quantities} for definitions of the symbols.

\emph{Longitudinal modes} are characterized by $\phonl=0$ and the frequency equation
\begin{equation}\label{eqn: phonon frequency equation 3}
  \det \begin{pmatrix}\phononBCmatrixcomp^{\rpos a} & \phononBCmatrixcomp^{\rpos b}\\\phononBCmatrixcomp^{\zpos a} & \phononBCmatrixcomp^{\zpos b}\end{pmatrix} = 0~.
\end{equation}
They have zero azimuthal displacement $\wmodecomp_\phonindex^\phipos = 0$ and are indeed largely longitudinal excitations similar to sound waves, but they do have a small nonzero radial component, see \subcrefand{fig: phonon modes}{b}{d}. Longitudinal modes exist in the bulk, mixed, and surface mode sector, and cross both sound lines. The radial partial wave of the displacement field is given explicitly in \cref{tab: longitudinal eigenmodes} for all five cases. These expressions hold as long as $\phononBCmatrixcomp^{\phipos c}\neq0$ (i.e., provided the mode is not located at a crossing with a torsional band). Otherwise, the eigenmode can be obtained from the general expressions in \cref{tab: flexural eigenmodes} by solving the boundary conditions \cref{eqn: free body boundary conditions matrix notation}. The frequency equations obtained from \cref{eqn: phonon frequency equation 3} are listed in \cref{tab: longitudinal modes frequency equations}, and are known as \emph{Pochhammer equations}. The roots of the Pochhammer equations form bands $\Lmode_{0\phonn}$ in the $(\phonk,\phonfreq)$ plane, as shown in \subcref{fig: phonon bands}{b}.

The fundamental longitudinal band $\Lmode_{01}$ lies in the mixed-mode sector in the low-frequency limit. In this case, the radial partial wave of the strain modal field has components
\begin{equation}\label{eqn: longitudinal mixed strain modal fields}
  \begin{split}
    \strainmodercomp^{\rpos\rpos}_\phonindex(\rpos) &= - \phonbbR^2 \frac{A}{\rad} \spare{ \frac{2\phonkR^2\tilde{\eta}}{\phonkR^2 - \phonaR^2} \frac{\phonaR}{\phonbbR} \besselJ{1}'(\phonaR \rposR ) - \besselI{1}'(\phonbbR \rposR) }\\
    \strainmodercomp^{\phipos\phipos}_\phonindex(\rpos) &= - \phonbbR \frac{A}{\rad} \spare{ \frac{2\phonkR^2\tilde{\eta} }{\phonkR^2 - \phonaR^2} \frac{\besselJ{1}(\phonaR \rposR )}{\rposR} - \frac{\besselI{1}(\phonbbR \rposR)}{\rposR} }\\
    \strainmodercomp^{\zpos\zpos}_\phonindex(\rpos) &= \phonkR^2 \frac{A}{\rad} \spare{ \frac{2\phonaR \phonbbR\tilde{\eta}}{\phonkR^2 - \phonaR^2} \besselJ{0}(\phonaR \rposR ) - \besselI{0}(\phonbbR \rposR) }\\
    \strainmodercomp^{\rpos\zpos}_\phonindex(\rpos) &= - \im \phonbbR \phonkR \frac{A}{\rad}  \spare{ \tilde{\eta} \besselJ{1}(\phonaR \rposR) - \besselI{1}(\phonbbR\rposR) } ~,
  \end{split}
\end{equation}
while the remaining independent components vanish. Refer to \cref{tab: phonon quantities} for definitions of the symbols.

\emph{Flexural modes} appear for azimuthal orders $|\phonl|\geq0$. In this case, the frequency equation \cref{eqn: phonon frequency equation} does not in general factorize. There is then only one family of flexural modes, with a displacement field specified in \cref{eqn: phonon cylinder eigenmode partial wave decomposition,eqn: radial partial waves flexural eigenmodes}, as well as \cref{tab: flexural eigenmodes}.  A flexural mode with azimuthal order $|\phonl| = 1$ is shown in \subcrefand{fig: phonon modes}{c}{e}. The boundary conditions \cref{eqn: free body boundary conditions matrix notation} enable us to relate two of the three amplitudes $A,B,C$ to the third one. For instance,
\begin{equation}\label{eqn: dependent amplitudes phonon modes jneq0}
  \begin{split}
    B &= \frac{ \phononBCmatrixcomp^{\rpos c} \phononBCmatrixcomp^{\phipos a} - \phononBCmatrixcomp^{\rpos a} \phononBCmatrixcomp^{\phipos c} }{ \phononBCmatrixcomp^{\rpos b} \phononBCmatrixcomp^{\phipos c} - \phononBCmatrixcomp^{\rpos c} \phononBCmatrixcomp^{\phipos b} } A \\
    C &= \frac{ \phononBCmatrixcomp^{\rpos a} \phononBCmatrixcomp^{\phipos b} - \phononBCmatrixcomp^{\rpos b} \phononBCmatrixcomp^{\phipos a} }{ \phononBCmatrixcomp^{\rpos b} \phononBCmatrixcomp^{\phipos c} - \phononBCmatrixcomp^{\rpos c} \phononBCmatrixcomp^{\phipos b} } A ~,
  \end{split}
\end{equation}
and the remaining amplitude $A$ is fixed by the normalization condition \cref{eqn: normalization phonon radial partial wave} %
\footnote{%
\Cref{eqn: dependent amplitudes phonon modes jneq0} is only valid if all subdeterminants of $\phononBCmatrix$ are nonzero, as is usually the case for flexural modes. Otherwise, the flexural family decomposes into different independent mode families, with frequency equations determined by the respective subdeterminants.
}.

The roots of the frequency equation $\det \phononBCmatrix=0$ with matrix components listed in \cref{tab: stress tensor components} form bands $\Fmode_{|\phonl|\phonn}$ in the $(\phonk,\phonfreq)$ plane. In \subcrefand{fig: phonon bands}{c}{d}, the $\Fmode_{1\phonn}$ and $\Fmode_{2\phonn}$ bands are shown. The fundamental flexural band $\Fmode_{01}$ lies in the surface mode sector. The radial partial waves of the strain modal field in this case are listed in \cref{tab: flexural strain tensor components}.

In summary, we can label the phononic eigenmodes of a fiber with indices
\begin{equation}\label{eqn: fiber phonon mode indices}
  \begin{aligned}
    \phonl &\in \Z~, \quad & \phonfam &\in \{\Lmode,\Tmode\}_{\phonl=0} \text{ or }  \{\Fmode\}_{\phonl\neq0}~, \\
    \phonk &\in \R~, & \phonn &\in \N~.
  \end{aligned}
\end{equation}
Following the conventions used for photonic fiber eigenmodes, we label the different phonon bands by their mode indices as $\phonfam_{|\phonl|\phonn}$. At azimuthal order $\phonl = 0$, there are then $\Tmode_{0\phonn}$ and $\Lmode_{0\phonn}$ bands, shown in \subcrefand{fig: phonon bands}{a}{b}. At azimuthal order $|\phonl| \geq 1$, there are $\Fmode_{|\phonl|\phonn}$ bands, plotted in \subcrefand{fig: phonon bands}{c}{d} for $|\phonl|=1,2$.

\medskip{}

\Cref{fig: phonon bands} shows that there are three fundamental bands without a finite minimum frequency: $\Lmode_{01}$, $\Tmode_{01}$, and $\Fmode_{11}$. Nanofiber-based cold atom traps have trap frequencies on the order of $\SI{100}{\kilo\hertz}$ \cite{meng_near-ground-state_2018,ostfeldt_dipole_2017,corzo_large_2016,goban_demonstration_2012} as we discuss in \cref{sec: case study}, so only modes on the fundamental bands can resonantly couple to the atoms for typical parameters of the nanofiber. The fundamental bands are therefore of special importance in this article, and we provide approximate expressions for the dispersion relations and displacement fields in the low-frequency limit. The band $\Tmode_{01}$ (see \subcref{fig: phonon bands}{a}) lies on the transverse sound line and is thus given by the dispersion relation
\begin{equation}\label{eqn: dispersion fundamental torsional mode}
  \phonfreq_\Tmode(\phonk) = \ctrans |\phonk|~.
\end{equation}
The only nonzero component of the displacement eigenmode, normalized according to \cref{eqn: normalization phonon radial partial wave}, is
\begin{equation}\label{eqn: displacement T modes}
    \wmodercomp^\phipos_{\phonindex}(\rpos) = \frac{2}{\rad^2}\rpos~.
\end{equation}
The displacement induced by a $\Tmode_{01}$ mode is shown in \subcref{fig: phonon modes}{a}.
The band $\Lmode_{01}$ (see \subcref{fig: phonon bands}{b}) lies in the mixed-mode sector for low frequencies. The exact dispersion relation is the solution to the transcendental Pochhammer equation given in \cref{tab: longitudinal modes frequency equations}. It has the linear asymptote
\begin{equation}\label{eqn: dispersion fundamental longitudinal mode low freq limit}
  \phonfreq_\Lmode (\phonk) \simeq \cFL  |\phonk|~,
\end{equation}
with an effective hybrid speed of sound
\begin{equation}\label{eqn: effective fundamental longitudinal speed of sound}
  \cFL \equiv \sqrt{\frac{\YoungE}{\dens}}~.
\end{equation}
The components of the normalized radial partial waves approximated to linear order in $\phonk\rpos \ll 1$ for wavelengths much larger than the fiber radius are
\begin{equation}
  \begin{split}\label{eqn: displacement L modes low-frequency limit}
    \wmodercomp^\rpos_{\phonindex}(\rpos) &\simeq \frac{\sqrt{2}\Poissonnu\phonk}{\rad} \rpos\\
    \wmodercomp^\zpos_{\phonindex}(\rpos) &\simeq \im \frac{\sqrt{2}}{\rad}~,
  \end{split}
\end{equation}
while the azimuthal component vanishes. An $\Lmode_{01}$ mode is plotted in \subcrefand{fig: phonon modes}{b}{d}.
The band $\Fmode_{11}$ (see \subcref{fig: phonon bands}{c}) lies in the surface mode sector. It derives from the frequency equation \cref{eqn: phonon frequency equation}. Close to the origin, the band has a quadratic asymptote:
\begin{equation}\label{eqn: dispersion fundamental flexural mode low freq limit}
  \phonfreq_{F}(\phonk) \simeq \frac{\cFL\rad}{2} \phonk^2~.
\end{equation}
The density of states therefore diverges as $\phonk \to 0$, which leads to a strong coupling between $\Fmode_{11}$ modes and trapped atoms, as we discuss in \cref{sec: case study}. The normalized radial partial waves are
\begin{equation}
  \begin{split}\label{eqn: displacement F modes low-frequency limit}
      \wmodercomp^\rpos_{\phonindex}(\rpos) &\simeq \frac{1}{\rad}\\
      \wmodercomp^\phipos_{\phonindex}(\rpos) &\simeq \frac{\im \phonl}{\rad} \\
      \wmodercomp^\zpos_{\phonindex}(\rpos) &\simeq -\frac{\im\phonk}{\rad}\rpos
  \end{split}
\end{equation}
to linear order in $\phonk\rpos$.
\subCrefand{fig: phonon modes}{c}{e} show the displacement caused by a $\Fmode_{11}$ mode.

\medskip{}

As discussed in \cref{sec: case study}, there is experimental evidence that low-frequency $\Tmode_{01}$ modes are reflected at the tapered ends of the nanofiber, resulting in standing waves confined to the nanofiber region and discrete mechanical resonance frequencies. Likewise, we are interested in $\Fmode_{11}$ modes confined to the nanofiber as a way of reducing the atom heating. In order to obtain the corresponding phononic eigenmodes, it is in principle necessary to account for the two tapers that connect the nanofiber region to the regular macroscopic glass fiber and to solve the phononic eigenmode equation for this more complex geometry \cite{wuttke_optically_2013,pennetta_tapered_2016}. Here, we approximate the desired behavior by imposing periodic boundary conditions on the eigenmodes of an infinite cylinder and require the displacement to vanish at the beginning ($\zpos = 0$) and the end of the nanofiber ($\zpos = \len$). This condition can be met for all three fundamental modes when approximating the radial partial waves \cref{eqn: displacement L modes low-frequency limit,eqn: displacement F modes low-frequency limit} to constant order in $\phonk\rpos$, which is a good approximation for a nanofiber; compare \cref{fig: phonon modes}. The resulting eigenmodes and corresponding strain modal fields are then of the form
\begin{equation}\label{eqn: phonon cylinder discrete eigenmode partial wave decomposition}
  \begin{split}
    \wmode_\phonindex(\pos) &= \frac{\wmoder_\phonindex(\rpos) }{\sqrt{\pi\len}}\sin(\phonk_m \zpos) e^{\im\phonl \phipos} \\
    \strainmode_\phonindex(\pos) &= \frac{\strainmoder_\phonindex(\rpos)}{\sqrt{\pi\len}} \cos(\phonk_m \zpos) e^{\im\phonl \phipos}
  \end{split}
\end{equation}
instead of \cref{eqn: phonon cylinder eigenmode partial wave decomposition}, where $\len$ is the length of the nanofiber. Torsional resonances in particular appear at $\phonfreq_m = m\pi \ctrans/\len$. The propagation constant can take only the discrete values
\begin{align}\label{eqn: momentum resonator modes}
  \phonk_m &= m \pi/\len & m \in \N
\end{align}
which form a subset of each fundamental band. The normalization condition \cref{eqn: normalization phonon radial partial wave} for the radial partial waves is unchanged.

\section{Atom-Phonon Interaction}
\label{sec: interaction appendix}
There are two mechanisms that lead to atom-phonon interaction, as we discuss in \cref{sec: framework} of this article: displacement coupling (\radpress) and strain coupling (\strainopt). In consequence, the coupling functions  $\couplingfunct_{\phonindex}(\atpos)$ appearing in the interaction Hamiltonian \cref{eqn: general form atom-phonon interaction Hamiltonian}, as well as the coupling constants $\atphong_{\phonindex i}$ appearing in the linear-force interaction Hamiltonian \cref{eqn: linear force interaction Hamiltonians} have two contributions:
\begin{equation}\label{eqn: coupling function contributions}
  \begin{split}
    \couplingfunct_{\phonindex}(\atpos) &= \couplingfunct^\radpress_{\phonindex}(\atpos) + \couplingfunct^\strainopt_{\phonindex}(\atpos)\\
    \atphong_{\phonindex i} &= \atphong^\radpress_{\phonindex i} + \atphong^\strainopt_{\phonindex i}~.
  \end{split}
\end{equation}
In \cref{sec: displacement coupling,sec: strain coupling}, we model the dependence of the potential $\pot$ experienced by the atom on displacement $\ufield$ and strain $\straintens$ in the case of a two-color nanofiber-based atom trap. This model then enables us to derive explicit expressions for the coupling functions and the atom-phonon coupling constants. \Cref{sec: heating} provides additional details on how to calculate phonon-induced atom heating rates once the coupling constants are known.

\subsection{Displacement Coupling}
\label{sec: displacement coupling}

Only modes on the three fundamental phonon bands $\Tmode_{01}$, $\Lmode_{01}$, and $\Fmode_{11}$ of the nanofiber introduced in \cref{sec: phonon fiber eigenmodes} have frequencies comparable to those of an atom moving in a nanofiber-based trap, and will therefore interact significantly with the atom. Modes on the longitudinal $\Lmode_{01}$ band and on the flexural $\Fmode_{11}$ band lead to a displacement of the surface at first order in $\ufield$. Modes on the torsional $\Tmode_{01}$ band (\subcref{fig: phonon modes}{a}) lead to a change of the fiber radius of second order because $\ufield$ is orthogonal to the surface normal. In consequence, only the longitudinal and flexural modes will interact with the atom through displacement coupling in the linearized Hamiltonian \cref{eqn: atom-phonon interaction Hamiltonian contributions}.

The $\Lmode_{01}$ modes (\subcref{fig: phonon modes}{b}) lead to a $\zpos$-dependent modulation of the fiber radius by the radial displacement on the fiber surface $\ufieldcomp^\rpos(\rpos=\rad,\zpos)$ without displacing the fiber axis. The change in radius has two effects: First, it shifts the surface of the fiber together with the electromagnetic fields surrounding it relative to the trapped atom. Second, it leads to new photonic eigenmodes and therefore deforms the electromagnetic fields. As discussed in \cref{sec: framework}, we neglect the second effect and assume that both optical and surface potentials are shifted radially by $\ufieldcomp^\rpos(\rad,\zpos)\ervec$ without being deformed.

The $\Fmode_{11}$ modes (\subcref{fig: phonon modes}{c}) displace the entire fiber cross section in the plane orthogonal to the fiber axis by $\ufieldcomp^\rpos(\rad,\varphi,\zpos)\ervec + \ufieldcomp^\phipos(\rad,\varphi,\zpos)\ephivec$, without changing the fiber radius. Since the wavelengths of the relevant vibrations are much larger than the optical wavelengths, the fiber appears approximately unchanged on length scales relevant for the photon modes. We can therefore again neglect deformations of the photon eigenmodes and model the effect of the flexural mode as a displacement of the entire potential along with the fiber cross section.

The effect of the fundamental modes is thus to shift the potential at position $\pos$ by a vector $\Delta \ufield(\pos)$ that depends on the phonon field on the fiber surface. The direct dependence of the potential on the displacement can then be modeled as \cite{le_kien_phonon-mediated_2007}
\begin{equation}\label{eqn: displacement coupling model}
  \begin{split}
    \pot[\ufield,\straintens](\pos) \equiv \pot[\zerovec,\straintens](\pos - \Delta \ufield(\pos))~.
  \end{split}
\end{equation}
The entire potential is shifted due to displacement of the fiber surface in addition to any changes to the potential that arise from the strain $\straintens$ caused by displacement inside the fiber. This model allows us to evaluate the displacement coupling term in \cref{eqn: atom-phonon interaction Hamiltonian contributions}: The functional derivative reduces to conventional partial derivatives of the unperturbed potential $\pot_0$ and
\begin{equation}
  \partialfrechetD_{\ufield}\pot_{(\zerovec,\zerovec)}[\ufieldop](\atposop) = - \Delta \ufieldop(\atposop) \cdot \grad \pot_0(\atposop)~.
\end{equation}
By expanding the displacement field in terms of the fiber eigenmodes, see \cref{eqn: phonon field operator mode expansion}, the shifts due to the fundamental phonon modes can be summarized as
\begin{multline}
  \Delta \ufieldop(\pos) \equiv \sum_\phonindex \ufieldmodedens_\phonindex \{ [\wmodecomp^\rpos_\phonindex(\rad,\phipos,\zpos) \ervec \\+ \kronecker_{\phonfam \Fmode} \wmodecomp^\phipos_\phonindex(\rad,\phipos,\zpos) \ephivec ] \phonaop_\phonindex + \hc\}~.
\end{multline}
Here, $\ufieldmodedens_\phonindex$ is the displacement mode density and $\wmode_\phonindex$ the phonon eigenmodes \cref{eqn: phonon cylinder eigenmode partial wave decomposition}. The Kronecker symbol $\kronecker_{\phonfam \Fmode}$ selects the flexural mode family $\phonfam = \Fmode$. This equation holds for all three fundamental bands since $\wmodecomp^\rpos(\pos) = \wmodecomp^\phipos(\pos) = 0$ for torsional modes. The resulting displacement coupling function in \cref{eqn: general form atom-phonon interaction Hamiltonian,eqn: coupling function contributions} is
\begin{multline}\label{eqn: displacement coupling function}
  \couplingfunct^\radpress_\phonindex(\pos) = - \ufieldmodedens_\phonindex \Big[ \wmodecomp_\phonindex^\rpos (\rad,\phipos,\zpos)  \delr \pot_0(\pos)  \\
  +  \kronecker_{\phonfam   \Fmode}   \wmodecomp_\phonindex^\phipos (\rad,\phipos,\zpos)\frac{\delphi}{\rpos}\pot_0(\pos) \Big]~.
\end{multline}
The corresponding displacement coupling constants for nanofiber-trapped atoms obtained from \cref{eqn: definition coupling constants,eqn: displacement coupling function} are
\begin{align}\label{eqn: 3d trap radiation pressure coupling constants}
    \atphong^\radpress_{\phonindex \rpos} &= - \frac{ \rtrapfreq}{2} \spare{ \frac{\ufieldmodedens_\phonindex \wmodecomp^{\rpos}_0}{\atrpossd} + \kronecker_{\phonfam \Fmode} \pare{\frac{\rphitrapfreq}{\rtrapfreq}}^2 \frac{\ufieldmodedens_\phonindex \wmodecomp^{\phipos}_0}{\atrpossd} } \nonumber\\
    \couplingfunct^{\radpress}_{\phonindex\phipos} &= - \frac{ \phitrapfreq}{2} \spare{ \kronecker_{\phonfam \Fmode} \frac{\ufieldmodedens_\phonindex \wmodecomp^{\phipos}_0}{ \trapr \atphipossd} +  \pare{\frac{\rphitrapfreq}{\phitrapfreq}}^2 \frac{\ufieldmodedens_\phonindex \wmodecomp^{\rpos}_0}{ \trapr \atphipossd} } \\
    \couplingfunct^{\radpress}_{\phonindex\zpos} &= - \frac{ \ztrapfreq}{2} \spare{ \pare{\frac{\rztrapfreq}{\ztrapfreq}}^2 \frac{\ufieldmodedens_\phonindex \wmodecomp^{\rpos}_0}{ \atzpossd} + \kronecker_{\phonfam \Fmode} \pare{\frac{\phiztrapfreq}{\ztrapfreq}}^2 \frac{\ufieldmodedens_\phonindex \wmodecomp^{\phipos}_0}{ \atzpossd} }~, \nonumber
\end{align}
where
\begin{align}
    \wmodecomp^\rpos_0 &\equiv \wmodecomp^\rpos_\phonindex(\rad,\trapphi,\trapz) &
    \wmodecomp^\phipos_0 &\equiv \wmodecomp^\phipos_\phonindex(\rad,\trapphi,\trapz)
\end{align}
are the displacement modal fields evaluated on the fiber surface. Note that the model predicts only coupling between phonons and the axial motion of the atom if $\phiztrapfreq \neq 0$ or $\phiztrapfreq \neq 0$, that is, if the potential has symmetries misaligned with the cylindrical coordinate axes.

We can derive explicit expressions for the displacement coupling constants by using the approximate expressions for the displacement field of modes on the fundamental phonon bands $\Lmode_{01}$ and $\Fmode_{11}$ given in \cref{sec: phonon fiber eigenmodes}. In the case when the cross-couplings $\ijtrapfreq$ are negligible, the coupling constant of the radial atomic motion to an $\Lmode_{01}$ phonon mode \cref{eqn: displacement L modes low-frequency limit} of frequency $\phonfreq_\phonindex$ is
\begin{equation}
  |\atphong^\radpress_{\Lmode \rpos}| \equiv \frac{\Poissonnu}{2\pi\cFL}\sqrt{\frac{M\rtrapfreq^3\phonfreq_\phonindex}{2 \dens}}~,
\end{equation}
and there is no coupling to the azimuthal and axial motion, $\atphong^\radpress_{\Lmode \phipos} = \atphong^\radpress_{\Lmode \zpos}  = 0$. The coupling constant of the radial and azimuthal motion to a $\Fmode_{11}$ phonon mode \cref{eqn: displacement F modes low-frequency limit} is
\begin{align}\label{eqn: flexural mode displacement coupling constant}
  |\atphong^\radpress_{\Fmode i}| &= \frac{1}{4\pi\rad}\sqrt{\frac{M \itrapfreq^3}{\dens \phonfreq_\phonindex}}& i &\in \{\rpos,\phipos\}~,
\end{align}
and there is no coupling to the axial motion, $\atphong^\radpress_{\Fmode \zpos} = 0$.

The model \cref{eqn: displacement coupling model} relies on the simple geometrical shape of the nanofiber and the symmetries of its fundamental mechanical modes. Nanophotonic structures that have more complex geometries in general require a more careful analysis of the change of optical and dispersion potentials. The variation of the optical potential, for instance, can be modeled more generally by perturbatively calculating the new photonic eigenmodes in the presence of shifted boundaries of the nanostructure \cite{johnson_perturbation_2002}. We choose a similar approach in the next section to obtain the perturbed eigenmodes in the presence of a modified permittivity but unchanged boundaries.

\subsection{Strain Coupling}
\label{sec: strain coupling}

All three fundamental phonon bands $\Tmode_{01}$, $\Lmode_{01}$, and $\Fmode_{11}$ of the nanofiber induce strain in the fiber. In order to evaluate the strain coupling term $\partialfrechetD\pot_{(\zerovec,\zerovec)}[\straintens]$ in \cref{eqn: atom-phonon interaction Hamiltonian contributions}, we model how each phonon mode changes the potential through the strain it causes. We neglect the influence of strain on the surface forces $\partialfrechetD_{\straintens} \pot_{\adsorption \,{(\zerovec,\zerovec)}}[\straintens] = 0$, as we discuss in \cref{sec: framework}. The strain dependence then arises only from changes of the red- and blue-detuned optical potentials. A nonzero strain changes the electromagnetic properties of the fiber due to the photoelastic effect, which we model through a strain-dependent permittivity tensor $\prtrbd\relpermitttens[\straintens]$ \cite{nelson_theory_1971,narasimhamurty_photoelastic_2012,wuttke_optically_2013,wuttke_thermal_2013}. A modified permittivity leads to new photonic eigenmodes and electric modal fields $\prtrbd\emode_\photindex[\prtrbd\relpermitttens]$, and therefore to modified electric fields $\prtrbd\Efield_0[\{\prtrbd\emode_\photindex\}]$ surrounding the fiber. In consequence, the optical potential $\pot_\optical^\reddetuned[\prtrbd\Efield_0^\reddetuned] + \pot_\optical^\bluedetuned[\prtrbd\Efield_0^\bluedetuned]$ created by the red- and blue-detuned light field is changed, and the potential $\pot[\ufield,\straintens]$ ultimately depends on strain.

\medskip{}

The photoelastic effect can be quantified by a tensor $\pockelstens$ of fourth rank, called the \emph{photoelastic tensor}, which phenomenologically describes how the optical properties of a material change under strain \cite{nelson_theory_1971}:
\begin{align}
 (\prtrbd\relpermitttens^{-1})\indices{^i^j}[\straintens] &= \spare{(\relpermitttens^{-1})^{ij} + \sum_{kl}\pockelstenscomp^{ijkl}  \straintenscomp^{kl}}~,
\end{align}
where the exponent $({-1})$ indicates the inverse tensor. The photoelastic tensor has symmetries $\pockelstenscomp^{ijkl} = \pockelstenscomp^{jikl} = \pockelstenscomp^{ijlk}$ and therefore possesses at most 36 independent components \cite{nelson_theory_1971}. We use a compact index notation to group the first pair of indices $ij \equiv (N)$ and the second pair of indices $kl \equiv (M)$ according to $11 \equiv (1)$, $22 \equiv (2)$, $33 \equiv (3)$, $12, 21\equiv (4)$, $13, 31 \equiv (5)$, and $23, 32 \equiv(6)$. The independent components can then be arranged in a $6 \times 6$ matrix $(\pockelstens)$, where $N$ corresponds to the row and $M$ to the column number. For materials like silica that exhibit a homogeneous and isotropic photoelastic effect, the components of the photoelasticity tensor $\pockelstens$ in both Cartesian and cylindrical coordinates are \cite{holmes_direct_2009}
\begin{equation}\label{eqn: homogeneous isotropic photoelastic tensor}
  (\pockelstens) =%
    \begin{pmatrix}
      \photela & \photelb & \photelb & 0 & 0 & 0\\
      \photelb & \photela & \photelb & 0 & 0 & 0\\
      \photelb & \photelb & \photela & 0 & 0 & 0\\
      0 & 0 & 0 & \photelc & 0 & 0\\
      0 & 0 & 0 & 0 & \photelc & 0\\
      0 & 0 & 0 & 0 & 0 & \photelc
    \end{pmatrix}~,
\end{equation}
where $\photela, \photelb \in \R$ and $\photelc = (\photela-\photelb)/2$.

We are interested in the strain-induced variation $\Delta \relpermitttens \equiv \prtrbd\relpermitttens[\straintens] - \relpermitttens $ of the permittivity tensor. To linear order in the strain,
\begin{equation}
  \Delta \relpermitttenscomp^{ij} \simeq (\frechetD \prtrbd\relpermitttenscomp^{ij})_{\zerovec}[\straintens] = - \relpermitt^2 \sum_{kl}\pockelstenscomp^{ijkl}  \straintenscomp^{kl}
\end{equation}
for a medium that is isotropic while unperturbed, $\relpermitttens = \relpermitt \id$.

\medskip{}

The new photonic eigenmodes $\prtrbd\amode_\photindex[\prtrbd\relpermitttens]$ in the presence of a modified permittivity $\prtrbd\relpermitttens$ are solutions to the photonic eigenmode equation
\begin{equation}\label{eqn: photon eigenmode equation perturbed}
  \Dphot \prtrbd\amode_\photindex = -\prtrbd\photev_\photindex \, \prtrbd\relpermitttens  \prtrbd\amode_\photindex~;
\end{equation}
compare \cref{eqn: eigenvalue equation}, where $\prtrbd\photev_\photindex = \prtrbd\photfreq^2_\photindex/\cvac^2$ and $\prtrbd\photfreq_\photindex$ are the frequencies of the perturbed eigenmodes. We are interested in the new eigenmodes $\prtrbd\amode_\photindex$ and eigenvalues $\prtrbd\photev_\photindex$ in the presence of a perturbation $\Delta \relpermitttens$ of the permittivity tensor. To this end, we perturbatively expand both eigenmodes and eigenvalues in orders $n$ of $\Delta \relpermitttens$, analogous to time-independent perturbation theory in quantum mechanics \cite{sakurai_modern_2011}:
\begin{align}
  \prtrbd\photev_\photindex &\equiv \photev_\photindex + \Delta\photev_\photindex  &  \Delta\photev_\photindex &\equiv \sum_n \photev_\photindex^{(n)}\\
  \prtrbd\amode_\photindex &\equiv c \amode_\photindex + \Delta\amode_\photindex &  \Delta\amode_\photindex &\equiv \sum_n \amode_\photindex^{(n)}~.
\end{align}
The normalization constant $c \in \C$ is found by normalizing the perturbed eigenmode $\prtrbd\amode_\photindex$. This expansion, in conjunction with \cref{eqn: photon eigenmode equation perturbed} and the orthogonality relation \cref{eqn: normalization photon a modes}, leads to the relations
\begin{equation}\label{eqn: eigenvalue perturbative}
  \Delta\photev_\photindex = - \photev_\photindex \frac{\inner{\amode_\photindex}{\Delta\relpermitttens \, \prtrbd\amode_\photindex}}{\inner{\amode_\photindex}{(\relpermitttens + \Delta\relpermitttens)\prtrbd\amode_\photindex}}
\end{equation}
and
\begin{multline}\label{eqn: eigenvector perturbative}
    \Delta\amode_\photindex = \sum_{\photindexb \neq \photindex} \frac{\relpermitttens^{-1}}{\photev_\photindexb - \photev_\photindex} \Big [ \photev_\photindex \inner{\amode_\photindexb}{\relpermitttens \,  \Delta\relpermitttens \,  \prtrbd\amode_\photindex} \\
    + \Delta \photev_\photindex \inner{\amode_\photindexb}{\relpermitttens( \relpermitttens + \Delta\relpermitttens ) \prtrbd\amode_\photindex} \Big ] \amode_\photindexb
\end{multline}
for the corrections to eigenvalues and eigenmodes. The bracket indicates the $L^2$ scalar product,
\begin{equation}
  \inner{\vec{a}}{\vec{b}} = \int \cconj{\vec{a}} \cdot \vec{b}\dd\pos~.
\end{equation}
\Cref{eqn: eigenvector perturbative} holds provided the perturbed eigenmode $\prtrbd\amode_\photindex$ does not overlap with modes $\amode_\photindexb$ degenerate with the unperturbed mode $\amode_\photindex$, that is, $\photev_\photindexb = \photev_\photindex$ only if $\inner{\amode_\photindexb}{ \relpermitttens \Delta\relpermitttens  \prtrbd\amode_\photindex} = 0 = \inner{\amode_\photindexb}{\relpermitttens( \relpermitttens + \Delta\relpermitttens ) \prtrbd\amode_\photindex}$. The corrections $\Delta\photev_\photindex$ and $\Delta\amode_\photindex$ can be obtained order by order in $\Delta \relpermitttens$. The first-order correction to the eigenvalue is
\begin{equation}\label{eqn: 1st order frequency shift}
 \photev_\photindex^{(1)} = - \photev_\photindex \frac{\inner{\amode_\photindex}{\Delta\relpermitttens \, \amode_\photindex}}{\inner{\amode_\photindex}{\relpermitttens \amode_\photindex}}~,
\end{equation}
which reduces to the known formula for first-order corrections of the eigenfrequency in the case of isotropic permittivities $\relpermitttens$, $\prtrbd\relpermitttens$ \cite{joannopoulos_photonic_2011}. The first-order correction to the eigenmode is
\begin{multline}\label{eqn: 1st order eigenmode correction}
   \amode_\photindex^{(1)} = \sum_{\photindexb \neq \photindex} \frac{\relpermitttens^{-1}}{\photev_\photindexb - \photev_\photindex} \Big [ \photev_\photindex\inner{\amode_\photindexb}{ \relpermitttens \, \Delta\relpermitttens \, \amode_\photindex} \\
   -  \photev_\photindex^{(1)} \inner{\amode_\photindexb}{\relpermitttens^2\amode_\photindex}   \Big ] \amode_\photindexb~,
\end{multline}
where we use that $c = 1 + \order(\Delta\relpermitttens)$ \cite{sakurai_modern_2011}.

In the case study in \cref{sec: case study} of this article, we are concerned with the case $\relpermitttens = \relpermitt \id$. The first-order correction to the eigenvalue then simplifies to
\begin{equation}\label{eqn: 1st order frequency shift main text}
 \photev_\photindex^{(1)} = - \photev_\photindex \frac{ \int \cconj\amode_\photindex \cdot ( \Delta\relpermitttens \, \amode_\photindex ) \dd\pos }{\int \cconj\amode_\photindex \cdot \amode_\photindex \dd \pos}~.
\end{equation}
Moreover, one can show that the first-order shift is zero for perturbations of the permittivity caused by nanofiber phonons of propagation constant $\phonk\neq0$, so
\begin{equation}\label{eqn: 1st order eigenmode correction main text}
    \amode_\photindex^{(1)} = \sum_{\photindexb \neq \photindex} \amode_\photindexb \frac{\photev_\photindex}{\photev_\photindexb - \photev_\photindex} \int \cconj\amode_\photindexb \cdot( \Delta\relpermitttens \, \amode_\photindex) \dd\pos  ~.
\end{equation}
Consider, for example, a mode on the $\HEmode_{11}$ band of a fiber, see \subcref{fig: photon bands}{b}. Fiber phonon modes with azimuthal order $\phonl = 0$ lead to the population of photon modes on the same band, at slightly different propagation constants $\photk$. Phonon modes with azimuthal order $\phonl =\pm1$, on the other hand, can populate modes on the $\TEmode_{01}$, $\TMmode_{01}$, and $\HEmode_{21}$ bands shown \subcrefand{fig: photon bands}{a}{c}. We neglect coupling to radiative modes (leading to phonon-induced transmission losses), since radiative fields are extended, with low amplitudes, and interact only very weakly with the atom.

Having obtained the first-order correction to the photonic eigenmodes, we can now approximate variation $\Delta\emode_\photindex$ of the electric modal field due to the modified permittivity. Using \cref{eqn: electric and magnetic modal fields}, $\Delta\emode_\photindex \simeq \im (\frechetD\prtrbd\amode_\photindex)_{\relpermitttens}[\Delta\relpermitttens]/ \vacpermitt = \im \amode_\photindex^{(1)}/\vacpermitt$ to linear order in the permittivity. The variation of the electric modal field is thus
\begin{equation}\label{eqn: 1st order eigenmode correction simplified}
  \Delta \emode_\photindex \simeq \sum_{\photindexb \neq \photindex} \emode_\photindexb \frac{\photfreq^2_\photindex}{\photfreq^2_\photindexb- \photfreq^2_\photindex}
  \int \amode_\photindexb \cdot (\Delta\relpermitttens \, \amode_\photindex) \dd \pos
\end{equation}
for nanofiber eigenmodes.

\medskip{}

The light coupled into the fiber determines the frequencies at which photonic modes are populated. Since there are no frequency shifts of the eigenmodes at first order in $\Delta \relpermitttens$, we can assume that the amplitude $\photnormvar_{\photindex}$ of each photonic mode remains unchanged, while its spatial form $\prtrbd\emode_{\photindex}$ is periodically modified by the vibrations. The modified complex field profile of a monochromatic light field is therefore
\begin{equation}
  \prtrbd\Efield_0[\{\prtrbd\emode_\photindex\}] \simeq \sum_{\photindex} \photnormvar_{\photindex} \prtrbd\emode_{\photindex}~.
\end{equation}
To linear order in the modal fields, the variation of the field profile $\Delta \Efield_0 \equiv \prtrbd\Efield_0 - \Efield_0$ is
\begin{equation}
    \Delta \Efield_0 \simeq \sum_\photindex (\partialfrechetD_{\prtrbd\emode_{\photindex}} \prtrbd\Efield_{0}) _{\emode_{\photindex}}[\{\Delta\emode_\photindex\}] =  \sum_{\photindex} \photnormvar_{\photindex} \Delta\emode_{\photindex}~.
\end{equation}

\medskip{}

The changed electric fields lead to a changed optical potential $\pot_\optical^\reddetuned[\prtrbd\Efield_0^\reddetuned] + \pot_\optical^\bluedetuned[\prtrbd\Efield_0^\bluedetuned]$. We can now use the chain rule to express the strain coupling term in the interaction Hamiltonian \cref{eqn: atom-phonon interaction Hamiltonian contributions} through the derivative of the optical potential with respect to the electric fields~%
\footnote{%
The chain rule states that the derivative of a composition of functionals $(F \circ G)[\vec{x}] \equiv F[G[\vec{x}]]$, evaluated at $\vec{x} = \vec{a}$ and in direction $\vec{n}$, is the derivative of the outer functional $F$ in direction of the derivative of the inner functional $G$ \cite{werner_funktionalanalysis_2011},
$$ \frechetD (F\circ G)_\vec{a}[\vec{n}] = \frechetD F_{G[\vec{a}]}[\frechetD G_\vec{a}[\vec{n}]]~,$$
analogous to the chain rule for ordinary functions.
}:%
\begin{equation}
  \partialfrechetD_{\straintens}\pot_{(\zerovec,\zerovec)}[\straintens] = (\frechetD\pot_\optical^\reddetuned)_{\Efield_0^\reddetuned}[\Delta \Efield_0^\reddetuned] +
  (\frechetD\pot_\optical^\bluedetuned)_{\Efield_0^\bluedetuned}[\Delta \Efield_0^\bluedetuned]~.
\end{equation}
Each optical potential is the sum of  scalar, vector, and tensor light shift; see \cref{eqn: total light shift}. Thus,
\begin{multline}
  \partialfrechetD_{\straintens}\pot_{(\zerovec,\zerovec)}[\straintens] = \sum_{j}
  (\frechetD\pot_j^\reddetuned)_{\Efield^\reddetuned_0}[\Delta\Efield^\reddetuned_0] \\
  + (\frechetD\pot_j^\bluedetuned)_{\Efield^\bluedetuned_0}[\Delta\Efield^\bluedetuned_0]~.
\end{multline}
where $j \in \{\scalarshift,\vectorshift,\tensorshift\}$ for the scalar, vector, and tensor contributions given in \cref{eqn: scalar light shift,eqn: vector light shift,eqn: tensor light shift}. The functional derivatives of the light shifts reduce to conventional derivatives. For each of the two colors, the derivative of the scalar light shift is
\begin{equation}\label{eqn: variation scalar light shift}
  (\frechetD\pot_\scalarshift)_{\Efield_0}[\Delta\Efield_0] = - \HFSscalarpolarizab  \spare{\cconj\Efield_0(\pos) \cdot \Delta \Efield_0(\pos) + \cc}~,
\end{equation}
the derivative of the vector light shift
\begin{equation}\label{eqn: variation vector light shift}
  (\frechetD\pot_\vectorshift)_{\Efield_0}[\Delta\Efield_0] = -\frac{\HFSvectorpolarizab}{2\im} \frac{\atMF}{\atF} \spare{\cconj\Efield_0(\pos) \times \Delta\Efield_0(\pos) - \cc } \cdot \ezvecB~,
\end{equation}
and the derivative of the tensor light shift
\begin{multline}\label{eqn: variation tensor light shift}
  (\frechetD\pot_\tensorshift)_{\Efield_0}[\Delta\Efield_0] = - 3\HFStensorpolarizab \frac{3\atMF^2 - \atF(\atF+1)}{2\atF(2\atF-1)}  \\
  \spare{ {\cconj\Efieldcomp_0}^{\zposB}(\pos) \Delta \Efieldcomp_0^{\zposB}(\pos) + \cc }~.
\end{multline}

\medskip{}

Finally, the strain coupling functions $\couplingfunct^\strainopt_{\phonindex}(\atpos)$ are obtained by making the dependence on strain explicit in \cref{eqn: variation scalar light shift,eqn: variation vector light shift,eqn: variation tensor light shift} and by expanding the strain operator \cref{eqn: strain operator mode expansion} in terms of the strain modal fields $\strainmode_\phonindex$ of a nanofiber. The strain coupling function then has contributions from the three light shifts $j \in \{\scalarshift,\vectorshift,\tensorshift\}$ for both light colors $\{\reddetuned, \bluedetuned\}$:
\begin{equation}\label{eqn: strain coupling function}
  \couplingfunct^\strainopt_{\phonindex}(\pos) = \sum_{j} [ \couplingfunct^{\reddetuned j}_{\phonindex}(\pos) + \couplingfunct^{\bluedetuned j}_{\phonindex}(\pos)].
\end{equation}
The contributions of the three light shifts to the coupling functions are listed in \cref{tab: strain coupling functions} for a monochromatic light field. Note that the difference between running waves \cref{eqn: phonon cylinder eigenmode partial wave decomposition} and standing waves \cref{eqn: phonon cylinder discrete eigenmode partial wave decomposition} leads to different coupling functions for discrete modes on the $\Tmode_{01}$ band on one hand, and modes on the continuous $\Lmode_{01}$ and $\Fmode_{11}$ bands on the other hand. The corresponding strain coupling constants $\atphong^\strainopt_{\phonindex i}$ are obtained using the definition \cref{eqn: definition coupling constants}.

\begin{table*}
  \setlength{\extrarowheight}{3pt}
  \newcolumntype{A}{>{\begin{math}\displaystyle}r<{\end{math}}}
  \newcolumntype{B}{>{\begin{math}\displaystyle}c<{\end{math}}}
  \newcolumntype{C}{>{\begin{math}\displaystyle}l<{\end{math}}}
  \begin{tabular*}{\textwidth}{ABC}
    \toprule
    \multicolumn{3}{l}{Continuous phonon modes: $\phonfam = \Lmode, \Fmode$}\\
    \couplingfunct^\scalarshift_{\phonindex}(\pos) &=& - \polarizab_\scalarshift  \spare{ \cconj\Efield_0(\pos)  \cdot \Delta \Efield^-_\phonindex(\pos) + \Efield_0(\pos)  \cdot \Delta \Efield^{+\ccsymbol}_\phonindex(\pos)} \\    \couplingfunct^\vectorshift_{\phonindex}(\pos) &=& - \frac{\polarizab_\vectorshift}{2\im} \frac{\atMF}{\atF} \spare{ \cconj\Efield_0(\pos)  \times  \Delta \Efield^-_\phonindex(\pos) -  \Efield_0(\pos)  \times  \Delta \Efield^{+\ccsymbol}_\phonindex(\pos) } \cdot \ezvecB \\
    \couplingfunct^\tensorshift_{\phonindex}(\pos) &=& - 3 \polarizab_\tensorshift \frac{3\atMF^2 - \atF(\atF-1)}{2\atF(2\atF+1)} \spare{ \Efieldcomp^{\ccsymbol\ezvecBcomp}_0(\pos)  \Delta \Efieldcomp^{-\ezvecBcomp}_\phonindex(\pos) + \Efieldcomp^{\ezvecBcomp}_0(\pos)  \Delta \Efieldcomp^{+\ccsymbol\ezvecBcomp}_\phonindex(\pos) }\\
    \Delta\Efield^\pm_\phonindex(\pos) &=& - \sum_{\photindex \,\photfamb\photnb} \photnormvar_\photindex \photonoverlap^\pm_{\photindexb \phonindex \photindex} \left. \emode_\photindexb(\pos) \right\rvert_{\substack{%
         \photlb = \photl \pm \phonl\\\photkb = \photk \pm \phonk}}\\
    \photonoverlap^{-}_{\photindexb\phonindex\photindex} &=& \frac{\photfreq_0^2}{\photfreq^2_\photindexb-\photfreq_0^2} \frac{\ufieldmodedens_\phonindex}{2\pi} \relpermitt^2 \int_0^\rad \spare{ \rpos \cconj\amoder_{\photindexb}(\rpos)  \pockelstens  \strainmoder_\phonindex(\rpos)  \amoder_{\photindex}(\rpos)} \dd\rpos\\
    \photonoverlap^{+}_{\photindexb\phonindex\photindex} &=& \frac{\photfreq_0^2}{\photfreq^2_\photindexb-\photfreq_0^2} \frac{\ufieldmodedens_\phonindex}{2\pi} \relpermitt^2 \int_0^\rad \spare{ \rpos \cconj\amoder_{\photindexb}(\rpos)  \pockelstens  \cconj\strainmoder_\phonindex(\rpos)  \amoder_{\photindex}(\rpos)} \dd\rpos\\
    \midrule
    \multicolumn{3}{l}{Discrete phonon modes: $\phonfam = \Tmode$}\\
    \couplingfunct^\scalarshift_{\phonindex}(\pos) &=& - 2 \polarizab_\scalarshift  \Re \spare{ \cconj\Efield_0(\pos)  \cdot \Delta \Efield_\phonindex(\pos) }  \\
    \couplingfunct^\vectorshift_{\phonindex}(\pos) &=& - \polarizab_\vectorshift \frac{\atMF}{\atF}  \Im \spare{ \cconj\Efield_0(\pos)  \times  \Delta \Efield_\phonindex(\pos)  } \cdot \ezvecB \\
    \couplingfunct^\tensorshift_{\phonindex}(\pos) &=& - 6 \polarizab_\tensorshift \frac{3\atMF^2 - \atF(\atF-1)}{2\atF(2\atF+1)}  \Re\spare{ \Efieldcomp^{\ccsymbol\ezvecBcomp}_0(\pos)  \Delta \Efieldcomp^{\ezvecBcomp}_\phonindex(\pos)} \\
    \Delta \Efield_\phonindex(\pos) &=& \frac{1}{2} \spare{ \Delta \Efield^-_\phonindex(\pos) + \Delta \Efield^+_\phonindex(\pos) }\\
    \Delta\Efield^\pm_\phonindex(\pos) &=& - \sum_{\photindex \, \photfamb\photnb} \photnormvar_\photindex \photonoverlap_{\photindexb \phonindex \photindex} \left. \emode_\photindexb(\pos) \right\rvert_{\substack{%
         \photlb = \photl \pm \phonl\\\photkb = \photk \pm \phonk}}\\
    \photonoverlap_{\photindexb\phonindex\photindex} &=& \frac{\photfreq_0^2}{\photfreq^2_\photindexb-\photfreq_0^2} \frac{\ufieldmodedens_\phonindex}{\sqrt{\pi\len}} \relpermitt^2 \int_0^\rad \spare{ r \cconj\amoder_{\photindexb}(\rpos)  \pockelstens  \strainmoder_\phonindex(\rpos)  \amoder_{\photindex}(\rpos)} \dd\rpos \\
    \midrule
    \cconj\amoder_{\photindexb} \pockelstens  \strainmoder_\phonindex  \amoder_{\photindex} &=&
     \amodercomp_{\photindex}^\rpos \amodercomp_{\photindexb}^{\rpos\ccsymbol} \spare{ \photela \strainmodercomp_\phonindex^{\rpos\rpos} + \photelb \pare{\strainmodercomp_\phonindex^{\phipos\phipos} + \strainmodercomp_\phonindex^{\zpos\zpos}} }
    + \amodercomp_{\photindex}^\phipos \amodercomp_{\photindexb}^{\phipos\ccsymbol} \spare{ \photela \strainmodercomp_\phonindex^{\phipos\phipos} + \photelb \pare{\strainmodercomp_\phonindex^{\rpos\rpos} + \strainmodercomp_\phonindex^{\zpos\zpos}} }
    + \amodercomp_{\photindex}^\zpos \amodercomp_{\photindexb}^{\zpos\ccsymbol} \spare{ \photela \strainmodercomp_\phonindex^{\zpos\zpos} + \photelb \pare{\strainmodercomp_\phonindex^{\rpos\rpos} + \strainmodercomp_\phonindex^{\phipos\phipos}} } \\
    &&+ \pare{\amodercomp_{\photindex}^\rpos \amodercomp_{\photindexb}^{\phipos\ccsymbol} + \amodercomp_{\photindex}^\phipos \amodercomp_{\photindexb}^{\rpos\ccsymbol}} \pare{\photela - \photelb} \strainmodercomp_\phonindex^{\rpos\phipos}
    + \pare{\amodercomp_{\photindex}^\rpos \amodercomp_{\photindexb}^{\zpos\ccsymbol} + \amodercomp_{\photindex}^\zpos \amodercomp_{\photindexb}^{\rpos\ccsymbol}} \pare{\photela - \photelb} \strainmodercomp_\phonindex^{\rpos\zpos}
    + \pare{\amodercomp_{\photindex}^\phipos \amodercomp_{\photindexb}^{\zpos\ccsymbol} + \amodercomp_{\photindex}^\zpos \amodercomp_{\photindexb}^{\phipos\ccsymbol}} \pare{\photela - \photelb} \strainmodercomp_\phonindex^{\phipos\zpos}\\
    \bottomrule
  \end{tabular*}
  \caption{Atom-phonon coupling functions due to strain in a nanofiber-based atom trap. The coupling functions correspond to scalar, vector, and tensor light shift induced by a single monochromatic light field of frequency $\photfreq_0$ and complex field profile $\Efield_0$. All symbols are defined in \cref{sec: photon appendix,sec: phonon appendix}, and \ref{sec: interaction appendix}. In particular, the photon mode indices are $\photindex = (\photl, \photfam, \photk, \photn)$ as defined in \cref{eqn: fiber photon mode indices}, and the phonon mode indices $\phonindex = (\phonl, \phonfam, \phonk, \phonn)$ as defined in \cref{eqn: fiber phonon mode indices}. The index $\photindex$ designates an unperturbed eigenmode, while the primed index $\photindexb$ labels modes perturbatively populated due to strain.}
  \label{tab: strain coupling functions}
\end{table*}

\begin{figure}
  \centering
  \setlength{\widthFigA}{238.0056pt}
  \includegraphics[width = \widthFigA]{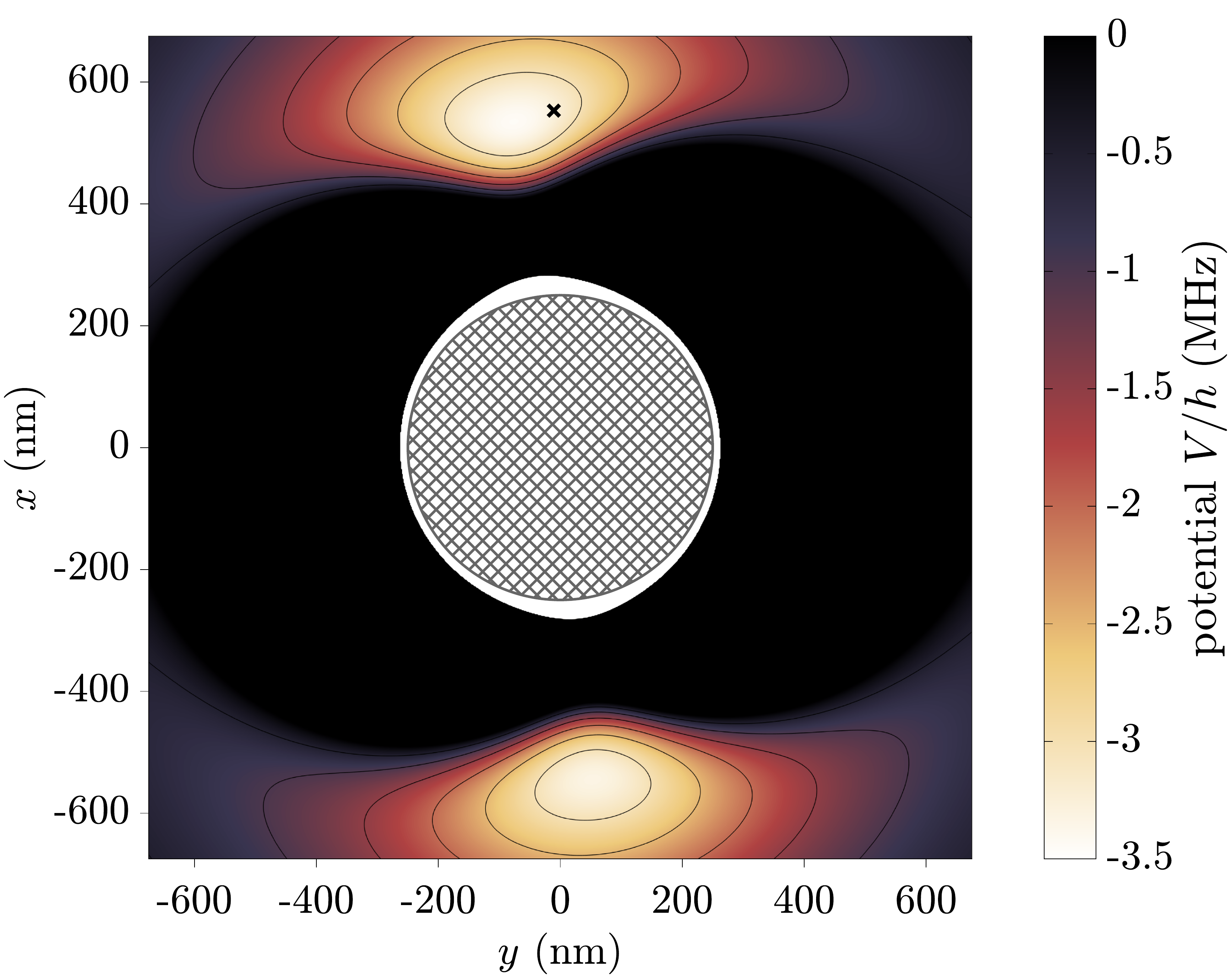}
  \caption{Potential $\pot_0 (\pos) + 2 \phonnormvar_{\phonindex} \Re\couplingfunct^\strainopt_{\phonindex}(\pos)$ perturbed by strain due to the single, coherently excited torsional mode $\phonindex$ of a nanofiber. The potential is evaluated at the trap minimum $\trapz$ close to the center of the nanofiber. The mode $\phonindex$ is the discrete torsional mode closest to resonance with the atom trap in the azimuthal direction; see \cref{sec: case study appendix} for the parameters. The amplitude $\phonnormvar_{\phonindex}$ of the coherent excitation is exaggerated to an unphysical value such that the effect is visible at the given scales. It is apparent that the torsional mode couples to the atomic motion both in the radial and azimuthal direction. However, the contribution to the atom heating rate turns out to be negligible; see \cref{sec: case study}.}
  \label{fig: discrete T modes blue and red scalar coupling r phi}
\end{figure}

To conclude the derivation of the strain coupling, let us illustrate the strain-induced change of the potential due to the photoelastic effect. To first order, the change is given by the strain coupling term $\partialfrechetD_\straintens\pot_{(\zerovec,\zerovec)}[\straintensop](\pos)$. We consider the nanofiber to undergo macroscopic vibrations, described by the multimode coherent state $\ket{\beta}$ with amplitudes $\phonnormvar_\phonindex = |\phonnormvar_\phonindex|e^{\im \phi_\phonindex} \in \C$ for each mode. The expectation value of the change caused by strain is then
\begin{multline}
  \braket{\beta,\tm|\partialfrechetD_\straintens\pot_{(\zerovec,\zerovec)}[\straintensop](\pos)|\beta,\tm} \\
  = 2 \sum_\phonindex |\phonnormvar_\phonindex| \Big[ \cos(\phonfreq_\phonindex \tm - \phi_\phonindex ) \Re\couplingfunct^\strainopt_\phonindex(\pos)\\
  - \sin(\phonfreq_\phonindex \tm - \phi_\phonindex ) \Im\couplingfunct^\strainopt_\phonindex(\pos) \Big]~.
\end{multline}
The coupling function $\couplingfunct^\strainopt_\phonindex(\pos)$ therefore describes the change to the potential $\pot$ due to a phonon mode $\phonindex$. We plot $\pot_0 (\pos) + 2 \phonnormvar_{\phonindex} \Re\couplingfunct^\strainopt_{\phonindex}(\pos)$ in \cref{fig: discrete T modes blue and red scalar coupling r phi} as an example of how strain due to a single torsional mode $\phonindex$ perturbs the potential. The torsional mode qualitatively leads to rotation of the potential around the fiber axis, which results in a coupling between the torsional mode and the atom motion in the azimuthal and radial direction.

\subsection{Atom Heating}
\label{sec: heating}

Let us consider an atom trapped in the optical near field of a nanofiber of temperature $\temp$. We assume that the atom is in the motional pure quantum ground state $\atdensop_0$ of the harmonic trap at time $\tm=0$. At the same time, the phonon field of the fiber is in the thermal quantum state $\phondensop_\thermal \equiv \exp(-\Hamilop_\vibrational/\boltzmann \temp) / \tr[\exp(- \Hamilop_\vibrational/\boltzmann \temp)]$ \cite{gerry_introductory_2005}. Over time, the atom acquires energy by absorbing phonons from the fiber, reflected in the increase of the expected number of motional quanta (population) $n_i(\tm) \equiv \tr [ \densop(\tm) \hconj\ataop_i \ataop_i]$ along the spatial direction $i \in \{\rpos,\phipos,\zpos\}$. Here, $\densop(\tm)$ is the state operator of the coupled atom-phonon system at times $\tm$, and $\tr$ is the trace. The population is initially zero. The evolution of $\densop(\tm)$ from the initial state $\densop_0 = \atdensop_0 \otimes \phondensop_\thermal$ is governed by the full Hamiltonian \cref{eqn: total Hamiltonian}. Provided the atom-phonon coupling is weak, $\atphong_{\phonindex i} \ll \itrapfreq, \phonfreq_\phonindex$, the population grows linearly for sufficiently short times $\tm>0$, with phonon-induced ground-state heating rate $\scatrate_i^\thermal$ in trap direction $i$. As discussed in \cref{sec: framework}, we distinguish the contributions of continuous phonon bands and discrete mechanical resonances to the atom heating rate,
\begin{equation}\label{eqn: heating rate contributions appendix}
 \scatrate^{\thermal}_{i} = \scatrate^{\continuous}_{i} + \scatrate^{\discrete}_{i}~.
\end{equation}

Fermi's golden rule \cref{eqn: Fermis golden rule} can be used to calculate the contribution of the continuous phonon bands. For a nanofiber, there are only two continuous phonon bands resonant with the trapped atom: the longitudinal $\Lmode_{01}$ band and the flexural  $\Fmode_{11}$ band. In consequence,
\begin{equation}\label{eqn: contributions continuous mode heating rate}
  \scatrate_{i}^\continuous = \scatrate_{\Lmode i} + \scatrate_{\Fmode i}~.
\end{equation}
The $\Lmode_{01}$ band has dispersion relation $\phonfreq_\phonindex = \cFL |\phonk|$ in the low-frequency limit, see \cref{eqn: dispersion fundamental longitudinal mode low freq limit}, resulting in a constant density of state $\DOS_{\Lmode} \equiv 1/\cFL$. There are two resonant longitudinal modes $\phonindex_{\zpol}$ at each trap frequency $\itrapfreq$, with propagation constants $\phonk = \zpol \itrapfreq/\cFL$ and $\zpol=\pm$.
The contribution of longitudinal phonon modes to the ground-state heating rate it thus
\begin{equation}\label{eqn: heating rate fundamental longitudinal modes}
  \scatrate_{\Lmode i} = \frac{2\pi \nbar_i}{\cFL} \sum_{\zpol} \left| \atphong_{\phonindex_\zpol i} \right|^2 ~.
\end{equation}
The $\Fmode_{11}$ band has dispersion relation $\phonfreq_\phonindex =  \cFL \rad\phonk^2 /2$ in the low-frequency limit, see \cref{eqn: dispersion fundamental flexural mode low freq limit}, resulting in the density of states $\DOS_{\Fmode i} \equiv 1/\sqrt{2 \itrapfreq \cFL\rad}$. There are four resonant flexural modes propagating in direction $\zpol=\pm$ and of $\phipol=\pm$ circular polarization. The corresponding azimuthal order is $\phonl = \phipol\zpol$, and the propagation constant $\phonk = \zpol \sqrt{2\itrapfreq/\cFL \rad}$. The ground-state heating rate due to the fundamental flexural phonon modes then simplifies to
\begin{equation}\label{eqn: heating rate fundamental flexural modes}
  \scatrate_{\Fmode i} =  \frac{2\pi\nbar_i}{\sqrt{2 \itrapfreq \cFL \rad}} \sum_{\phipol,\zpol} \left| \atphong_{\phonindex_{\phipol\zpol} i}\right|^2~.
\end{equation}
The contribution of each mode $\phonindex_i$ on the continuous phonon bands to the heating rate \cref{eqn: Fermis golden rule} is proportional to the density of states $\DOS_{\phonindex_i}$. Since the $\Fmode_{11}$ band is asymptotically quadratic, see \subcref{fig: phonon bands}{c}, the density of states of the flexural modes diverges as $\phonfreq_i \to 0$. This dependence is reflected by $\scatrate_{\Fmode i} \propto 1/\sqrt{\phonfreq_i}$ in \cref{eqn: heating rate fundamental flexural modes}. On the other hand, the density of states of the longitudinal modes is constant because the $\Lmode_{01}$ has a linear asymptote in the low-frequency limit, see \subcref{fig: phonon bands}{b}. The effect of flexural modes is therefore enhanced in comparison with longitudinal modes for atom-trap frequencies $\itrapfreq$ that are small compared to the frequency scale of the phonon bands. Moreover, the contribution of strain coupling is negligible for flexural modes. Using the displacement coupling constants \cref{eqn: flexural mode displacement coupling constant} in \cref{eqn: heating rate fundamental flexural modes} yields the formula \cref{eqn: final heating formula} for the atom heating rate due to flexural modes, which is sufficient to explain heating rates observed in experiments; see \cref{sec: case study}.

The discrete torsional modes are not reflected perfectly at the end of the nanofiber and therefore have a finite lifetime corresponding to decay rates $\phondecayrate_\phonindex$. This behavior can be modeled by including dissipation in the dynamics of the phonon field. The evolution of the density matrix $\densop$ of the coupled atom-phonon system is then governed by the master equation \cite{breuer_theory_2002}
\begin{multline}
  \fd{}{\tm}\densop(\tm) = \frac{1}{\im\hbar} \com{\Hamilop}{\densop} + \sum_\phonindex \phondecayrate_\phonindex ( \nbar_\phonindex + 1) \dissipator_{\phonaop_\phonindex}(\densop) \\
  + \sum_\phonindex \phondecayrate_\phonindex  \nbar_\phonindex  \dissipator_{\hconj\phonaop_\phonindex}(\densop)
\end{multline}
with dissipator
\begin{equation}
  \dissipator_{\aop}(\densop) \equiv \aop \densop \hconj\aop - \frac{1}{2} \acom{\hconj\aop\aop}{\densop}~.
\end{equation}
Here, the sum runs over all discrete phonon modes $\phonindex$, $\com{\cdot\,}{\cdot}$ indicates the commutator, and $\acom{\cdot\,}{\cdot}$ the anticommutator.
This model captures the essential features of the discrete phonon modes: The steady state of the phonon modes in the absence of atom-phonon interaction is a thermal state $\phondensop_\thermal$ with thermal phonon occupation $\nbar_\phonindex$ for each mode \cite{breuer_theory_2002}. Furthermore, if a phonon mode initially has occupation $n_\phonindex^0$, it decays with rate $\phondecayrate_\phonindex$ back to the thermal phonon occupation, $n_\phonindex(\tm) = \nbar_\phonindex + (n_\phonindex^0 - \nbar_\phonindex) e^{-\phondecayrate_\phonindex \tm}$. We are interested in the effective dynamics of the atom density operator $\atdensop$ under the assumption that the phonons remain in a thermal state. The total density operator is then $\densop(\tm) \simeq \atdensop(\tm) \otimes \phondensop_\thermal$. Adiabatic elimination of the phonon degrees of freedom in the limit of weak coupling compared to the phonon decay rates and atom and phonon frequencies, $\phondecayrate_\phonindex, \phonfreq_\phonindex, \itrapfreq \gg \atphong_{\phonindex i}$, yields the master equation
\begin{multline}\label{eqn: torsional master equation}
  \fd{}{\tm}\atdensop(\tm) =  \frac{1}{\im\hbar} \com{\Hamilop_\atomic}{\atdensop} + \sum_i \scatrate_i^- \dissipator_{\ataop_i}(\atdensop) \\
  + \sum_i \scatrate_i^+ \dissipator_{\hconj\ataop_i}(\atdensop)
\end{multline}
for the motion of the atom \cite{cirac_laser_1992,wilson-rae_cavity-assisted_2008}. Here, $\Hamilop_\atomic$ contains a small Lamb shift of the trap frequencies that is not relevant to our discussion. The phonon-induced decay ($-$) and heating ($+$) rates are
\begin{equation}\label{eqn: torsional heating general}
  \begin{split}
    \scatrate_i^\pm &\equiv 2 \sum_\phonindex |\atphong_{\phonindex i}|^2 \spare{ \nbar_\phonindex G^\mp_{\phonindex i} + (\nbar_\phonindex + 1) G^\pm_{\phonindex i} }~,\\
    G^\pm_{\phonindex i} &\equiv \frac{2\phondecayrate_\phonindex}{\phondecayrate_\phonindex^2 + 4(\itrapfreq\pm\phonfreq_\phonindex)^2}~.
  \end{split}
\end{equation}
If the atom is initially in the motional ground state and there is no cross-coupling between the motional directions ($\ijcrosscoupling=0$), the population of its motion in direction $i$ evolves as $n_i(\tm) = n_i^\infty (1- e^{-\scatrate \tm})$, with $\scatrate \equiv \scatrate_i^- - \scatrate_i^+$
and $n_i^\infty \equiv \scatrate_i^+/\scatrate$. At times $\tm \ll \scatrate$, the population grows linearly, with ground-state heating rate $\scatrate^\discrete_i = \scatrate_i^+$. The total phonon-induced heating rate $\scatrate_i^\thermal$ of the atomic motion along direction $i$ is then obtained according to \cref{eqn: heating rate contributions} by summing $\scatrate_{i}^\discrete$ with the contribution $\scatrate_{i}^\continuous$ of the continuous modes in \cref{eqn: contributions continuous mode heating rate}.

\medskip{}

One approach to reducing the atom heating rate is to optimize the nanofiber such that flexural modes are also reflected at the its ends; see \cref{sec: case study}. The flexural eigenmodes then become standing waves \cref{eqn: phonon cylinder discrete eigenmode partial wave decomposition} with frequency spectrum $\phonfreq_m$, $m \in \N$, given in \cref{eqn: flexural resonator spectrum}, and decay rates $\phondecayrate_m$. If the spacing between resonator frequencies is sufficiently large, the trap frequencies $\itrapfreq$ can be detuned from resonance with the flexural modes $|\itrapfreq-\phonfreq_m| \gg \phondecayrate_m$. The spacing of phonon frequencies close to the trap frequency is approximately $2 \sqrt{\itrapfreq\phonfreq_1} + \phonfreq_1$, where $\phonfreq_1$ is the fundamental frequency of the resonator defined in \cref{eqn: flexural resonator spectrum}; the shorter the resonator and the larger the fiber radius, the easier it is to detune the trap from resonance. Provided the coupling rates $\atphong_{mi}$ between phonon mode $m$ and atomic motion in direction $i \in \{\rpos,\phipos\}$ are smaller than phonon decay rates $\phondecayrate_m$ and atom and phonon frequencies, $\atphong_{mi} \ll \phondecayrate_m, \phonfreq_m, \itrapfreq$, the effective dynamics of the atom is described by a master equation of the form \cref{eqn: torsional master equation}. The heating rate in the radial and azimuthal direction due to the flexural resonator modes of an atom at position $\trapz$ is then
\begin{equation}\label{eqn: general heating discrete flexural modes}
  \scatrate_i^\thermal \simeq 4 \sum_{m\in \N} |\atphong_{mi}(\trapz)|^2 \spare{\nbar_m G^-_{mi} + (\nbar_m +1)G^+_{mi})}~,
\end{equation}
with a position-dependent displacement coupling constant
\begin{equation}\label{eqn: flexural resonator coupling constants}
  \atphong_{mi}(z) = -\frac{\itrapfreq}{2\rad} \sqrt{\frac{\atmass}{\pi \clamplen \dens}\frac{\itrapfreq}{\phonfreq_m}} \sin(\phonk_m \zpos)
\end{equation}
and $G^\pm_{mi}$ as defined in \cref{eqn: torsional heating general}. The atom heating rate due to flexural resonances \cref{eqn: general heating discrete flexural modes} can be explicitly evaluated in different limiting cases, yielding the heating rates \cref{eqn: heating off-resonant 1,eqn: heating off-resonant 2,eqn: resonant heating} that we discuss in \cref{sec: case study}.

\section{Case Study Parameters}
\label{sec: case study appendix}
In nanofiber-based two-color atom traps, different combinations of linearly and circularly polarized trapping light fields are commonly used, both as propagating or standing waves \cite{le_kien_state-dependent_2013}. In \cref{sec: monochromatic fields}, we summarize the corresponding shapes of the electric field required for the atom heating case study in \cref{sec: case study}. In \cref{sec: parameters}, we provide a listing of the parameters used in the case study based on ref.~\cite{albrecht_fictitious_2016}.

\subsection{Monochromatic Guided Fields}
\label{sec: monochromatic fields}

\begin{table*}
  \setlength{\extrarowheight}{3pt}
  \newcolumntype{A}{>{\begin{math}}r<{\end{math}}}
  \newcolumntype{B}{>{\begin{math}}c<{\end{math}}}
  \newcolumntype{C}{>{\begin{math}}l<{\end{math}}}
  \begin{tabular*}{\textwidth}[t]{l @{\quad} ABC @{\qquad} ABC }
    \toprule
    Case & \multicolumn{3}{l}{Field profile} & \multicolumn{3}{l}{Field}\\
    \midrule
    \multicolumn{7}{l}{(1) circular polarized running wave}\\&
    \Efieldcomp^\rpos_0 &=& \zpol \photnormvar \emodercomp^\rpos_{\photl}(\photk;\rpos) e^{\im (\photl_\polindex\phipos + \photk_\polindex\zpos + \theta)} &
    \Efieldcomp^\rpos &=& -\zpol 2\photnormvar \Im\spare{\emodercomp^\rpos_{\photl}(\photk;\rpos)} \sin\pare{ \photl_\polindex \phipos + \photk_\polindex \zpos - \photfreq\tm + \theta } \\&
    \Efieldcomp^\phipos &=& \phipol \photnormvar \emodercomp^\phipos_{\photl}(\photk;\rpos) e^{\im (\photl_\polindex\phipos + \photk_\polindex\zpos + \theta)} &
    \Efieldcomp^\phipos &=& \phipol 2\photnormvar \emodercomp^\phipos_{\photl}(\photk;\rpos) \cos\pare{ \photl_\polindex \phipos + \photk_\polindex \zpos - \photfreq\tm + \theta } \\&
    \Efieldcomp^\zpos &=& \photnormvar \emodercomp^\zpos_{\photl}(\photk;\rpos) e^{\im (\photl_\polindex\phipos + \photk_\polindex\zpos + \theta)} &
    \Efieldcomp^\zpos &=& 2\photnormvar \emodercomp^\zpos_{\photl}(\photk;\rpos) \cos\pare{ \photl_\polindex \phipos + \photk_\polindex \zpos - \photfreq\tm + \theta } \\
    \midrule
    \multicolumn{7}{l}{(2)  non-rotating running wave}\\&
    \Efieldcomp^\rpos_0 &=& \zpol 2 \photnormvar \emodercomp^\rpos_{\photl}(\photk;\rpos) \cos\pare{\photl \phipos + \phiphase} e^{\im (\photk_\polindex \zpos+ \theta)} &
    \Efieldcomp^\rpos &=& - \zpol 4 \photnormvar \Im\spare{\emodercomp^\rpos_{\photl}(\photk;\rpos)} \cos\pare{\photl \phipos + \phiphase} \sin\pare{\photk_\polindex \zpos - \photfreq\tm + \theta } \\&
    \Efieldcomp^\phipos_0 &=& \zpol 2 \im \photnormvar \emodercomp^\phipos_{\photl}(\photk;\rpos) \sin\pare{\photl \phipos + \phiphase} e^{\im (\photk_\polindex \zpos+ \theta)} &
    \Efieldcomp^\phipos &=& - \zpol 4  \photnormvar \emodercomp^\phipos_{\photl}(\photk;\rpos) \sin\pare{\photl \phipos + \phiphase} \sin\pare{\photk_\polindex \zpos - \photfreq\tm + \theta} \\&
    \Efieldcomp^\zpos_0 &=& 2\photnormvar \emodercomp^\zpos_{\photl}(\photk;\rpos) \cos\pare{\photl \phipos + \phiphase} e^{\im (\photk_\polindex \zpos+ \theta)} &
    \Efieldcomp^\zpos &=& 4 \photnormvar \emodercomp^\zpos_{\photl}(\photk;\rpos) \cos\pare{\photl \phipos + \phiphase} \cos\pare{\photk_\polindex \zpos - \photfreq\tm + \theta} \\
    \midrule
    \multicolumn{7}{l}{(3) non-rotating standing wave}\\&
    \Efieldcomp^\rpos_0 &=& 4\im \photnormvar \emodercomp^\rpos_{\photl}(\photk;\rpos) \cos\pare{\photl \phipos + \phiphase} \sin\pare{\photk \zpos + \zphase} e^{\im  \theta } &
    \Efieldcomp^\rpos &=& -8 \photnormvar \Im\spare{\emodercomp^\rpos_{\photl}(\photk;\rpos)} \cos\pare{\photl \phipos + \phiphase} \sin\pare{\photk \zpos + \zphase} \cos\pare{\photfreq\tm + \theta } \\&
    \Efieldcomp^\phipos_0 &=& - 4\photnormvar \emodercomp^\phipos_{\photl}(\photk;\rpos) \sin\pare{\photl \phipos + \phiphase} \sin\pare{\photk \zpos + \zphase} e^{\im  \theta } &
    \Efieldcomp^\phipos &=& - 8 \photnormvar \emodercomp^\phipos_{\photl}(\photk;\rpos) \sin\pare{\photl \phipos + \phiphase} \sin\pare{\photk \zpos + \zphase}\cos\pare{\photfreq\tm + \theta } \\&
    \Efieldcomp^\zpos_0 &=& 4\photnormvar \emodercomp^\zpos_{\photl}(\photk;\rpos) \cos\pare{\photl \phipos + \phiphase} \cos\pare{\photk \zpos + \zphase} e^{\im  \theta } &
    \Efieldcomp^\zpos &=& 8 \photnormvar \emodercomp^\zpos_{\photl}(\photk;\rpos) \cos\pare{\photl \phipos + \phiphase} \cos\pare{\photk \zpos + \zphase}\cos\pare{\photfreq\tm + \theta } \\
    \bottomrule
  \end{tabular*}
  \caption{Electric fields of monochromatic waves of frequency $\photfreq$ on the $\HEmode_{|\photl|\photn}$ and $\EHmode_{|\photl|\photn}$ bands. The sign $\zpol$ indicates the propagation direction along the fiber axis, the sign $\phipol$ the rotation direction around the axis. The propagation constant is $\photk_\polindex = \zpol \photk$ with $\photk>0$, the azimuthal order $\photl_\polindex = \phipol \zpol \photl$ with $\photl>0$, and the amplitude $\photnormvar \in \R$ is determined by the transmitted power. The quantities $\theta, \phiphase, \zphase \in \R$ are phases explained in the text. The radial partial waves $\emodercomp^i_{\photl}(\photk;\rpos)$ are listed in \cref{tab: guided modes modal fields}.}
  \label{tab: monochromatic E fields}
\end{table*}

\begin{table*}
    \newcolumntype{L}{>{\raggedright\arraybackslash}X}
    \begin{tabularx}{\textwidth}{r@{${}={}$}l l c r X r@{${}={}$}l l c r}
      \toprule
      \multicolumn{2}{l}{Parameter} & Description & \multicolumn{2}{l}{Source} && \multicolumn{2}{l}{Parameter} & Description & \multicolumn{2}{l}{Source} \\
      \midrule
      \multicolumn{2}{l}{mechanical}\\
      $\rad$ & $\SI{250}{\nano\meter}$ & fiber radius & \cite{albrecht_fictitious_2016} &&&
        $\lamemu$ & $\SI{31.2}{\giga\pascal}$ & 2nd Lamé coeff. & (\ref{eqn: Lame coefficients calculated}) & $\star$ \\
      $\YoungE$ & $\SI{72.6}{\giga\pascal}$ & Young's modulus & \cite{bass_handbook_2001} &&&
        $\clong$ & $\SI{5.94E3}{\meter/\second}$ & long. sound speed &(\ref{eqn: definition sound speeds}) & $\star$ \\
      $\Poissonnu$ & $0.164$ & Poisson's ratio & \cite{bass_handbook_2001}  &&&
        $\ctrans$ & $\SI{3.76E3}{\meter/\second}$ & trans. sound speed & (\ref{eqn: definition sound speeds}) & $\star$ \\
      $\dens$ & $\SI{2.20}{\gram/\centi\meter^3}$ & mass density & \cite{bass_handbook_2001} &&&
        $\cFL$ & $\SI{5.74E3}{\meter/\second}$ & eff. sound speed & (\ref{eqn: effective fundamental longitudinal speed of sound}) & $\star$  \\
      $\lamelambda$ & $\SI{15.2}{\giga\pascal}$ & 1st Lamé coefficient & (\ref{eqn: Lame coefficients calculated}) & $\star$ \\
      \midrule
      $\phonfreq_\Tmode/2\pi$ & $\SI{258}{\kilo\hertz}$ & fundamental frequency &&&&
      $\kappa/2\pi$ & $\SI{48.0}{\hertz}$ & decay rate & \\
      $Q$ & $\num{5380}$ & quality factor &&  $\star$ &&
      $\len$ & $\SI{7.29}{\milli\meter}$ & eff.\ nanofiber length & (\ref{eqn: effective taper length}) & $\star$ \\
      \midrule
      \multicolumn{2}{l}{optical}\\
      $\relpermitt$ & $\num{2.1}$ & rel. permittivity & \cite{vetsch_optical_2010,bass_handbook_2001} \\
      $\wavelen_\reddetuned$ & $\SI{1064}{\nano\meter}$ & free-space w.-length & \cite{albrecht_fictitious_2016} &&&
        $\wavelen_\bluedetuned$ & $\SI{783}{\nano\meter}$ & free-space w.-length & \cite{albrecht_fictitious_2016} \\
      $\photfreq_\reddetuned/2\pi$ & $\SI{282}{\tera\hertz}$ & ang. frequency && $\star$ &&
        $\freq_\bluedetuned/2\pi$ & $\SI{383}{\tera\hertz}$ & ang. frequency && $\star$ \\
      $|\photk_\reddetuned|$ & $\SI{6.31}{\micro\meter^{-1}}$ & propagation const. & (\ref{eqn: guided mode frequency equation}) & $\star$ &&
        $|\photk_\bluedetuned|$ & $\SI{9.41}{\micro\meter^{-1}}$ & propagation const. & (\ref{eqn: guided mode frequency equation}) & $\star$ \\
      $|\power_\reddetuned|$ & $\SI{1.25}{\milli\watt}$ & power & \cite{albrecht_fictitious_2016} &&&
        $|\power_\bluedetuned|$ & $\SI{17.8}{\milli\watt}$ & power & \cite{albrecht_fictitious_2016} \\
      $\photnormvar_\reddetuned$ & $\SI{1.41}{\pico\ampere.\second/\meter}$ & amplitude & (\ref{eqn: power general expression}) & $\star$ &&
        $\photnormvar_\bluedetuned$ & $\SI{5.70}{\pico\ampere.\second/\meter}$ & amplitude & (\ref{eqn: power general expression}) & $\star$ \\
      $\phiphase^\reddetuned$ & $0$ & $\phipos$ phase shift &&&&
        $\phiphase^\bluedetuned$ & $\pi/2$ & $\phipos$ phase shift && $\star$ \\
      $\zphase^\reddetuned$ & $\pi/2 - \photk_\reddetuned \len/2$ & $\zpos$ phase shift \\
      $\photela$ & $\num{0.100}$ & photoelasticity & \cite{vedam_elastic_1950,holmes_direct_2009} &&&
      $\photelb$ & $\num{0.285}$ & photoelasticity & \cite{vedam_elastic_1950,holmes_direct_2009} \\
      \midrule
      \multicolumn{2}{l}{atomic}\\
      $\atmass$ & $\SI{2.21e-25}{\kilo\gram}$ & mass & \cite{meija_atomic_2016} & $\star$ \\
      $\atF$ & $4$ & HFS state & \cite{albrecht_fictitious_2016} &&&
      $\atMF$ & $-4$ & Zeeman substate & \cite{albrecht_fictitious_2016} \\
      \midrule
      $\FSscalarpolarizab^\reddetuned$ & \SI{1164}{au} & FS scalar pol. at $\freq_\reddetuned$ & \cite{le_kien_dynamical_2013} & $\star$ &&
        $\FSscalarpolarizab^\bluedetuned$ & \SI{-1761.6}{au} & FS scalar pol. at $\freq_\bluedetuned$& \cite{le_kien_dynamical_2013} & $\star$\\
      $\FSvectorpolarizab^\reddetuned$ & \SI{-198.64}{au} & FS vector pol. at $\freq_\reddetuned$& \cite{le_kien_dynamical_2013} & $\star$ &&
        $\FSvectorpolarizab^\bluedetuned$ & \SI{-479.96}{au} & FS vector pol. at $\freq_\bluedetuned$ & \cite{le_kien_dynamical_2013} & $\star$\\
      $\FStensorpolarizab^\reddetuned$ & 0 & FS tensor pol. at $\freq_\reddetuned$& \cite{le_kien_dynamical_2013}& $\star$ &&
        $\FStensorpolarizab^\bluedetuned$ & 0 & FS tensor pol. at $\freq_\bluedetuned$& \cite{le_kien_dynamical_2013} & $\star$\\
      \midrule
      $\CPpotstrength/h$ & $\SI{1.178}{\tera\hertz.\nano\meter^3}$ & strength disp. force & \cite{stern_simulations_2011} \\
      \midrule
      \multicolumn{2}{l}{trap}\\
      $\trapr$ & $\SI{553}{\nano\meter}$ & trap position & & $\star$ &&
        $\trapz$ & $\len/2$ & & & $\star$ \\
      $\trapphi$ & $\SI{-0.0190}{\radian}$ & & & $\star$ &&
        $\pot_0/h$ & $\SI{-3.21}{\mega\hertz}$ & trap depth & & $\star$ \\
      \midrule
      $\rtrapfreq/2\pi$ & $\SI{123}{\kilo\hertz}$ & trap frequency & (\ref{eqn: definition 3d trap frequencies}) & $\star$ &&
        $\atrpossd$ & $\SI{17.6}{\nano\meter}$ & zero-point motion & (\ref{eqn: zero-point fluctuations atom position 3d trap}) & $\star$ \\
      $\phitrapfreq/2\pi$ & $\SI{71.8}{\kilo\hertz}$ & & (\ref{eqn: definition 3d trap frequencies}) & $\star$ &&
        $\trapr\atphipossd$ & $\SI{23.0}{\nano\meter}$ &  & (\ref{eqn: zero-point fluctuations atom position 3d trap}) & $\star$  \\
      $\ztrapfreq/2\pi$ & $\SI{193}{\kilo\hertz}$ &  & (\ref{eqn: definition 3d trap frequencies}) & $\star$ &&
        $\atzpossd$ & $\SI{14.0}{\nano\meter}$ & & (\ref{eqn: zero-point fluctuations atom position 3d trap}) & $\star$ \\
      \bottomrule
  \end{tabularx}
  \caption{Parameters for the case study in \cref{sec: case study}. A star ($\star$) indicates that a parameter depends on previously chosen parameters.}
  \label{tab: parameters}
\end{table*}

For each frequency $\photfreq$ on the $\HEmode_{|\photl|\photn}$ and $\EHmode_{|\photl|\phonn}$ bands, there are four degenerate eigenmodes propagating in positive ($\zpol = 1$) or negative ($\zpol = -1$) direction along the $\zpos$ axis and rotating with positive ($\phipol=1$) or negative ($\phipol=-1$) orientation around the fiber axis. The corresponding propagation constant is $\photk_\polindex = \zpol \photk$ with $\photk > 0$, and the azimuthal order $\photl_\polindex = \zpol \phipol \photl$ with $\photl>0$. The multi-index $\polindex = (\phipol,\zpol)$ is used to distinguish the propagation and polarization state. Superposition of these four modes yields a monochromatic electromagnetic field
\begin{equation}\label{eqn: monochromatic light field 2}
  \begin{split}
    \Efield(\post) &= \Efield_0(\pos) e^{-\im\photfreq\tm} + \cc\\
    \Bfield(\post) &= \Bfield_0(\pos) e^{-\im\photfreq\tm} + \cc
  \end{split}
\end{equation}
with complex field profiles
\begin{equation}
  \begin{split}
      \Efield_0(\pos) &= \sum_\polindex \photnormvar_\polindex \emode_{\photl_\polindex}(\photk_\polindex;\pos)\\
      \Bfield_0(\pos) &= \sum_\polindex \photnormvar_\polindex \bmode_{\photl_\polindex}(\photk_\polindex;\pos)
  \end{split}
\end{equation}
Here, $\photnormvar_\polindex \in \C$ are amplitudes, and $\emode_{\photl_\polindex}(\photk_\polindex;\pos)$, $\bmode_{\photl_\polindex}(\photk_\polindex;\pos)$ are modal fields \cref{eqn: vector wave equation solutions} with radial partial waves listed in \cref{tab: guided modes modal fields}. We drop all irrelevant mode indices, keeping $\photl_\polindex$ and $\photk_\polindex$. The overall magnitude of the amplitudes is related to the power transmitted along the fiber axis $\ezvec$,
\begin{equation}\label{eqn: power general expression}
    \power = \int_0^{2\pi}  \int_0^\infty \intensity(\rpos,\phipos) \dd\rpos\dd \phipos~.
\end{equation}
Here, $\intensity = \braket{\poynting}_\tm\cdot \ezvec$ is light intensity in direction $\ezvec$, and $\braket{\poynting}_\tm = 2 \Re\spare{ \Efield_0(\pos) \times \cconj\Bfield_0(\pos) } /\vacpermeab$ is the Poynting vector averaged over an oscillation period. The star indicates the complex conjugate, and $\vacpermeab$ is the vacuum permeability.

A light field rotating with orientation $\phipol$ around the fiber axis and propagating in direction $\zpol \ezvec$ is realized by the amplitudes
\begin{equation}\label{eqn: amplitudes rotating propagating waves}
  \photnormvar_\polindexb = (\phipol \zpol)^{\photl}\, 2\pi  \photnormvar e^{\im\theta} \, \kronecker_{\phipol \phipolb} \kronecker_{\zpol \zpolb}~,
\end{equation}
where $\photnormvar \in \R$, and $\theta \in R$ is the overall phase of the wave. We include a factor of $2\pi$ and the sign $(\phipol \zpol)^{\photl}$ for later convenience. The field profile and resulting electric field are given by \mbox{case (1)} in \cref{tab: monochromatic E fields}. The power transmitted along the fiber axis can be expressed as
\begin{multline}\label{eqn: power circular polarized propagating wave}
  \power = - \zpol \frac{4\pi\photnormvar^2}{\vacpermeab}
    \int_0^\infty \rpos \big[ \emodercomp^\rpos_{\photl}(\photk;\rpos) \bmodercomp^\phipos_{\photl}(\photk;\rpos) \\
    + \emodercomp^\phipos_{\photl}(\photk;\rpos) \bmodercomp^\rpos_{\photl}(\photk;\rpos) \big] \dd\rpos
\end{multline}
using the symmetries \cref{eqn: HE EH modes symmetries}.

A nonrotating light field, propagating in direction $\zpol$ corresponds to the choice of amplitudes
\begin{equation}\label{eqn: amplitudes nonrotating propagating waves}
  \photnormvar_\polindexb = (\phipolb \zpol)^{\photl}\, 2\pi\photnormvar e^{\im (\phipolb \zpol \phiphase + \theta)} \,\kronecker_{\zpol \zpolb} ~,
\end{equation}
and the resulting electric field is given by \mbox{case (2)} in \cref{tab: monochromatic E fields}. The phase $\phiphase \in \R$ determines the orientation of the wave in the $(\xpos,\ypos)$ plane. For azimuthal order $|\photl|=1$ in particular, the electric field is mainly oriented along an axis in $(\xpos,\ypos)$ plane that encloses the angle $\phiphase$ with the $\xpos$ axis. These waves are therefore called \emph{quasilinear polarized}. The transmitted power is $2\power$ as given in \cref{eqn: power circular polarized propagating wave}.

Two counterpropagating quasilinear waves create a standing wave, corresponding to the amplitudes
\begin{equation}\label{eqn: amplitudes nonrotating standing waves}
  \photnormvar_\polindex = (\phipol \zpol)^{\photl}\, 2\pi \photnormvar e^{\im (\phipol \zpol \phiphase + \zpol \zphase + \theta)}~.
\end{equation}
The phase $\zphase \in \R$ determines the position of nodes of the standing wave along the fiber axis. The electric field is given by \mbox{case (3)} in \cref{tab: monochromatic E fields}. The power transmitted along the fiber axis vanishes, but each counterpropagating wave has again power $2\power$ as given in \cref{eqn: power circular polarized propagating wave}.

\subsection{Physical Parameters}
\label{sec: parameters}

In \cref{tab: parameters} we list the parameters used in the case study \cref{sec: case study} based on the setup described in \cite{albrecht_fictitious_2016}. Citations are given in square brackets and references to equations used to calculate dependent parameters in parentheses. A star indicates that a parameter depends on previously chosen parameters.

The mechanical properties of the silica fiber are determined by the choice of material. The frequencies of the discrete torsional modes confined to the nanofiber are determined experimentally, as we describe in \cref{sec: case study}. The (effective) length $\len$ of the nanofiber for torsional modes is inferred from the measured frequency $\phonfreq_\Tmode$ of the fundamental torsional mode using \cref{eqn: dispersion fundamental torsional mode}:
\begin{equation}\label{eqn: effective taper length}
  \len = \ctrans \pi / \phonfreq_\Tmode~.
\end{equation}

The optical properties of the fiber are determined by permittivity and radius. The power of the red-detuned field quoted below corresponds to each beam separately. The coordinate system is chosen such that the red-detuned laser beam is polarized along the $\xpos$ axis. This is reflected in the choice of azimuthal phase shifts $\phiphase$. Moreover, the axial phase shift $\zphase$ is chosen such that there is a trapping site at $\zpos = \len/2$ in the middle of the fiber. The magnetic offset field $\extBfield$ is oriented perpendicular to the fiber axis, along
\begin{equation}
 \ezvecB = \cos(\phi) \exvec + \sin(\phi) \eyvec
\end{equation}
with $\phi = \SI{66}{\degree}$. The potential experienced by the atom depends on its mass and the polarizability of its internal hyperfine-structure state; see \cref{sec: atom appendix}. The values in \cref{tab: parameters} correspond to a \ce{^{133}_{55}Cs} atom in the ground state \ce{6^2S_{1/2}} interacting with the red-detuned (\reddetuned) and blue-detuned (\bluedetuned) light field. The atomic unit of polarizability is $\SI{1}{au} = (4\pi\vacpermitt)^4 \hbar^6  / (\mel^3 \elcharge^6) =  \SI{1.65E-41}{\ampere^2 . \second^4 /\kilogram}$. The resulting atom trapping site $\trappos$ indicated in \cref{fig: trap} is slightly shifted away from the $\xpos$ axis by vector light shifts due to the orientation of the magnetic offset field.

\end{document}


\selectlanguage{english}

\title{Supplemental Material for `Heating in Nanophotonic Traps for Cold Atoms'}

\author{Daniel Hümmer}
\affiliation{Institute for Quantum Optics and Quantum Information of the Austrian Academy of Sciences, 6020 Innsbruck, Austria}
\affiliation{Institute for Theoretical Physics, University of Innsbruck, 6020 Innsbruck, Austria}
\author{Philipp Schneeweiss}
\affiliation{Atominstitut, TU Wien, 1020 Vienna, Austria}
\author{Arno Rauschenbeutel}
\affiliation{Atominstitut, TU Wien, 1020 Vienna, Austria}
\affiliation{Department of Physics, Humboldt-Universität zu Berlin, 10099 Berlin,  Germany}
\author{Oriol Romero-Isart}
\affiliation{Institute for Quantum Optics and Quantum Information of the Austrian Academy of Sciences, 6020 Innsbruck, Austria}
\affiliation{Institute for Theoretical Physics, University of Innsbruck, 6020 Innsbruck, Austria}
\date{\today}

\maketitle

We discuss additional heating mechanisms for the motion of atoms in nanofiber-based optical dipole traps. In particular, we provide estimates for the setup described in \cite{albrecht_fictitious_2016} and \cite{meng_near-ground-state_2018}, which is the setup the case study in our manuscript refers to.

\section*{Power fluctuations of the trapping laser fields}
The two-color optical dipole trap used in \cite{albrecht_fictitious_2016} is formed by a blue-detuned running-wave light field and a red-detuned field which is in a standing-wave configuration. Beyond polarization fluctuations, which are already treated in the main text of our manuscript, only power fluctuations are relevant for the blue-detuned field and would lead to resonant heating (via a shift of the trap center) of the trap's radial DOF, and to parametric heating (via a change of the trap frequency) of the azimuthal DOF, while no coupling occurs for the axial DOF. For the red-detuned light field, power fluctuations would lead to resonant heating along the radial direction, as well as parametric heating along azimuthal and axial directions. We characterized the power fluctuations of the trap lasers used in experiment~\cite{albrecht_fictitious_2016}. The measured power noise characteristics cannot explain the large heating rates we observe. We estimate that heating due this technical noise amounts to a heating rate of about~$\SI{0.02}{quanta/\milli\second}$.

\section*{Off-resonant scattering}
In the two-color dipole trapping scheme, the blue-detuned light field leads to an increase of the potential energy of the electronic atomic ground state while the red-detuned light field lowers this energy. The trap is formed as the sum of both potentials, and the trap depth is smaller than the magnitudes of the individual shifts. For this reason, the nanofiber-based two-color trap shows a larger off-resonant scattering rate for a given trap depth compared to, e.g., single-color free-space dipole traps. The expected heating rate due off-resonant scattering amounts to roughly $\SI{2}{quanta/s}$ (calculated for the z-axis), much less than what is experimentally observed.
\begin{figure}
	\centering
	\includegraphics[width=0.8\columnwidth]{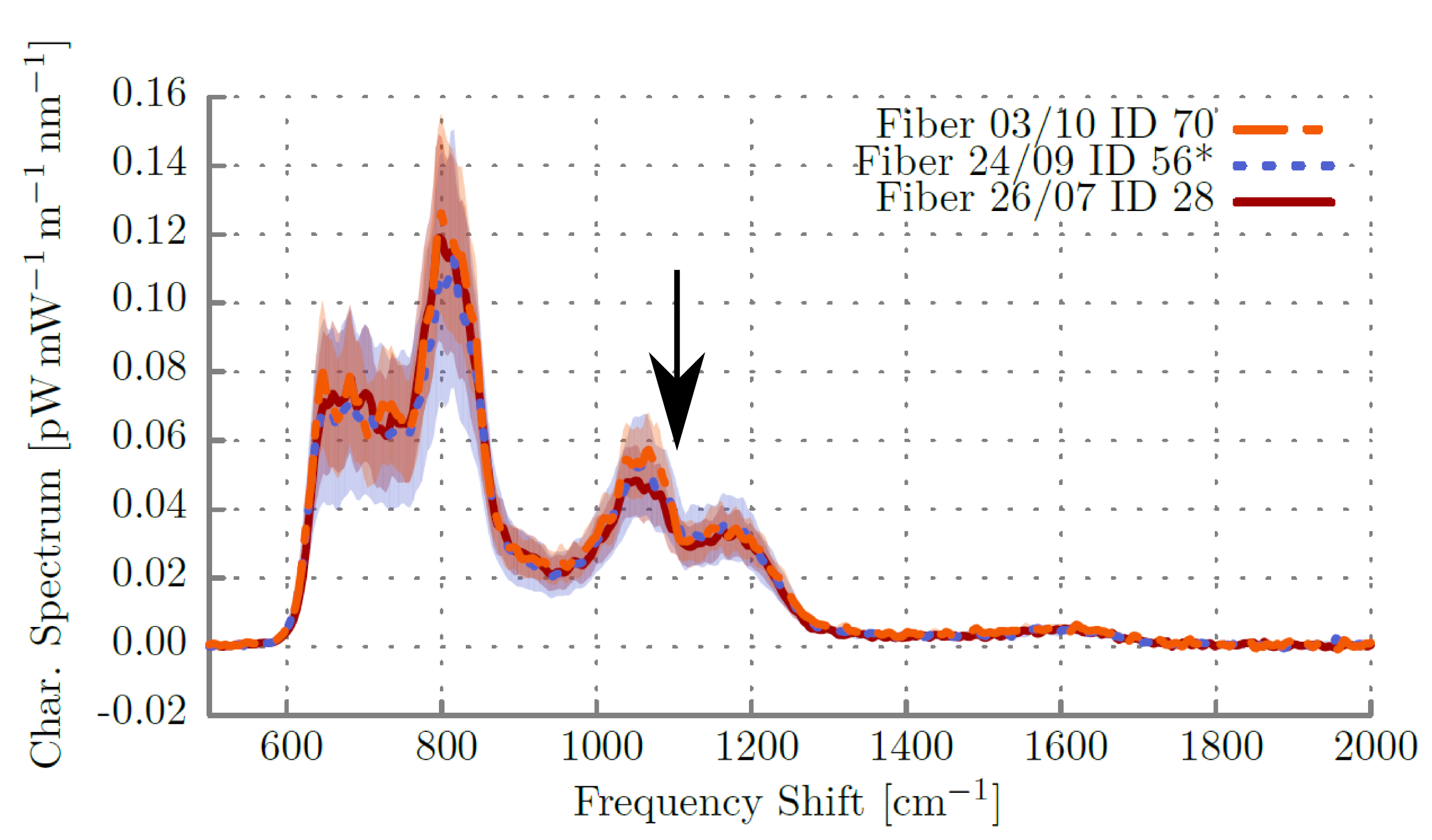}
	\caption{Measured Stokes Raman scattering spectrum of three pieces of Liekki Passive-6-125 optical fiber. The wavelength of the fiber-guided pump light field is $\lambda=\SI{780}{nm}$. The black arrow indicates the Cesium $D_2$ resonance at $\SI{852}{nm}$.}
	\label{fig:Raman}
\end{figure}

\section*{Raman scattering}
The trapping light fields propagating in the tapered optical fiber give rise to Raman scattering in the fiber material, which is fused silica. The contribution of the blue-detuned light field to the relevant wavelength-range of the Raman-scattered light is much larger than that of the red-detuned light field and, therefore, we focus the discussion on the effect of that laser only. We measured the Raman scattering induced by a fiber-guided laser field of $\SI{780}{\nano\meter}$ wavelength and recorded the Stokes-scattered light using a spectrometer. The test fiber was a Liekki Passive-6-125, of the same type as the one used in~\cite{albrecht_fictitious_2016}. Figure~\ref{fig:Raman} shows the measured power spectral density of the Raman signal over the frequency difference from the pump laser frequency for three different pieces of fiber. The signal below $\SI{600}{\centi\meter^{-1}}$ is blocked by two edge filters that are required to sufficiently extinguish the pump light. The $D_2$ line of Cesium corresponds to a Stokes shift of about $\SI{1100}{\centi\meter^{-1}}$ where there is still a sizable Raman signal. However, while the Raman-scattered light is many THz wide, the absorption linewidth of cold Cesium atoms is only a few MHz. In order to get a precise estimate, we measure the Raman-scattered signal induced by the blue-detuned trap light field using a SPCM. The fiber including the tapered section as used in the experiment~\cite{albrecht_fictitious_2016} has a length of about $\SI{5}{\meter}$. All the light exiting the fiber passes a filter that removes the trapping light field. It then passes a narrow (FWHM $\SI{0.12}{\nano\meter}$) bandpass filtering stage centered around the Cesium's $D_2$ line, which isolated the fraction of the Raman-scattered light which is (near-)resonant with the atoms' dominant optical transition. From that and further system parameters, we estimate that a single nanofiber-trapped Cesium atom absorbs about $\SI{50}{photons/\second}$ of Raman light, which yields a heating rate even lower than that arising from off-resonant scattering of trap light fields described above.

\section*{Brillouin scattering}
Guided light fields such as these that generate the optical dipole trap in experiment~\cite{albrecht_fictitious_2016} can experience Brillouin scattering. For this, a phase matching condition between the optical mode and an acoustic mode of the nanofiber has to be fulfilled. For a nanofiber of $\SI{250}{nm}$ radius and a guided light field of $\SI{780}{nm}$ wavelength, the first matching for Brillouin backscattering occurs for an acoustic wave with a frequency of about $\SI{11}{GHz}$ (calculations analogous to \cite{florez_brillouin_2016}. The next resonances occur at higher acoustic frequencies. The spectral width of the Brillouin-scattered light amounts to a few $\SI{10}{MHz}$, i.e., much narrower than for Raman scattering. The Brillouin-scattered light together with the pump light can, in principle, drive stimulated Raman transitions between different internal states of the nanofiber-trapped cold atoms. When the trapping-potential has a dependence on the internal atomic state (as it can be the case for nanofiber-based traps, see \cite{le_kien_state-dependent_2013}, this could lead to heating. The only internal atomic states with a comparable energy separation are the two hyperfine ground-state manifolds of the Cesium atom (HFS splitting: about $\SI{9}{GHz}$). However, the two-photon detuning for this hyperfine state-changing Raman transition is still about $\SI{2}{GHz}$. Such a large detuning, combined with the low power of the Brillouin-scattered light that we estimate to be at the sub-nW level, and given the narrow width of the Brillouin signal, suggests that Brillouin scattering in the nanofiber is irrelevant for the observations in the manuscript. A similar estimation can be made for the standard fiber part. Here, the Brillouin shift amounts to about $\SI{22.3}{GHz}$, i.e., an even larger two-photon detuning. Again, a scattered power in the sub-nW level is found. In the view of these numbers, we consider also Brillouin scattering in the bulk fiber to have a negligible effect on heating.

For the discussion of another mechanism related to Brillouin scattering, we consider the following scenario: Each of the two red-detuned trapping light fields gives rise to a Brillouin scattered (BS) light field that also propagates in the fiber. These two BS fields form an additional standing wave (SW) that has a randomly fluctuating phase with respect to the original trapping SW. This SW formed from BS results in a stochastic modulation a) of the axial trapping frequency and b) of the position of the minima of the trapping potential along the fiber. The former gives rise to parametric heating while the latter leads to resonant heating.

We first consider parametric heating resulting from BS-induced fluctuations of the axial trap frequency, $\omega_z$. When a potential minimum of the trapping SW coincides with a potential minimum of the SW formed from BS, the resulting trap frequency is increased. When the relative phase between the two SWs is such that a potential minimum of the trapping SW falls onto a potential maximum of the SW formed from BS, the resulting trap frequency is decreased. Due to the relatively small detuning of the BS light (about 10~GHz vs. about 300~THz), we can neglect the mismatch between the periodicity of the SW formed from BS and the trapping SW. The rate of parametric heating then depends on the power spectral density (PSD) of the relative phase fluctuations of the two SWs evaluated at twice the axial trap frequency, $2\omega_z$. The fluctuations of the relative phase between the two BS light fields is determined by the individual spectral widths of the two BS fields. The spectrum of BS light for a situation comparable to ours has been published in \cite{beugnot_brillouin_2014}. An approximately Gaussian spectrum with a FWHM of about $\SI{25}{MHz}$ was found. In order to apply the formalism by~\cite{savard_laser-noise-induced_1997} to the calculation of heating rates, we normalize the frequency-integrated PSD of the individual BS light fields to the root-mean square value of the relative intensity noise. For the latter, the ratio of the powers of the trapping and the BS light fields serves as an upper estimate. The PSD of the SW formed from BS then follows from the (appropriately normalized) convolution of the two PSDs of the two BS fields. Using this approach, we compute an exponential increase of the temperature with a time constant $\tau \approx 7 \times 10^7\,\mathrm{s}$. This corresponds to a heating rate for the axial degree of freedom on the order of $\SI{e-8}{quanta/\second}$, i.e., ten orders of magnitude smaller than the heating rate measured for the azimuthal degree of freedom.

We now consider resonant heating that originates from BS-induced position fluctuations of the trapping potential minimum. The heating rate depends on the PSD of the position fluctuations evaluated at the trap frequency. The displacement reaches a maximum value,  $\Delta z_{\rm max}$, when the SW formed from BS is shifted by a quarter of the wavelength of the trapping light with respect to the trapping SW. Again, we assume the PSD of the position fluctuations to have the above-mentioned Gaussian spectrum. We use $\Delta z_{\rm max}$ as a worst-case estimate for the RMS position fluctuation. For our setting, we estimate a maximal displacement of $\Delta z_\mathrm{max} \approx 10^{-4}\,\mathrm{nm}$. This yields a heating rate in the axial direction on the order of $4 \times 10^{-5}\,\mathrm{quanta/s}$. Thus, we exclude BS-induced fluctuations of the axial trap frequency, or of the axial position of the trapping potential minimum, as mechanisms that currently limit the lifetime of atoms in our nanofiber-based trap.

\section*{Blackbody radiation and Johnson--Nyquist noise}
Blackbody radiation and Johnson--Nyquist noise are fundamental processes which have been studied extensively in the context of atom chips. For example, the authors of \cite{henkel_loss_1999} quantitatively estimate the expected heating rate for a spin confined in a harmonic potential with $\SI{100}{kHz}$ trap frequency in close proximity to a material half space. While the surface-induced heating rates can be comparably large for atoms close to a conductive material such as copper (on the order of $\SI{10}{quanta/s}$), the rates are only on the order of $\SI{e-14}{quanta/s}$ for an atom $\SI{200}{nm}$ away from a glass surface, mainly thanks to the low electrical conductivity of glass. Thus, these heating mechanisms are negligible for our experimental conditions.